\documentclass[aps,twoside,reprint,amsmath,amssymb,pre]{revtex4-1}

\usepackage{amsfonts}%
\usepackage{amsmath}%
\usepackage{subfigure}
\usepackage[colorlinks=true]{hyperref}
\usepackage{amssymb}%
\usepackage{graphicx}
\usepackage{color}

\newcommand{\mn}[1]{\left\langle #1 \right\rangle}

\begin{document}

\title{Algorithms for optimized maximum entropy and diagnostic tools for analytic continuation}

\author{Dominic Bergeron$^1$}
\email{dominic.bergeron@usherbrooke.ca}
\author{A.-M. S. Tremblay$^{1,2}$}
\email{andre-marie.tremblay@usherbrooke.ca}
\affiliation{$^1$ D\'epartement de physique, Regroupement Qu\'eb\'ecois sur les Mat\'eriaux de Pointe, Universit\'e de Sherbrooke, Qu\'ebec, Canada \\
$^2$ Quantum Materials Program, Canadian Institute for Advanced Research, Toronto, Ontario,  M5G 1Z8, Canada  }

\pacs{71.10.-w,71.20.-b,02.50.Tt}

\begin{abstract}
Analytic continuation of numerical data obtained in imaginary time or frequency has become an essential part of many branches of quantum computational physics. It is, however, an ill-conditioned procedure and thus a hard numerical problem. The maximum-entropy approach, based on bayesian inference, is the most widely used method to tackle that problem. Although the approach is well established and among the most reliable and efficient ones, useful developments of the method and of its implementation are still possible. In addition, while a few free software implementations are available, a well-documented, optimized, general purpose and user-friendly software dedicated to that specific task is still lacking. Here we analyze all aspects of the implementation that are critical for accuracy and speed, and present a highly optimized approach to maximum-entropy. Original algorithmic and conceptual contributions include (1) numerical approximations that yield a computational complexity that is almost independent of temperature and spectrum shape (including sharp Drude peaks in broad background for example) while ensuring quantitative accuracy of the result whenever precision of the data is sufficient, (2) a robust method of choosing the entropy weight $\alpha$ that follows from a simple consistency condition of the approach and the observation that information- and noise-fitting regimes can be identified clearly from the behavior of $\chi^2$ with respect to $\alpha$,  and (3) several diagnostics to assess the reliability of the result. Benchmarks with test spectral functions of different complexity and an example with an actual physical simulation are presented. Our implementation, which covers most typical cases for fermions, bosons and response functions, is available as an open source, user friendly software.
\end{abstract}

\maketitle

\section{Introduction}
Most calculations in quantum statistical mechanics, as applied in condensed matter physics, nuclear physics, and particle physics, rely on the Matsubara approach. The procedure is in principle well defined: Compute the appropriate correlation function in imaginary time or frequency with a given method, then perform the analytic continuation to obtain physically meaningful results. However, when the calculations are numerical, the latter step is not straightforward, and is in fact an ill-conditioned problem for most worthwhile cases. This means that very high precision data is needed to obtain a reasonable precision in the final result. Therefore, although very simple qualitative information, such as the insulating or conducting character, can be deduced from imaginary time or frequency representation, obtaining quantitative information contained in spectral functions is a difficult task. 

For very accurate data, fitting an analytic interpolating function and substituting the Matsubara frequency with a real frequency can yield good results. Rational polynomials, also called Pad\'e approximants, are usually the best choice for that procedure.\cite{Vidberg_Serene:1977,Deisz:1996, Gunnarsson_Haverkort:2010} However, for noisy data, obtained from quantum Monte Carlo for example, Pad\'e approximants often fail to give sensible results. Even for non-stochastic results, finite precision round-off errors may yield unphysical results.\cite{BeachGooding:2000} Other direct fitting schemes based on known analytical properties of the result can also be used.\cite{HauleYeeFullPotential:2010}

To extract as much information as possible from noisy Matsubara data, one must restrict the space of possible spectral functions by using what is known \textit{a priori} about the exact result. This type of approach, called Bayesian inference, has proven very successful for the analytic continuation problem.\cite{MaximumEntropySilver:1990,quantum_gubernatis_1991,Jarrell:1996,Jarrell_MaxEnt_code} In that context, instead of an interpolating scheme like the Pad\'e approximants approach, the \textit{maximum entropy} (MaxEnt) procedure amounts to a smoothing of the data involving the minimization of a ``free energy'' $Q=\frac{1}{2}\chi^2-\alpha S$, where $\chi^2$ is the quadratic distance between the data and the fit, $S$ is an entropy relative to a default model, and $\alpha$ is an adjustable parameter.\cite{Jarrell:1996} The choice of $\alpha$ is critical since it controls the distance between the fit and the data, which has a very large effect on the resulting spectrum because of the bad conditioning of the problem. A few different approaches to determine $\alpha$ have been used so far. Although some are more reliable, it is not clear which one is the best approach.

Other approaches use \textit{stochastic} sampling over likely spectra for analytic continuation.~\cite{SandvikMaxEnt:1998, MischchenkoProkofevMaxEnt:2000, Beach:2004, VafayiMaxEnt:2007, Fuchs_Pruschke:2010, Gunnarsson_Haverkort:2010} The most widely used stochastic analytic continuation approach (SAC), computes the average spectrum according to the distribution $\exp(-\chi^2/\Theta)$.\cite{SandvikMaxEnt:1998, Beach:2004, VafayiMaxEnt:2007, Fuchs_Pruschke:2010, Gunnarsson_Haverkort:2010,Sandvik:2015} Therefore, $\chi^2$ in this approach is treated as the energy of a fictitious physical system and $\Theta$ as the temperature. SAC can also be formulated using the bayesian principle, \cite{Fuchs_Pruschke:2010} but it does not, in principle, require a default model, or a grid choice~\cite{MischchenkoProkofevMaxEnt:2000}, although one can be introduced to yield the same limits as MaxEnt at large and small $\alpha$.\cite{Beach:2004} However, it has been pointed out recently that the choice of a grid can be equivalent to the choice of a default model.~\cite{Koch:2014}\footnote{See also Appendix \ref{sec:Heuristics} for the link between grid choice and default model in SAC.} As in the MaxEnt approach, there is no unique way of choosing the temperature $\Theta$ in SAC. Moreover, it typically takes several hours to compute a single spectrum with SAC, hundreds of times longer than for MaxEnt.\cite{Fuchs_Pruschke:2010} It is also not clear if it systematically produces improved results over MaxEnt. Indeed, for reasonably precise data, both approaches should yield the same result. Considering how much faster MaxEnt is, it is clearly the best method for these cases. For moderate or low precision data, only qualitative results can be obtained, whatever the approach used. The main question about SAC, which remains to be answered, is therefore whether it is better than MaxEnt to extract all the available information from the data in those cases. Note that recent improvements related to an optimal choice of bounds~\cite{Sandvik:2015} should be considered in MaxEnt as well. 

MaxEnt thus remains the most practical approach available so far for analytical continuation, and although it has been used for more than two decades in statistical physics, we show here that relevant new developments can still be made for the approach itself and its implementation. In particular, we introduce (1) techniques that minimize errors due to numerical approximations and make them negligible compared to typical noise magnitudes, and at the same time yield a computational complexity almost independent of temperature and spectrum complexity, which therefore ensure quantitative accuracy of the result, whenever data precision is sufficient; (2) a different approach to choose the optimal $\alpha$ that produces optimal results under some reasonable conditions, and that follows merely from internal consistency of the MaxEnt approach and from the observation that information- and noise-fitting regimes can be identified clearly when $\chi^2$ is plotted as a function of $\alpha$ on a log-log scale. An analogous idea~\cite{Beach:2004}, or variations~\cite{SandvikMaxEnt:1998}, have been discussed before in the context of SAC and analogies to phase transitions, but here we justify why this procedure works in the context of MaxEnt without invoking phase transitions and we also show how to verify its consistency; (3) graphical diagnostic tools to assess consistency of the above choice of $\alpha$, the quality of the fit and the reliability of the resulting spectrum. 

To summarize, we present specific algorithms to optimize all aspects of the maximum entropy approach, with the aim of extracting as much information as possible from the data, while also keeping computation time low. Those algorithms are implemented in a user friendly software, freely available under the GNU \textit{general public license} (GPL). The software can take as input data a fermionic or a bosonic Green function, correlation function, or self-energy, given as a function of Matsubara frequency or of imaginary time, with diagonal or general covariance matrices. It treats normal Green functions, namely those for which the spectral weight $A$ is positive, namely $A(\omega)>0$ for fermions, while for bosons, $A(\omega)/\omega>0$. \footnote{By defining suitable auxiliary Green functions, cases where the spectral weight is not positive, like that of the Gorkov function in the theory of superconductivity, can also be handled~\protect\cite{Reymbaut:2015}} The code \cite{OmegaMaxEnt_url} was written in C++ using the Armadillo C++ linear algebra library.\cite{Conrad_Sanderson:2010}

The paper is organized as follows. A short review of Bayesian inference and the basis of maximum entropy is presented in Sec. \ref{sec:Bayes_ME}, then our algorithms are summarized in Sec. \ref{sec:algos}, which contains in particular subsections \ref{sec:kernel} on the kernel matrix definition, \ref{sec:alpha} on the choice of the coefficient $\alpha$ of the entropy and \ref{sec:diagn_tools} on what we call ``diagnostic tools'' to check the level of confidence in the results.  The step-by-step procedure is described in Sec. \ref{step_by_step}, which is divided into subsections \ref{sec:preproc} on the extraction of information directly from Matsubara data, \ref{sec:TFGtau} on the case of imaginary-time data, \ref{eq:grids} on the grid and default model definition, \ref{sec:kerneldef} on the kernel, \ref{sec:minimization} on solving the minimization problem, and \ref{sec:optimalalpha} on choosing the optimal $\alpha$. Section \ref{sec:benchmarks} presents benchmarks, an application to a real physical problem, and the diagnostic tools. Finally, the discussion and conclusion are in sections \ref{sec:discussion} and \ref{sec:conclusion}, respectively. Mathematical details on all aspects of the various algorithms are provided in nine appendices.

\section{Bayesian inference and maximum entropy for analytic continuation}\label{sec:Bayes_ME}

Before addressing the algorithms, let us review the basic equations of the maximum entropy (MaxEnt) approach. More details and discussion about those equations can be found in the classic review of Jarrell and Gubernatis, Ref. \onlinecite{Jarrell:1996}. Appendix~\ref{sec:Heuristics} presents an independent heuristic derivation that suggests the connection with stochastic analytic continuation. 

The main equation of the analytic continuation problem for the Matsubara Green function is
\begin{equation}\label{eq:spectr_G_om_n}
G(i\omega_n)=\int d\omega K(i\omega_n,\omega) A(\omega)
\end{equation}
where $G(i\omega_n)$ is the input Green function (or correlation function) known numerically for a certain number of Matsubara frequencies $i\omega_n$, and $A(\omega)$ is the spectral function to be determined. The function $K(i\omega_n,\omega)$, to be defined for a specific problem, is usually called the \textit{kernel}. We assume here that $A(\omega)$ is a positive function. This covers the cases of ``normal'' Green functions, for which the true spectral function $A^*(\omega)$ is positive for fermions, while the condition $A^*(\omega)/\omega>0$ is satisfied for bosons. The Green function may also be known as a function of imaginary time, in which case the Fourier transform of Eq.~\eqref{eq:spectr_G_om_n} may also be used:
\begin{equation}\label{eq:spectr_G_tau}
G(\tau)=\int d\omega K(\tau,\omega) A(\omega).
\end{equation}

The kernels in those expressions are typically very sharp functions with large ratios between largest and smallest values. This makes it impossible to obtain a reasonable accuracy by numerically solving those equations in a standard way, namely by simply discretizing $\omega$ with as many points as there are values of $\tau$ (or of $i\omega_n$) and solving the resulting linear system. That system would be too ill-conditioned. Nevertheless, the problem is not hopeless. Although it remains very difficult, using Bayesian inference makes it tractable.

Consider a Green function $G^*$ related to a spectrum $A^*$ through \eqref{eq:spectr_G_om_n} and another $G$, which is an approximation to $G^*$ known with accuracy $\sigma$. Because of the bad conditioning of the system $G=\mathbf{K}A$ obtained from discretizing \eqref{eq:spectr_G_om_n}, the space of possible spectra $A$ that can produce a function $G'$ within $G\pm \sigma$ contains very different spectra. The main idea behind Bayesian inference is to use known information about the true spectrum $A^*$ other than the constraint $G-\sigma<\mathbf{K}A<G+\sigma$ to restrict that space to only the most probable ones according to that \textit{prior} information. The starting point to do that is Bayes' rule,
\begin{equation}
P(A|G)=\frac{P(G|A)P(A)}{P(G)}.
\end{equation}
$P(A|G)$, the probability that, given the data $G$, we obtain the spectral weight $A$, is called the \textit{posterior probability}; $P(G|A)$, the probability that, given $A$, we obtain the data $G$, is the \textit{likelihood}; and $P(A)$ is the \textit{prior} probability for $A$. For the present case, $P(G)$ is a constant that can be ignored since the data are fixed. The most probable $A$, given the data $G$, is thus obtained by maximizing $P(G|A)P(A)$. If the data are generated using a stochastic method, the likelihood is obtained from the \textit{central limit} theorem. For an element of $G$, we have
\begin{equation}\label{eq:P_Gi}
P(G_i| \bar{G}_i)\propto e^{-\frac{(G_i-\bar{G}_i)^2}{2\sigma_i^2}}\,,
\end{equation}
where $\bar{G}_i$ is the expected value of $G_i$ and $\sigma_i^2$ is the variance. For all the elements of $G$, we thus have
\begin{equation}
P(G|\bar{G})\propto e^{-\frac{\chi^2}{2}}\,,
\end{equation}
where
\begin{equation}
\chi^2=\sum_i \frac{(G_i-\bar{G}_i)^2}{\sigma_i^2}\,.
\end{equation}
If the covariance is not diagonal, one must instead use
\begin{equation}\label{eq:chi2_C}
\chi^2=\left(G-\bar{G}\right)^T \mathbf{C}^{-1}\left(G-\bar{G}\right)\,,
\end{equation}
here in matrix form, where $\mathbf{C}$ is the covariance matrix. Now, given a spectrum $A$, $\bar{G}=\mathbf{K}A$, where $\mathbf{K}$ is the kernel matrix, and thus
\begin{equation}\label{eq:P_G_si_A}
P(G|A)\propto e^{-\frac{\chi^2}{2}}\,,
\end{equation}
with
\begin{equation}\label{eq:chi2_KA}
\chi^2=\left(G-\mathbf{K}A\right)^T \mathbf{C}^{-1}\left(G-\mathbf{K}A\right)\,.
\end{equation}

Now, the basic assumptions when using \eqref{eq:P_G_si_A} as the likelihood are: 
\begin{enumerate}
\item $\bar{G}=\mathbf{K}A$ is a good approximation to the exact Green function. It is therefore a smooth function.
\item The elements of $\Delta G_U=\mathbf{U}^\dagger (G-\bar{G})=\mathbf{U}^\dagger (G-\mathbf{K}A)$, where $\mathbf{U}$ contains the eigenvectors of the covariance matrix $\mathbf{C}$, are \textit{uncorrelated} random variables. 
\end{enumerate}
We come back to those assumptions at end of the present section and in section \ref{sec:alpha}. 

The prior probability $P(A)$ is more difficult to define. Its main property should be to favour the prior information we have about the spectrum, without introducing any correlation not present in that information or in the data.~\cite{Shanon:1949,Jaynes:1957} In the present case, we want to enforce positivity and maybe a few global properties such as normalization and the first two moments of $A$. A form that satisfies those requirements is
\begin{equation}\label{eq:P_A}
P(A)= \frac{e^{\alpha S}}{Z_\alpha^S}\,,
\end{equation}
where
\begin{equation}
Z^S_\alpha=\int \mathcal{D}A\, e^{\alpha S}
\end{equation}
and
\begin{equation}\label{eq:S_diff}
S=-\int \frac{d\omega}{2\pi} A(\omega) \ln \frac{A(\omega)}{D(\omega)}
\end{equation}
is the relative entropy (also known as the Kullback-Leibler divergence), which becomes
\begin{equation}\label{eq:S_sum}
S=-\sum_i \frac{\Delta\omega_i}{2\pi} A(\omega_i) \ln \frac{A(\omega_i)}{D(\omega_i)}
\end{equation}
after discretizing $\omega$.\cite{Gull_Skilling:1984} $D(\omega)$, called the \textit{default model}, is the solution of the optimization problem without any data (for that form, the solution is actually $e^{-1}D(\omega)$), and should thus contain merely the prior information about the spectrum. 

With this definition, the posterior probability becomes
\begin{equation}\label{eq:P_A_si_G}
P(A|G)\propto e^{\alpha S-\frac{\chi^2}{2}}\,.
\end{equation}
This quantity must be maximized to obtain the spectrum $A$, hence the name \textit{maximum entropy}.

We still need to fix the value of $\alpha$. There is however no unique prescription to do that. Three different approaches have mainly been used to this day. Those are the \textit{historic} aproach, the \textit{classic} one,\cite{Skilling1989,Gull1989} and Bryan's approach.\cite{Bryan1990}

In the \textit{historic} approach, $\alpha$ is chosen such that $\chi^2=N$, where $N$ is the number of elements in $G$. This is indeed the expected value of $\chi^2$ if the elements of $G$ are randomly distributed according to \eqref{eq:P_G_si_A}. However, as such, $N$ is only an average and not the value to expect for a single sample $G$. In addition, the covariance matrix is not known exactly in practice, its elements being also random variables, which adds uncertainty to the value of $\chi^2$. The criteria $\chi^2=N$ at the optimal spectrum should therefore be a good choice only if $N$ is large and the covariance is known accurately. Indeed, this approach is very sensitive to bad estimates of the error on the data. For example, if it is overestimated by a factor of 10, $\chi^2$ calculated for the optimal spectrum, obtained with the correct error, will now be reduced by a factor of 100. The approach will also fail if the error is not well estimated for some of the data points, for example if some errors are overestimated in a frequency range with respect to another. In conclusion, the \textit{historic} approach works only in ideal cases.

In the \textit{classic} approach, Bayes inference is used again to find the most probable $\alpha$. Then Eq.~\eqref{eq:P_A_si_G} is maximized with $\alpha$ set to that value.\cite{Skilling1989,Gull1989,Jarrell:1996,JarrellJulichPavarini:2012} In addition to the above, this involves the definition of a prior probability for $\alpha$, $P(\alpha)$. According to Bayes rule, we have
\begin{equation}
\begin{split}
P(A,\alpha | G) &= \frac{P(G| A,\alpha)P(A,\alpha)}{P(G)}\\
&= \frac{P(G | A,\alpha)P(A|\alpha)P(\alpha)}{P(G)}\\
&= \frac{P(G | A)P(A|\alpha)P(\alpha)}{P(G)}\,,
\end{split}
\end{equation}
which, according to \eqref{eq:P_G_si_A} and \eqref{eq:P_A}, becomes
\begin{equation}
P(A,\alpha | G)\propto \frac{1}{Z_\alpha^S}e^{\alpha S-\frac{\chi^2}{2}} P(\alpha)\,.
\end{equation}
Now, by functionally integrating that expression with respect to all possible spectra $A$, one obtains the distribution 
\begin{equation}\label{eq:P_alpha_si_G}
P(\alpha | G) \propto \frac{P(\alpha)}{Z^S_\alpha} \int \mathcal{D}A \;e^{\alpha S-\frac{\chi^2}{2}}\,,
\end{equation}
The value of $\alpha$ is then taken as the most probable one. To find the maximum, one needs a guess for the prior $P(\alpha)$. The most commonly used is $1/\alpha$.

The classic approach assumes that $P(\alpha | G)$ is a narrow distribution centered around the maximum. When this assumption is not valid, the above reasoning leads to Bryan's approach,\cite{Bryan1990} where the spectrum is given by the average spectrum
\begin{equation}
A=\int d\alpha\; A_\alpha\, P(\alpha | G)\,,
\end{equation}
where $A_\alpha$ is the spectrum at the value $\alpha$.

To perform the functional integral in \eqref{eq:P_alpha_si_G}, it is necessary to use a Gaussian approximation for the posterior probability $P(A,\alpha | G)$.\cite{Skilling1989,Gull1989} This approximation is valid when the default model and the spectrum are close to each other and thus $\alpha$ remains large.\footnote{$P(G|A)$ is Gaussian in $A_i$, but Gull and Skilling use $\sqrt{A_i}$ as the integration variables, instead of $A_i$. Thus both $P(A,\alpha)$ and $P(G|A)$ are approximate in the integral.} When this condition is not satisfied, $P(\alpha | G)$ in \eqref{eq:P_alpha_si_G} is peaked at a value of $\alpha$ so small that it leads to overfitting. Therefore, in the classic and Bryan's approaches, the default model must be chosen carefully.\cite{Jarrell:1996}

Our approach to choose $\alpha$ consists in finding the spectrum that satisfies assumptions 1 and 2 given above. These assumptions are implicit when we use \eqref{eq:P_G_si_A} as the likelihood, with $\chi^2$ given by \eqref{eq:chi2_KA}. Since assumption 1 sets a lower bound on $\alpha$ and assumption 2 an upper bound, they define the region where the optimal $\alpha$ is located. Our criterion to choose $\alpha$ is equivalent to assuming that, at the optimal $\alpha$, the spectrum contains as much of the information present in the data as possible, without containing its noise. If one can identify a range of $\alpha$ where essentially only information is fitted and a range where no more information, but only noise is fitted, then we can define the range where the optimal $\alpha$ is located, namely in the crossover between the two regions. Under reasonable conditions, which are discussed below, those ranges of $\alpha$ do exist. The default model is not required to be close to the spectrum in this approach. The procedure to locate the crossover region and choose precisely the value of $\alpha$ within it is described in more details in section \ref{sec:alpha}.

\section{Algorithms}\label{sec:algos}

The following subsections describe the most important aspects of our maximum entropy implementation. More technical details are given in the appendices.

\subsection{The kernel matrix}\label{sec:kernel}

In the present discussion, we assume the thermodynamic limit, namely, that the spectrum $A(\omega)$ is a continuous function.

The kernel matrix $\mathbf{K}$ in $\chi^2$, Eq. \eqref{eq:chi2_KA}, is obtained from a numerical approximation to
\begin{equation}\label{eq:G_omega_n_K}
G(i\omega_n)=\int_{-\infty}^{\infty}\frac{d\omega}{2\pi}\,\frac{A(\omega)}{i\omega_n-\omega}\,\,,
\end{equation}
or 
\begin{equation}\label{eq:G_tau_K}
G(\tau)=-\int \frac{d\omega}{2\pi} \frac{e^{-\omega\tau} A(\omega)}{1\pm e^{-\beta\omega}}\,,
\end{equation}
where
\begin{equation}
\omega_n=
\begin{cases}
(2n+1)\pi T\,,&\quad fermions,\\
2n\pi T\,,&\quad bosons,
\end{cases}
\end{equation}
$n$ is an integer, $T$ is the temperature in the same units as the real frequencies, $\beta=1/T$, and $-\beta<\tau<\beta$ is the imaginary time. Now, the form to use seems at first to depend simply on which type of data is available, $G(i\omega_n)$ or $G(\tau)$. However, assuming that $G(i\omega_n)$ and $G(\tau)$ are equally accessible, the choice between \eqref{eq:G_omega_n_K} and \eqref{eq:G_tau_K} actually has important consequences numerically. Indeed, if we assume a piecewise polynomial approximation for $A(\omega)$ between the discrete points $\omega_j$ where the (trial) spectrum $A$ is defined, then integrating analytically the form \eqref{eq:G_omega_n_K} in each interval yields relatively simple and easy to evaluate expressions (see Appendix \ref{sec:Kernel_matrix_app}). In that case, the kernel $1/(i\omega_n-\omega)$ is treated exactly, and the accuracy of the result only depends on how well the ``true'' spectrum $A(\omega)$ is approximated by the piecewise polynomial. More specifically, it depends on the interpolation scheme --- linear, quadratic, cubic spline, etc --- and on how well the grid resolves the structures in $A(\omega)$. On the other hand, integrating \eqref{eq:G_tau_K} with the piecewise polynomial approximation for $A(\omega)$ produces quite cumbersome hypergeometric functions. The other possibility is to use a full numerical integration method, but then the grid must be adapted to both the spectrum $A(\omega)$ and to the kernel $e^{-\omega\tau}/(1\pm e^{-\beta\omega})$ to obtain a good accuracy for the integral. Since the kernel becomes increasingly sharp around $\omega=0$ as temperature decreases, numerical integration then becomes harder, regardless of the complexity of $A(\omega)$. Therefore, if both $G(i\omega_n)$ or $G(\tau)$ are accessible, it is much more convenient to use \eqref{eq:G_omega_n_K}, in combination with piecewise analytical integration, instead of \eqref{eq:G_tau_K} to compute $\mathbf{K}$. Now, as described in Appendix \ref{sec:TF_Gtau}, if $G(\tau)$ is the input data, $G(i\omega_n)$ can be computed accurately, using cubic splines, if the first two moments of the spectrum are known. These moments can be obtained by two different methods, i.e., by a commutator calculation, or by a polynomial fit to $G(\tau)$ if the step in $\tau$ is small enough. Hence, $G(i\omega_n)$ is relatively easy to obtain.

The accuracy of the approximation 
\begin{equation}\label{eq:Gwn_KA}
G(i\omega_n)\approx K(i\omega_n)A
\end{equation}
to \eqref{eq:G_omega_n_K}, written here in matrix notation with the line label explicit, depends on the choice of interpolation method, and depending on the noise level in the input data, this choice can be more or less critical. In general, the error introduced by using the approximation \eqref{eq:Gwn_KA} must be small compared to the noise in the input data $G_{in}(i\omega_n)$, so that the error on the spectrum $A$ is a consequence of the error on the data, and not of the error resulting from numerical approximations. Indeed, although the use of a maximum entropy approach renders the inversion of the form \eqref{eq:G_omega_n_K} tractable, the spectrum $A$ is still very sensitive to errors in $G$, and the integration error is equivalent to adding a systematic error to the data, which can distort the resulting spectrum. Gaussian quadratures are usually the best choice for numerical integration. However, they cannot be used in the present case because we want to treat exactly the kernel $1/(i\omega_n-\omega)$. We would thus use it as the weight function and would obtain grid points $\omega_i$ and corresponding weights $w_i$ that depend on $\omega_n$. However, the result we are looking for is a vector $A$ defined on a fixed grid. In that case, the best interpolation method for a smooth function is a cubic spline, which is what we use.

Another assumption on the spectrum $A(\omega)$ that is implicit is that it is bounded, namely, it decreases rapidly at high $|\omega|$. For this type of behavior, polynomials in $1/\omega$ are much better approximations than polynomials in $\omega$. For that reason, below a certain frequency $\omega_l$ and above $\omega_r$, we use splines that are cubic in $u=1/(\omega-\omega_\mu)$, where $\omega_\mu$ takes different values on the left and right sides and is determined by the cutoff frequencies. The spline we use to model the spectrum is thus a ``hybrid'' spline. In addition, instead of using a dense grid in $\omega$ in the high frequency regions, we use a grid density that is uniform in $u$, and thus highly non-uniform in $\omega$. For $\omega_l<\omega<\omega_r$, the grid density can be adapted to spectral features, but is uniform in $\omega$ by default. The grid and the spline are thus perfectly adapted to each other, and the combination of hybrid spline and non-uniform grid still yields highly accurate results for the integral \eqref{eq:G_omega_n_K} and also when computing moments of the spectrum (the use of moments is discussed below), although the frequency step $\Delta\omega_i=\omega_{i+1}-\omega_i$ increases rapidly in the high frequency regions, leading to a very sparse grid in those regions. Indeed, for example, when comparing the moments of a given spectrum, for which the moments are known exactly, computed with an ordinary spline and with the hybrid spline with the same non-uniform grid, the moments obtained with the ordinary spline have very low precision, especially high order ones, while the moments obtained with the hybrid spline are closer to the exact results by a few orders of magnitude. As a result, the combination hybrid spline and non-uniform grid greatly reduces the number of frequency grid points, which reduces computation time, while keeping high accuracy. Finally, the computation time can further be reduced by using a grid well adapted to the spectrum in the low frequency region (see section \ref{eq:grids} and Appendix \ref{sec:non_uniform_omega_grid} for more details).


Since the spectrum we are looking for is a function of frequency, it is also more convenient to work with $G(i\omega_n)$, instead of $G(\tau)$. This allows for a straightforward optimization of the Matsubara frequency grid when the number of Matsubara frequencies becomes large. (a) First, the part of the data that behaves as the asymptotic expansion of $G$ (see Eq. \eqref{eq:as_exp_G_app}), say for $\omega_n>\omega_{as}$, typically contains only information about a few moments of the spectrum. Therefore, the terms corresponding to frequencies larger than $\omega_{as}$ in $\chi^2$ can be replaced with a few terms constraining moments. Namely, all the terms of the form $(G_{in}(i\omega_n)-K(i\omega_n)A)/\sigma(\omega_n)$ with $\omega_n>\omega_{as}$ are replaced with a few terms of the form $(M_j-m_jA)^2/\sigma_{M_j}$, where $M_j$ is the $j^{th}$ moment of the true spectrum, $\sigma_{M_j}$ its standard deviations, and $m_j$ is a line vector such that $m_jA$ computes the $j^{th}$ moment of the trial spectrum $A$. In practice, the frequencies $\omega_n>\omega_{as}$ in the vector $G$ are replaced with the $M_j$'s and the lines corresponding to those frequencies in the matrix $\mathbf{K}$, replaced with the line vectors $m_j$'s. This substitution reduces the number of terms in $\chi^2$ essentially without losing information contained in the data. (b) The second optimization of the Matsubara frequency grid follows, first, from the fact that the low frequencies of $A(\omega)$ are related to the low Matsubara frequencies of $G(i\omega_n)$, and similarly for other frequency ranges, although not in a local manner, and second, from the fact that, in systems of interacting particles, quasiparticle lifetime decreases with frequency, and thus the structures in $A(\omega)$ are typically broader as $\omega$ increases. Consequently, most of the time, not all the terms corresponding to Matsubara frequencies below $\omega_{as}$ need to be kept in $\chi^2$, and the frequency grid can actually be made more sparse as $\omega_n$ increases, again while keeping the most relevant information contained in the data. The non-uniform Matsubara frequency grid we use is described in Appendix \ref{sec:non_uniform_Matsubara_grid_def}. Those two optimizations of the Matsuabra frequency grid greatly reduce computation time without noticeable effect on the result. In fact, they modify the scaling with temperature of the number of terms to be included in $\chi^2$, and once they are combined with the piecewise analytical integration of the form \eqref{eq:G_omega_n_K}, and the adapted non-uniform real frequency grid, the computational complexity becomes only weakly dependent on spectrum complexity and temperature.

Note that the spectral moments are mentioned in Ref. \cite{Jarrell:1996}, but only as prior information useful to define the default model. Although, this yields a good default model, if they are not included in $\chi^2$ as well, their importance in the problem then becomes smaller as $\alpha$ decreases. We also use the first and second moments to define a good default model, but because we use them in $\chi^2$ as well, they are treated in the same manner as the rest of the data, therefore as strong constraints on the spectrum. The moments have also been used as constraints in stochastic analytic continuation.\cite{SandvikMaxEnt:1998, Beach:2004,thesis_Fuchs,Fuchs_Pruschke:2010}

To summarize, we use the representation \eqref{eq:G_omega_n_K} with a piecewise polynomial approximation for $A(\omega)$ which allows easy evaluation of the integral by piecewise analytical integration. This operation solves the problem of difficult numerical integration at small $\omega_n$ that worsens as temperature decreases. In addition, we use a hybrid spline with a matching non-uniform real-frequency grid adapted to spectrum structure and a truncated non-uniform Matsubara frequency grid with constraints on moments included as terms in $\chi^2$. The combination of those techniques yields in the end a computation time that is only weakly dependent on temperature and spectrum complexity. Note that the hybrid spline and the piecewise analytical integration have already been used in a previous maximum entropy implementation, although only for the optical conductivity. Also, the frequency grid was less sophisticated and the other aspects described in the following sections where not present. This implementation is described in Appendix F of Ref. \cite{Bergeron:2011}.

The kernel matrix definition is described in details in Appendix \ref{sec:Kernel_matrix_app}. More discussion on the grid definition appears in section \ref{eq:grids} and details are given in appendices \ref{sec:non_uniform_omega_grid} and \ref{sec:non_uniform_Matsubara_grid_def}. Also, some preliminary steps are necessary before defining the grid and computing the kernel matrix. Those steps are discussed in section \ref{sec:preproc}.

\subsection{The optimal $\alpha$}\label{sec:alpha}

\begin{figure}
\includegraphics[width=0.9\columnwidth]{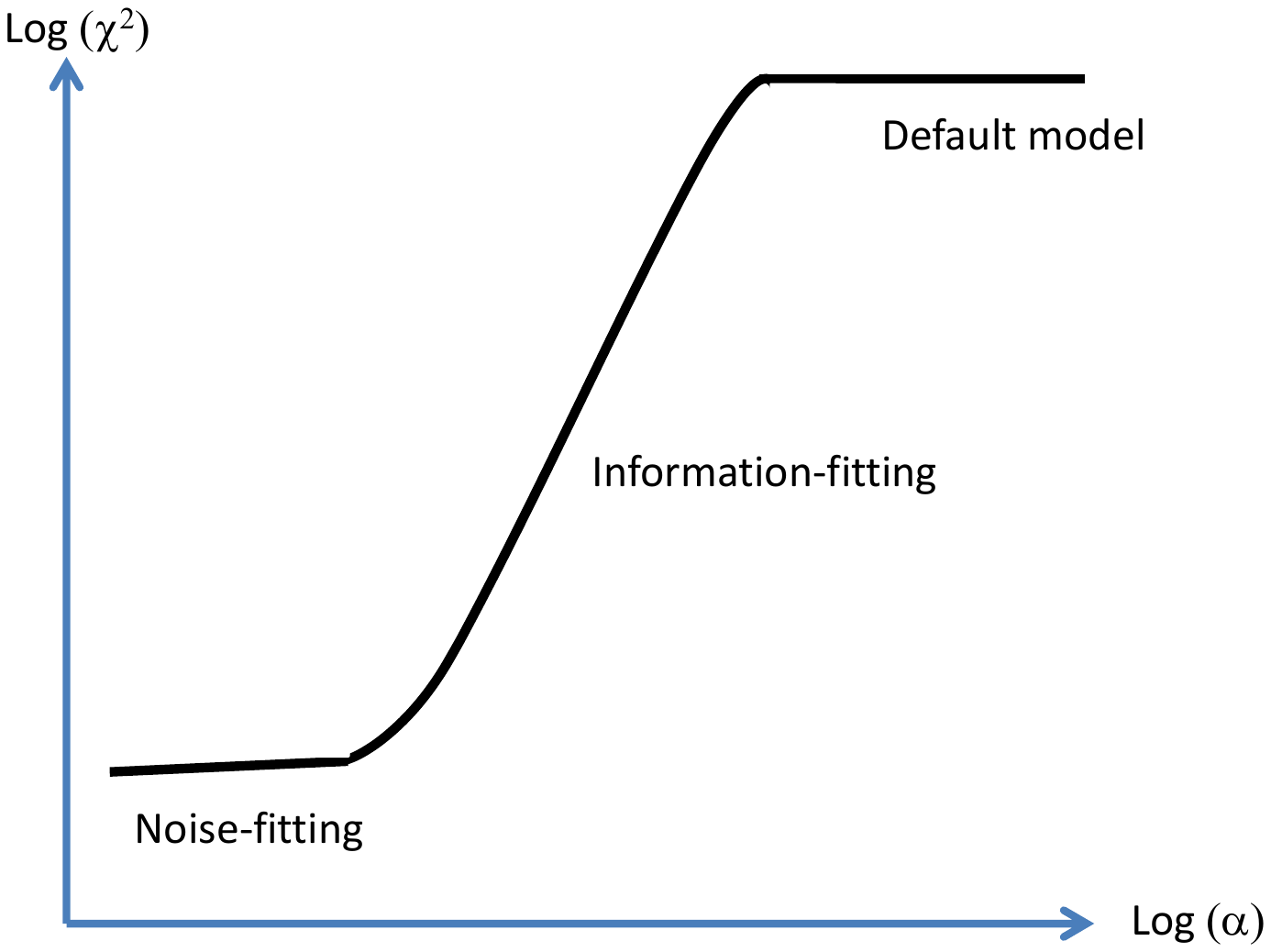}	
	\caption{Schematic behavior of $\log(\chi^2)$ vs $\log(\alpha)$. Three regimes can be identified.}
	\label{Fig0}
\end{figure}

Let us assume for now that the spectrum $A(\omega)$ has been computed for a wide range of $\alpha$, starting at large values, where the spectrum minimizing $\chi^2/2-\alpha S$ is essentially the default model $D(\omega)$ (which maximizes $S$ alone), to small values, where the term $\chi^2$ dominates. If we plot the function $\chi^2(\alpha)$ in log-log scale, we obtain a shape similar to the schematic curve of Fig.~\ref{Fig0}, where three different regimes are found: At large $\alpha$, $\chi^2$ does not change for a certain range of $\alpha$ where it is negligible compared to $\alpha S$, and thus the spectrum barely departs from the default model. Let us call that region the \textit{default-model} regime. Then, below a certain value of $\alpha$, while $\alpha S$ is still the largest term, $\chi^2$ is not negligible anymore and plays the role of a constraint. In that range, reducing $\alpha$ has a strong effect on $\chi^2$, as it directly controls how well that constraint is satisfied. We call that range of $\alpha$ the \textit{information-fitting} regime. Then, below some value of $\alpha$, the effect of decreasing that parameter has a much smaller effect on $\chi^2$. This is in fact the region where the noise in $G$ is also being fitted, which we call the \textit{noise-fitting} regime. 

The correspondence between the small slope region at low $\alpha$ and the range where the noise is being fitted can be made easily by looking at the function $\Delta G(i\omega_n)= G_{in}(i\omega_n)-K(i\omega_n)A$ (diagonal covariance case) at different values of $\alpha$. Above the value of $\alpha$ where the slope starts decreasing rapidly, $\Delta G(i\omega_n)$ has correlations between frequencies $i\omega_n$, while below that value, that function contains mainly noise. This indicates that, at the point of the drop, $G_{out}(i\omega_n)=K(i\omega_n)A$ is a good fit to the data $G_{in}(i\omega_n)$, but because $G_{out}(i\omega_n)$ does not yet contain noise, $\Delta G$ is essentially the noise in $G_{in}(i\omega_n)$. This will become more clear in section \ref{sec:benchmarks}.

Although the global behavior of $\chi^2(\alpha)$ in the default-model regime and the information-fitting one is intuitive, why the slope is much lower in the noise-fitting region is more subtle. A very small decreasing rate of $\chi^2$ with decreasing $\alpha$ is expected at very small $\alpha$, deep in the noise fitting region, when the term $\chi^2$ dominates and the entropy term mainly prevents $A(\omega)$ from becoming negative where it is close to zero. However, in the range of $\alpha$ where the slope decreases rapidly with decreasing $\alpha$, the entropy term is still in fact much larger than $\chi^2$, and $A$ and $G_{out}=\mathbf{K}A$ are smooth functions, namely, no noise from the data has been fitted yet. To understand the drop in the decreasing rate of $\chi^2$ with $\alpha$, expression \eqref{eq:delta_A_j},
\begin{multline}\label{eq:delta_A_j_main}
\delta A_{j}=\Delta\alpha_j\left[\tilde{\mathbf{K}}^T \tilde{\mathbf{K}}+\alpha_j\mathbf{\Delta\omega}\mathbf{A}_{j-1}^{-1}\right]^{-1}\\
\times\Big(\mathbf{\Delta\omega} \ln \left(\mathbf{D}^{-1}A_{j-1}\right)+\Delta\omega\Big)\,.
\end{multline}
from Appendix \ref{sec:alpha_opt} is very useful. Here, $\mathbf{\Delta\omega}$, $\mathbf{D}$, and $\mathbf{A}$ are the diagonal matrices with $\Delta\omega$, $D$ and $A$ as their diagonal, respectively. The spectrum at $\alpha_j=\alpha_{j-1}-\Delta\alpha_j$ is given by $A_j=A_{j-1}+\delta A_j$, assuming that the spectrum $A_{j-1}$ at $\alpha_{j-1}$ is known and that $\delta A_j$ is small. An important point about expression \eqref{eq:delta_A_j_main} is that it does not explicitly contain the data $G_{in}$, or $\Delta G=G_{in}-\mathbf{K}A$, which is essentially the noise in the data when most of the available information is contained in $A$. Instead, $\delta A$ is a smooth function at the crossover between the information- and noise-fitting regions since $A_{j-1}$ is smooth at that point, and therefore $\delta G=\mathbf{K}\delta A$ is necessarily also smooth and cannot reduce by much the amplitude in the noisy function $\Delta G$. To summarize, expression \eqref{eq:delta_A_j_main} tells us that, not only does the entropy term in $\chi^2/2-\alpha S$ controls the distance $\Delta G$, it also enforces smoothness on $A$, which prevents easily fitting the noise.



This behavior in $\chi^2$ as function of $\alpha$ is also observed in stochastic analytic continuation (SAC). In that context, where $\chi^2$ is the energy of the system, the drop in the rate of change in $\chi^2$ as a function of temperature corresponds to a drop in the specific heat, and thus to a phase transition.\cite{Beach:2004} Although the entropy does not appear explicitly in SAC, the statistical treatment in the canonical ensemble implies that it is maximal, given the value of the average ``energy'' $\mn{\chi^2}$. The drop in $d\chi^2/d\alpha$ observed in both approaches must therefore have a similar origin (the statistical entropy in SAC is equal to $\ln{Z}+\langle\chi^2\rangle/2\Theta$, where the partition function $Z=\text{Tr}[\exp(-\chi^2/2\Theta)]$ is a trace over all configurations $\{A(\omega_i)\Delta\omega_i\}$). The phase transition analogy of Beach~\cite{Beach:2004} that we just discussed can also be seen in the sudden entropy drop discussed by Sandvik~\cite{SandvikMaxEnt:1998}.

Now, because we want the spectrum to contain the information in $G$, but not its noise, it is clear that the optimal $\alpha$ is somewhere in the crossover region between the information- and the noise-fitting regimes. This is also where the basic assumptions for the likelihood, assumptions 1 and 2 in section \ref{sec:Bayes_ME}, become valid, and therefore where the maximum-entropy approach is consistent.

The crossover region is well delimited if the covariance matrix $\mathbf{C}$ is a good estimate of the actual one. This is not so on the other hand when the covariance is not well estimated in some frequency ranges. To illustrate this, let us first consider a diagonal covariance for simplicity. Then it is the standard deviation $\sigma$ that controls the rate at which $G_{out}$ becomes close to $G_{in}$ as $\alpha$ decreases. When all the information contained in $G_{in}$ is in the spectrum $A$,  but not its noise, the function $\Delta G(i\omega_n)=G_{in}(i\omega_n)-G_{out}(i\omega_n)= G_{in}(i\omega_n)-K(i\omega_n)A$  contains mostly noise. Indeed, if $G_{out}$ is smooth and is a good fit to $G_{in}$, which is the basic assumption here, then $\Delta G (i\omega_n)$ is approximately the noise in $G_{in}$. Now, if $\sigma(i\omega_n)$ is well estimated, $\Delta G(i\omega_n)$ becomes noisy around the same value of $\alpha$ for all values of $i\omega_n$. However, assume that $\sigma$ is underestimated in a given frequency range, then the corresponding $\Delta G(i\omega_n)$ becomes noisy at a larger $\alpha$ in that range of frequencies. Then, at the value of $\alpha$ where $\Delta G(i\omega_n)$ becomes noisy in the other frequencies ranges, the frequency range where $\sigma$ was underestimated will be overfitted. Therefore, when the ratio of the $\sigma(i\omega_n)$ between different frequency ranges is incorrect, some noise is fitted in some frequency ranges while some information has not yet been fitted in others. Large errors in $\sigma$ therefore reduce the range of $\alpha$ where only information is fitted, and create a large crossover region, where both noise and information are fitted at the same time. On the other hand, if $\sigma$ is a good estimate, the crossover region is narrow. When the covariance is not diagonal, the analysis is the same, but we have to consider $\sigma$ and $\Delta G$ expressed in the eigenbasis of the covariance $\mathbf{C}$.

Because of the rapid change of slope in $\log\chi^2$ as a function of $\log\alpha$ at the crossover between the information-fitting and noise-fitting regimes, there is a peak in the curvature of that function at the crossover. We choose the optimal $\alpha$, say $\alpha^*$, at the maximum curvature, and we look at the variation of the spectrum within the width of the peak to obtain an estimate of the uncertainty of the spectrum. No observable change to the spectrum within the crossover region points toward good quantitative accuracy of the results. Other diagnostic quantities, discussed in section \ref{sec:diagn_tools}, can be used to analyze systematically the results to assess the quality of the fit and the accuracy of the resulting spectrum.

Unlike in the classic and Bryan methods, the reliability of this approach to choose $\alpha$ does not depend on the proximity between the spectrum and the default model. Indeed, it yields the correct value for the optimal $\alpha$ even when the spectrum is extremely different from the default model, as illustrated by the second example of section \ref{sec:benchmarks} (Fig. \ref{fig:example_2}). This is an important advantage over the conventional methods since very little prior information is needed on the spectrum to obtain good results.

In the classic or Bryan's methods, the following procedure is often used to obtain a default model that is close to the final result: (a) One obtains data at high temperature, where the spectrum is simple and the analytic continuation easy. (b) The result for the spectrum is then used as the default model for a slightly lower temperature. (c) This annealing process is pursued until the lowest temperature.\cite{JarrellJulichPavarini:2012} We could also use this annealing procedure, but it is not necessary in our case. Note that a variation of $\alpha$ is different from annealing since the data and default model are not changed as $\alpha$ is varied.

The choice of $\alpha$ at the point where the slope of $\log\chi^2$ as a function of $\log\alpha$ drops has been used before with the stochastic approach.\cite{Beach:2004,Fuchs_Pruschke:2010} The justification for this choice was not clear however. The analysis in the context of the maximum entropy approach clarifies why $\alpha$ should be chosen in that region. 

To apply this approach to find the best $\alpha$, the spectrum must be computed typically for a few hundred values of $\alpha$. We thus need an efficient method to minimize $\chi^2/2-\alpha S$. This part of the calculation is discussed briefly in section \ref{sec:minimization} and details are given in Appendix \ref{sec:minimize_Q}.


\subsection{Diagnostic tools}\label{sec:diagn_tools}

As mentioned above, $G_{in}(i\omega_n)-G_{out}(i\omega_n)$ can be used to determine when most of the information has been fitted. Now, assuming again a diagonal covariance matrix, a more convenient quantity to analyze is the normalized function $\Delta \tilde{G} (i\omega_n)=[G_{in}(i\omega_n)-G_{out}(i\omega_n)]/\sigma(i\omega_n)$. Indeed, if $\sigma$ is accurate, $\Delta \tilde{G} (i\omega_n)$ should contain only noise of constant amplitude at $\alpha^*$, with a variance around unity. Its autocorrelation $\Delta \tilde{G}^2(\Delta n)\equiv (1/N)\sum_n\Delta \tilde{G}(\omega_n)\Delta \tilde{G}(\omega_{n+\Delta n})$ must therefore look like a Kronecker delta, although noisy because $N$ is finite. However, once the optimal $\alpha$ is reached, the shape of $\Delta \tilde{G}(i\omega_n)$ and $\Delta \tilde{G}^2(\Delta n)$ remains qualitatively similar at smaller $\alpha$, and their amplitude decreases very slowly. By contrast, at large values of $\alpha$, $\Delta \tilde{G}(i\omega_n)$ is a smooth function of $\omega_n$, since $G_{out}(i\omega_n)$ is smooth and farther away from $G_{in}(i\omega_n)$ than $\sigma(i\omega_n)$, and $\Delta \tilde{G}^2(\Delta n)$ is broad since long range correlations are present in $\Delta \tilde{G}(i\omega_n)$. Therefore, those two functions provide a way to determine whether $\alpha>\alpha^*$ or $\alpha\leq\alpha^*$, i.e. whether the optimal $\alpha$ has been reached.

The function $\Delta \tilde{G} (i\omega_n)$ (or $\Delta \tilde{G}_i$ where $i$ is a covariance matrix eigenvector index), is also very useful to tell whether the error estimate is good. Indeed, if it is the case, at values of $\alpha$ where $\Delta \tilde{G} (i\omega_n)$ is noisy, the noise amplitude will be comparable at all frequencies. If some ranges of frequencies in $\Delta \tilde{G} (i\omega_n)$ are noisy while others still contain correlations, this means that the ratio of the errors between those frequency ranges is off. In such cases, different frequency regions in $A(\omega)$ will ``converge'' at different values of $\alpha$, an undesirable state of affairs.

If $\mathbf{C}$ is not diagonal, the functions $\Delta\tilde{G}$ and $\Delta \tilde{G}^2$ to be analyzed must be expressed in the eigenbasis of the covariance matrix $\mathbf{C}$. Therefore, in that case we use $\Delta\tilde{G}=\tilde{G}_{in}-\tilde{G}_{out}$, where $\tilde{G}=\sqrt{\tilde{\mathbf{C}}^{-1}}\mathbf{U}^\dagger G$ and $\mathbf{U}$ contains the eigenvectors of $\mathbf{C}$, such that $\tilde{\mathbf{C}}=\mathbf{U}^\dagger \mathbf{C} \mathbf{U}$ is diagonal.

Finally, other quantities that are useful to estimate the quality of the analytic continuation are the curves of $A(\omega)$ at a few sample frequencies $\omega_i^{samp}$ as a function of $\log\alpha$. As we will see in specific examples treated in section \ref{sec:benchmarks}, if the quality of the data is sufficient, those curves have plateaus in a range of $\alpha$ around $\alpha^*$. If precision is not sufficient to observe plateaus, inflexion points or extrema will typically still be present in the curves around $\alpha^*$. In other words, there are local minima in $|dA(\omega)/d\log\alpha|$ around $\alpha^*$. Then, at a certain value of $\alpha$ below $\alpha^*$, the sample spectra $A(\omega_i^{samp})$ have unpredictable behavior, increasing or decreasing quickly, as the system's conditioning worsens. The $A(\omega_i^{samp})$ as a function of $\log\alpha$ curves also tell if the error is well estimated. If it is, the $A(\omega_i^{samp})$ at the different sample frequencies have the same optimal $\alpha$. Otherwise, if the ratio of the errors in different frequency ranges is off, the local minima in $|dA(\omega_i^{samp})/d\log\alpha|$ above the onset of unpredictable behavior will appear at different values of $\alpha$ for different frequencies. In those cases, and if improving the estimate of the covariance is impossible, the curves of $A(\omega_i^{samp})$ may be very useful to select the optimal $\alpha$ from a specific real frequency range instead of the average value obtained from the curvature of $\log\chi^2$ as a function of $\log\alpha$. This may also be preferable to averaging the spectrum over a range of $\alpha$, since there is less risk of ``contaminating'' the spectrum in the chosen frequency range with underfitted or overfitted spectra.

To summarize, in addition to the function $\chi^2(\alpha)$, three other quantities give information about the optimal value of $\alpha$ and more: $\Delta \tilde{G}(i\omega_n)$,  $\Delta \tilde{G}^2(\Delta n)$, and $A(\omega_i^{samp})$ versus $\log\alpha$. Those quantities also allow to evaluate the quality of the fit and the reliability of the result. Practical examples of this analysis are given in section \ref{sec:benchmarks}.

\section{Step-by-step procedure}\label{step_by_step}

Here we discuss briefly the steps in our maximum entropy implementation. The technical details are left in appendices.

\subsection{Extracting information directly from the input Green function}\label{sec:preproc}


Under certain conditions, we can obtain relatively easily two types of information directly from the Matsubara data: moments and a low frequency peak width and weight if present. This information is then used throughout the preprocessing stage, allowing important optimizations, some of which have been discussed in section \ref{sec:kernel}. The first step in the procedure is therefore to extract that information, when available.

\subsubsection{Extracting moments}

If the moments are not known, and if there are enough Matsubara frequencies, or imaginary time slices, such that the information on the moments is actually present in the data, then the moments can be extracted from the Matsubara data. At least the first two moments are generally well defined. The procedure to extract the moments basically involve fitting a Matsubara frequency, or imaginary time, asymptotic expansion to $G(i\omega_n)$, or $G(\tau)$, respectively, using a least-squares fitting method. The details for both cases are given in Appendix \ref{eq:extract_moments_app}. In the software, the moments can also be provided by the user. Note that, for a given model, the expressions for the first few moments as functions of Hamiltonian parameters can be obtained by straightforward commutator calculations. This is documented in standard many-body textbooks or here \cite{thesis_Blumer}. We do not discuss this calculation here, but only consider the analytical continuation procedure itself, starting from some numerical Matsubara data that can be represented by expression \eqref{eq:G_omega_n_K} or \eqref{eq:G_tau_K}, and useful additional data, such as a covariance matrix and spectral moments. The approach we use here is meant to apply in the most general case possible, and not only to specific Hamiltonians.

\subsubsection{Finding the width and weight of a quasiparticle or Drude peak}\label{sec:qp_peak}

Let us assume we know $G_{in}(i\omega_n)$ either directly, or from the Fourier transform of $G_{in}(\tau)$ (see Sec. \ref{sec:TF_Gtau} below). It can be very useful to be able to tell directly from $G_{in}(i\omega_n)$ if the system is metallic and to have an estimate of the width and the weight of the quasiparticle peak, or of the Drude peak in the case of optical conductivity. This information can then be used to determine the optimal frequency step around $\omega=0$.

When there is a well-defined peak near $\omega=0$, with weight essentially below a frequency $W_{qp}$ well separated from the rest of the spectrum, whose weight we assume is mostly above a frequency $W_{inc}$, then the Green function in the frequency range  $W_{qp}<\omega_n<W_{inc}$ can be expanded in a Laurent series
 \begin{multline}\label{eq:G_laurent_main}
G_{LS}(i\omega_n)\approx - (i\omega_n)^{L-1}M_{-L}^{inc}-\ldots- (i\omega_n)^2M_{-3}^{inc}\\
- (i\omega_n)M_{-2}^{inc}-M_{-1}^{inc}+\frac{M_0^{qp}}{i\omega_n}+\frac{M_1^{qp}}{(i\omega_n)^2}+\ldots+\frac{M_N^{qp}}{(i\omega_n)^{N+1}}\,.
\end{multline} 
In the above expression, the $M_{-j}^{inc}$s are the inverse moments of the high frequency part of the spectrum, and the $M_j^{qp}$s are moments of the quasiparticle peak. This expression is for a fermionic Green function. Those partial moments can be obtained by the same fitting procedure as the one used to extract the moments of the complete spectrum. The quasiparticle peak weight is then $M_0^{qp}$, its position is given by $M_1^{qp}/M_0^{qp}$, and its width is $\sqrt{M_2^{qp}/M_0^{qp}-(M_1^{qp}/M_0^{qp})^2}$. This procedure will give accurate values if the number of Matsubara frequencies satisfying $W_{qp}<\omega_n<W_{inc}$ is sufficient to ensure convergence of the partial moments. If that condition is not satisfied, it will nevertheless produce the right orders of magnitude. Knowing the width of the low frequency peak is useful to define the frequency step around $\omega=0$, and thus to define a frequency grid adapted to the spectrum's sharpest part. More details are given in Appendix \ref{sec:qp_peak_width}, with the discussion for the bosonic case.

\subsection{$G(\tau)$ as input data}\label{sec:TFGtau}

As mentioned in section \ref{sec:kernel}, we use the representation \eqref{eq:G_omega_n_K} as the relation between $G$ and $A$ during the calculation. Therefore, when the input data are given as a function of imaginary time $G_{in}(\tau)$, we need to Fourier transform it to obtain $G_{in}(i\omega_n)$. To do so, knowing some spectral moments becomes essential. Indeed, to obtain a $G_{in}(i\omega_n)$ as accurate as possible, i.e. without adding large errors, the best approach is to Fourier transform the cubic spline of $G_{in}(\tau)$, using the first and second moments to define the two additional equations required to define the spline. Using splines and partial integration for the transform enforces~\cite{Georges:1996,Bergeron:2011} the required~\cite{BaymMermin:1961} asymptotic behavior of $G_{in}(i\omega_n)$, and also produces more accurate results at low frequency than a discrete Fourier transform. The Fourier transforms of the data and the covariance matrix will add some noise, but the use of fast Fourier transforms minimizes that noise by minimizing the number of operations in the transforms. More details on this procedure, and how to obtain the covariance matrix of $G_{in}(i\omega_n)$ from the covariance of $G_{in}(\tau)$, are described in Appendix \ref{sec:TF_Gtau}. Once the moments, $G_{in}(i\omega_n)$, and its covariance matrix are computed, $G_{in}(\tau)$ is not used anymore in the calculation.

\subsection{Frequency grid and default model}\label{eq:grids}

Optimizing the real frequency and the Matsubara grids is critical for the efficiency of the maximum entropy calculation. As shown in Appendix \ref{sec:minimize_Q}, each iteration in the minimization of $Q=\frac{1}{2}\chi^2-\alpha S$ requires $O(N_{\omega_n}^2N_\omega)$ operations (for $N_{\omega_n}<N_\omega$). The number of values of $\alpha$ that need to be computed is typically a few hundreds. In addition, the grid should allow a good resolution of all the structures in the spectrum. Optimizing both the Matsubara and real frequency grids is thus necessary to keep computation time reasonable without affecting the quality of the result. 

To define a grid adapted to the spectrum, while minimizing the number of points, we use a dense grid in the main spectral range $[\omega_{l},\omega_{r}]$, where most of spectral weight is found, and a step size that increases with $|\omega|$ outside that range. By default, in the software, the step $\Delta\omega$ in the main spectral range is constant and defined by setting $(\omega_{r}-\omega_{l})/\Delta\omega$ to a preset value, which can be modified by the user.  At high frequency, the grid has a constant $\Delta u$ increment where
\begin{equation}\label{eq:u_grid}
u=
\begin{cases}
\frac{1}{\omega-\omega_{0r}}\,, & \omega> \omega_{r}\\
\frac{1}{\omega-\omega_{0l}}\,, & \omega< \omega_{l}\,.
\end{cases}
\end{equation}
The parameters $\omega_{0l}$ and $\omega_{0r}$ are determined by the high frequency cutoff, which can be modified by the user (see Appendix \ref{sec:non_uniform_omega_grid} for more details). The frequencies $\omega_{l}$ and $\omega_{r}$ that define the main spectral region are set by default to $M_1-\Delta\omega_{std}$ and $M_1+\Delta\omega_{std}$, respectively, where $\Delta\omega_{std}=\sqrt{M_2/M_0-(M_1/M_0)^2}$ is the standard deviation of the spectrum. This choice of grid at high frequency is the most natural one considering the type of interpolation method used to define the kernel matrix $\mathbf{K}$, and discussed in section \ref{sec:kernel}. It also efficiently reduces the number of points in the grid, which helps keep the computation time short. The grid in $u$ is defined such that the step in $\omega$ is continuous at the boundaries of the main spectral region. This is actually important to obtain smooth spectral functions.  Further details on the grid definition are provided in Appendix~\ref{sec:non_uniform_omega_grid}.

The software offers a few possibilities to define a more adapted grid in the main spectral range. The simplest choice given to the user is to choose the boundaries and the step $\Delta\omega$ in that range. There is also a tool to define a non-uniform grid with a smoothly varying step between intervals of different step sizes provided by the user in the form of a vector
\begin{equation}\label{Eq:Grid}
\begin{bmatrix}
\omega_1 & \Delta\omega_1 & \omega_2 & \Delta\omega_2 & \omega_3 & \ldots & \Delta\omega_{N-1} & \omega_N
\end{bmatrix},
\end{equation}
where the frequencies $\omega_i$ define the intervals' boundaries and $\Delta\omega_i$ the corresponding step size in each interval. Given this input by the user, the frequency step is defined to follow a hyperbolic tangent shape around the boundaries. An example of the resulting frequency grid density $1/\Delta\omega$ as a function of $\omega$ corresponding to three intervals of different step sizes is illustrated in Fig.\ref{fig:Delta_omega} of Appendix \ref{sec:non_uniform_omega_grid}. The reason for forcing a smooth variation of the grid between different intervals is that discontinuities in $\Delta\omega$ can cause spurious oscillations in the results. Finally, the user can also provide a customized grid for the main spectral region. The grid outside the main spectral range is always defined in the same way.

If a quasiparticle peak has been found and characterized using the Laurent series fit to $G(i\omega_n)$ described in section \ref{sec:qp_peak}, then the step $\Delta\omega$ around $\omega=0$ is set to a fraction of the width of the peak and a vector of the type \eqref{Eq:Grid} is generated automatically to define a non-uniform grid with a small number of points for the main spectral region. This feature allows the possibility of automatically computing the spectra for large sets of data.

As for the Matsubara frequency grid, if the moments are used as constraints, the Matsubara frequencies corresponding to the asymptotic part can be removed. This is the simplest modification of the Matsubara grid that can improve computational efficiency without losing accuracy. In addition, when the number of Matsubara frequencies below the cutoff becomes large at low temperature, a non-uniform grid can be used to further reduce the number of frequencies. This is because sharp features in the spectrum are usually located at low frequency, while the spectrum broadens as $|\omega|$ increases. Therefore, as the magnitude of $\omega_n$ increases, the grid can become sparser. The grid we use is given in Appendix \ref{sec:non_uniform_Matsubara_grid_def}.

The default model should contain the information known about the spectrum, but should otherwise be featureless, to ensure that any structure appearing in the spectrum comes necessarily from the data. In the program, the default model is defined as a Gaussian with the same first two moments as the spectrum, if they have either been extracted from the high frequencies of the data or else provided by the user. Otherwise, if the grid in the main spectral region is defined by the user, then the width and center of that region are used as the width and center of the default model, respectively. Other default models, such as a flat spectrum, are available and can be selected by the user. Finally, the default model can also be entirely provided by the user.

\subsection{Defining the kernel matrix}\label{sec:kerneldef}

The kernel matrix can be defined once the real frequency and Matsubara grids are defined. The general approach was discussed in section \ref{sec:kernel}. In practice, defining $\mathbf{K}$ with the hybrid spline interpolation method described in section \ref{sec:kernel} involves analytically integrating \eqref{eq:G_omega_n_K} in the case where $A(\omega)$ is a cubic polynomial in $\omega$ and the case where it is a cubic polynomial in $u$, defined by Eq.\eqref{eq:u_grid}. This yields a linear relation between the vector $G$ and a vector formed by the spline coefficients. Then, the linear relation between those coefficients and the vector $A$ are obtained by computing the general solutions for the linear systems of equations defining the splines in the different regions of the real frequency grid (low and high frequency regions). The details are given in Appendix \ref{sec:Kernel_matrix_app}.

\subsection{Computing the spectrum as a function of $\alpha$}\label{sec:minimization}

To obtain the spectrum as a function of $\alpha$ in the relevant range, we have to minimize $Q=\frac{\chi^2}{2}-\alpha S$, where $\chi^2$ is given by \eqref{eq:chi2_KA} and $S$, by \eqref{eq:S_sum}.
This function is bounded from below and is convex. It therefore has a unique minimum that is the solution to $\nabla Q=0$. Although that equation cannot be solved exactly because the entropy part is non-linear, we know that, for large $\alpha$, the solution must be close to the one minimizing the entropy term only, namely $A=e^{-1}D$. In addition, if the solution $A_\alpha$ at a given $\alpha$ is known, then the solution for a new value of $\alpha$, $A_{\alpha'}$, can be obtained by Newton's method, provided that $A_{\alpha'}$ is close to $A_\alpha$. In other words, in that case we can solve iteratively a linearized version of the system $\nabla Q=0$.

Therefore, the solutions for the whole $\alpha$ range can be obtained by starting from a large value of $\alpha$, using $e^{-1}D$ as the initial spectrum, and decreasing $\alpha$ at a rate such that $|A_{\alpha}-A_{\alpha'}|/A_{\alpha}$ is small, so that the minimization routine converges. The starting value of $\alpha$ is chosen to ensure that the first solution is very close to the default model. The calculation stops when the slope in $\log\chi^2$ as a function of $\log\alpha$ is smaller than a certain fraction of its maximum value in the whole $\alpha$ range (see the last two paragraphs of Appendix \ref{sec:minimize_Q}).

There are a few obstacles to overcome when solving the linearized system $\nabla Q=0$. On one hand, the system is ill-conditioned and, on the other hand, negative values of $A$ can appear because of linearization, even though the entropy term forbids it initially. The bad conditioning can be dealt with by using a preconditioning step, and by solving the system using a singular value decomposition (SVD).~\cite{Bryan1990,JarrellJulichPavarini:2012} Note that all the singular values are kept. The SVD is typically more efficient numerically than Gaussian elimination for that particular type of system. As for the problem of negative values of $A$, when the argument of $\log[A(\omega_j)/D(\omega_j)]$ in $\nabla S$ is below the smallest numerical value possible, the $\log$ is smoothly continued by a quadratic function. Although not strictly forbidding negative values, this strongly penalizes them and ensures stability of the algorithm. As a result, negative values of very small amplitude may still appear, but only in regions where the spectrum should actually vanish. The minimization approach is described in more details in Appendix \ref{sec:minimize_Q}.

\subsection{Finding the optimal $\alpha$}\label{sec:optimalalpha}

To find the optimal $\alpha$, we compute the curvature in $\log_{10}\chi^2$ as a function of $\gamma\log_{10}\alpha$, where $\gamma$ is an adjustable parameter smaller than 1. The local curvature is computed by fitting an arc of a circle to a section of the curve, and taking the inverse of the radius as the curvature. The sign is chosen to be positive if the circle center is above the curve and negative if below. Although fitting an arc is numerically more tedious than fitting a parabola, it was found to produce a smoother curvature as a function of $\alpha$. Then, as discussed in section \ref{sec:alpha}, the optimal $\alpha$ is chosen at the maximum of the curvature. The parameter $\gamma$ is used to increase the amplitude of the peak corresponding to the crossover region between the information- and the noise-fitting regions, compared to other peaks that may appear in the information-fitting region. A value of $\gamma$ smaller than 1 also shifts the location of the maximum toward lower $\alpha$, but not by a large amount if $\gamma$ is not too small. The value of $0.2$ was found to be a good compromise to identify clearly the correct peak in the curve, without shifting too much the value of $\alpha$. This parameter is considered as a fixed internal parameter in the program, but its default value can be modified easily by the user.


\subsection{Diagnostics}

At the end of the calculation, in addition to the optimal spectrum, the curves used to find the optimal $\alpha$, namely, $\log_{10}\chi^2$ as a function of $\log_{10}\alpha$ and its curvature, are displayed, along with the diagnostic quantities discussed in section \ref{sec:diagn_tools}. Analyzing those quantities yields information about the quality of the fit, the reliability of the result and, if necessary, how to improve the result. Section \ref{sec:benchmarks} gives examples of this analysis. The documentation on the software also explains how to perform the analysis.\cite{OmegaMaxEnt_url} Since the computation takes only between a few seconds and a few minutes, trying different sets of input parameters to improve the result is easy. For example, between iterations one can modify the frequency grid parameters to obtain the grid that optimally resolves all structures in the spectrum.

%

\section{Benchmarks, applications, and diagnostic tools for accuracy}\label{sec:benchmarks}

To test our approach, we first use the program to analytically continue Green functions obtained from inserting spectral functions in the spectral representation \eqref{eq:G_omega_n_K}. For the purpose of applying maximum entropy, Gaussian noise is added to those Green functions. By comparing with the exact spectrum we can then assess the accuracy that can be reached by the program. Finally, an example with a real physical Monte Carlo simulation is given. In all examples, frequency and temperature are in energy units.

\subsection{Two examples with known exact results}

We illustrate our approach first with a simple spectrum, and then with a difficult case that has both coherent and incoherent features. 

\begin{figure}
\includegraphics[width=0.8\columnwidth]{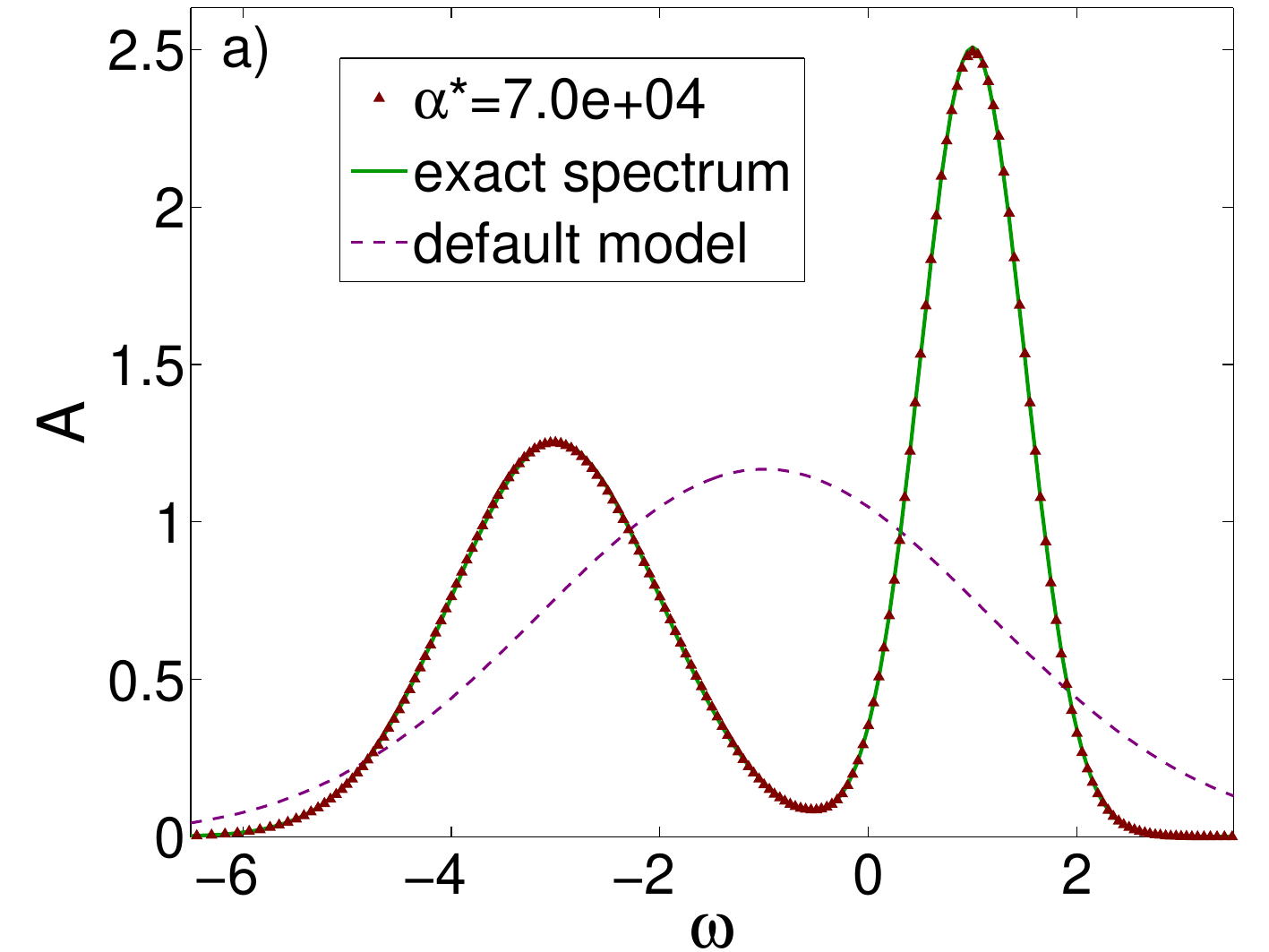}\\
\includegraphics[width=0.8\columnwidth]{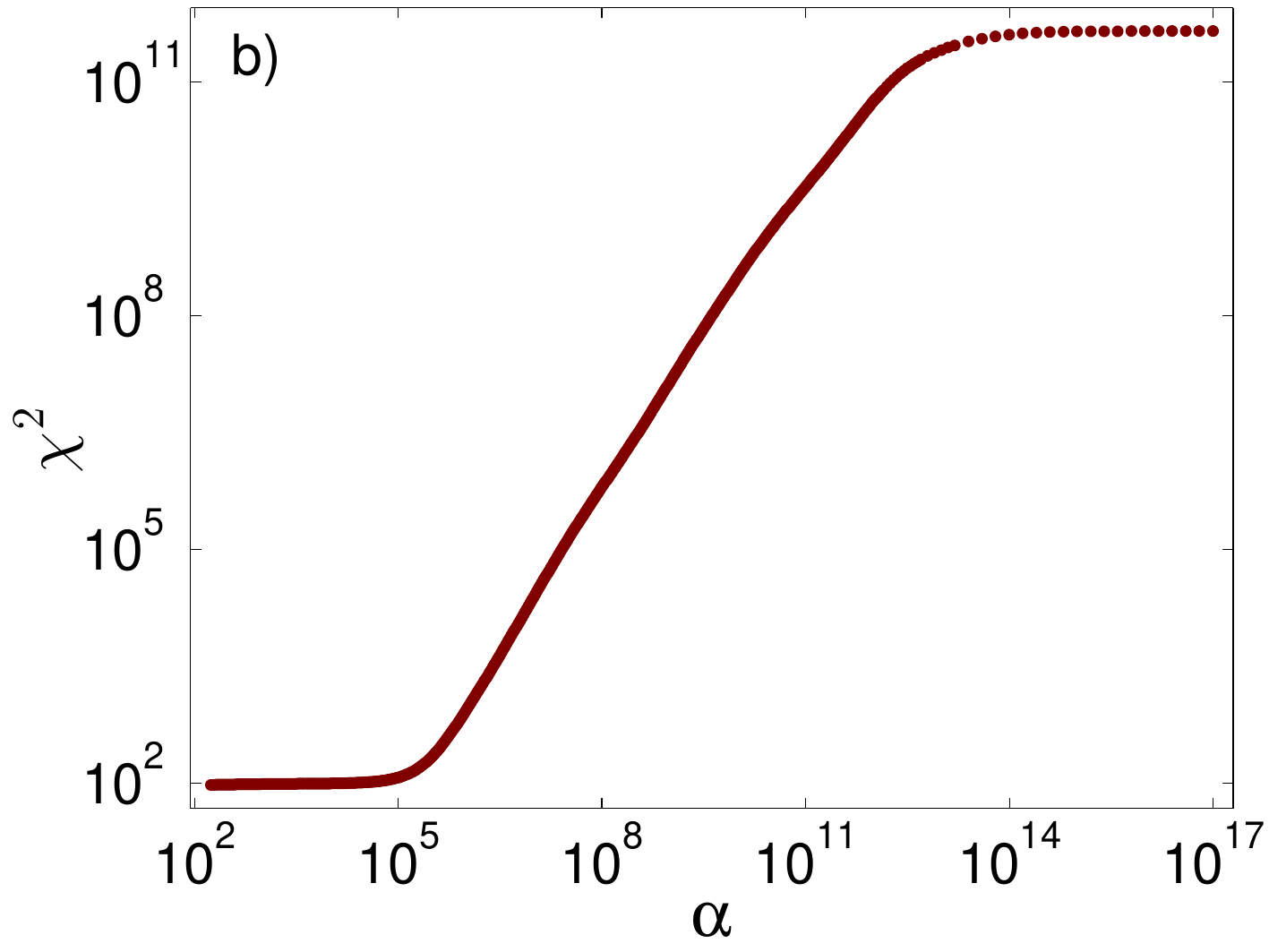}\\
\includegraphics[width=0.8\columnwidth]{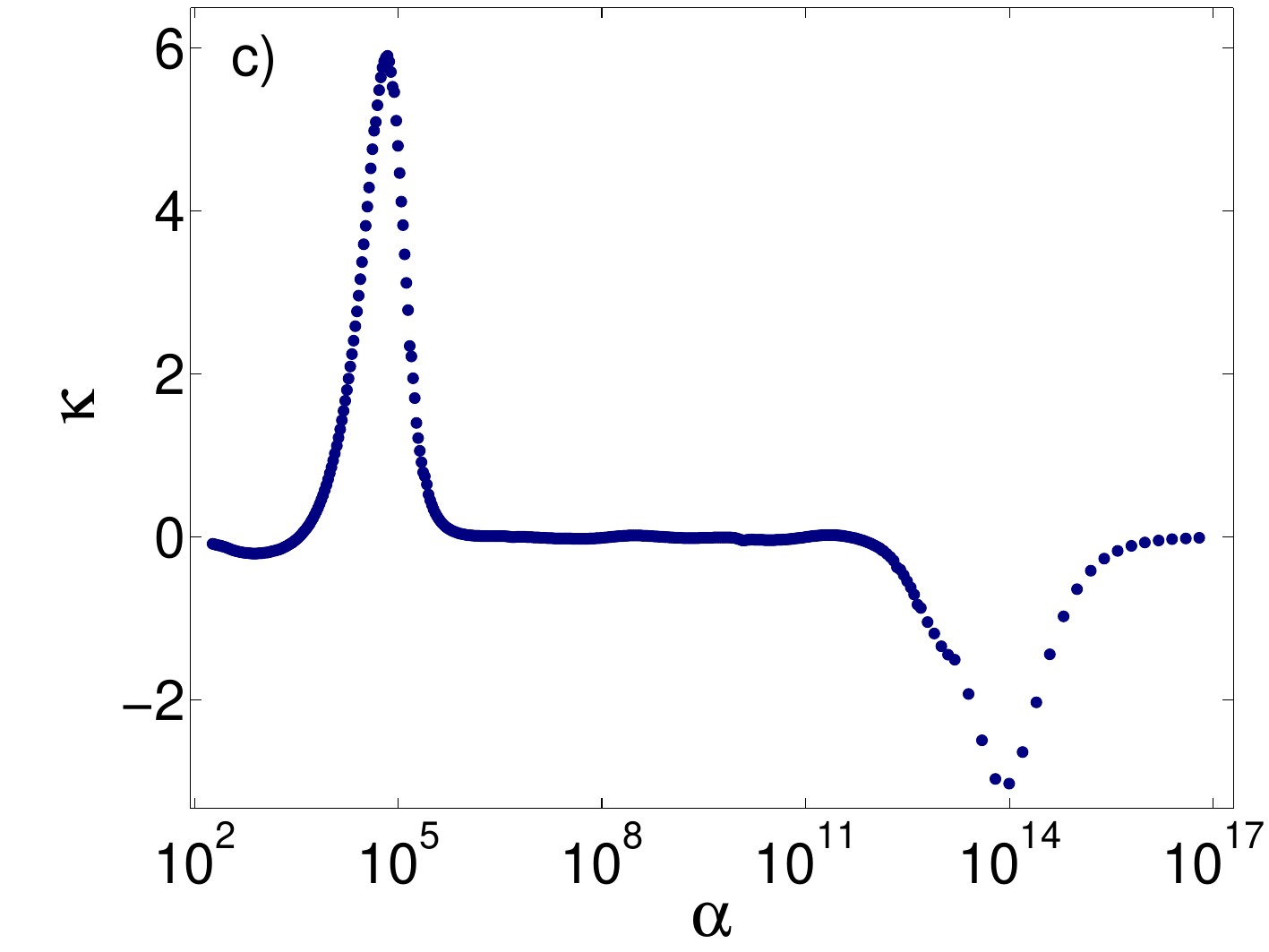}\\
\includegraphics[width=0.8\columnwidth]{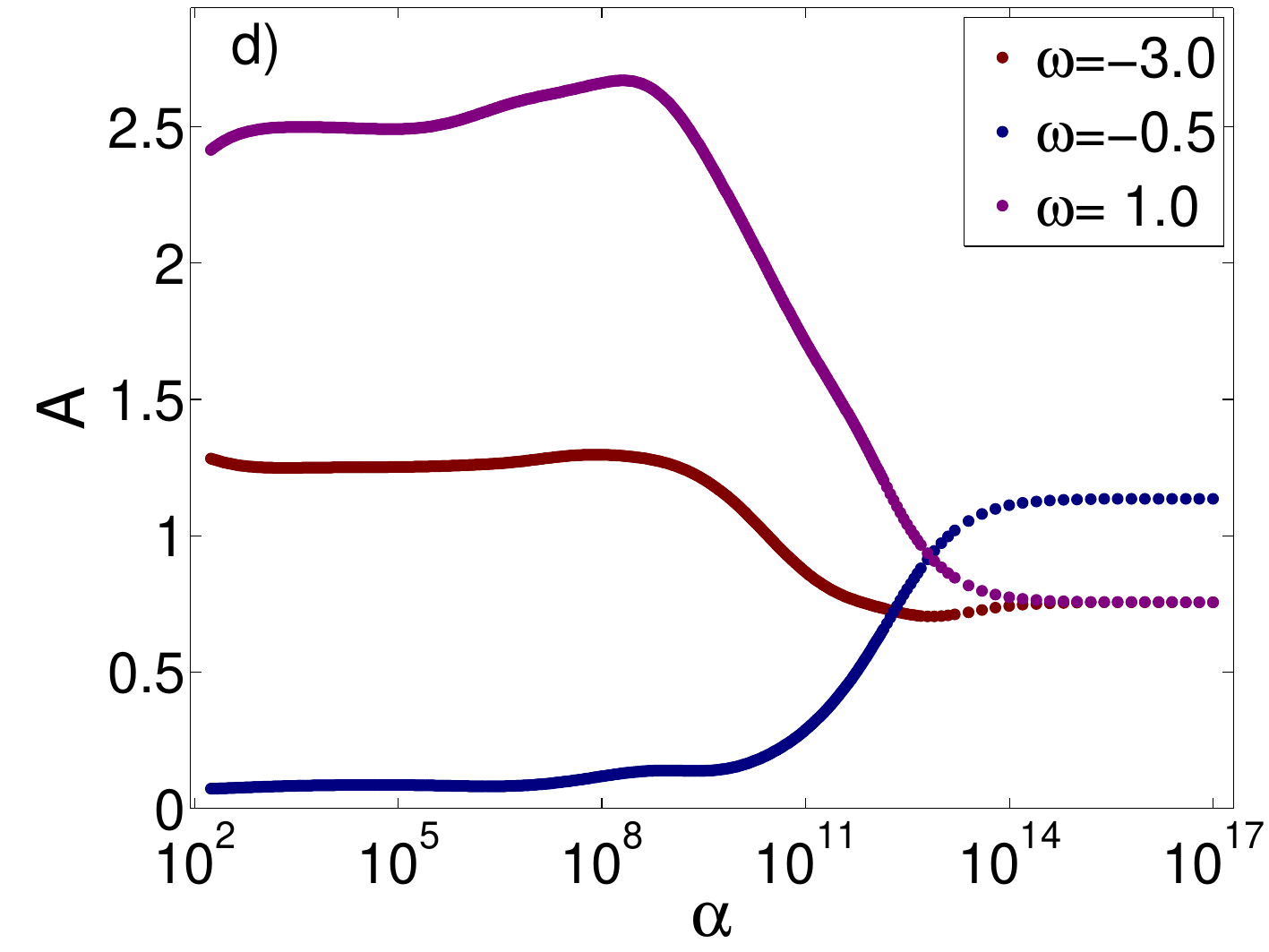}
\caption{a) Spectrum obtained with the program at a temperature $T=0.05$, with a constant relative diagonal error of $10^{-5}$, b) $\chi^2$ as a function of $\alpha$ for the example in a), c) curvature computed from $\log_{10}\chi^2$ as a function of $\gamma\log_{10}\alpha$, where $\gamma=0.2$, plotted as a function of $\log_{10}\alpha$, and d) spectrum as a function of $\alpha$ at sample frequencies corresponding to the three extrema of the spectrum in (a).}
\label{fig:example_1}
\end{figure}
The first example, shown in Fig. \ref{fig:example_1}, is the analytic continuation of a fermionic Green function given as a function of Matsubara frequency, with a diagonal covariance, a constant relative error of $10^{-5}$, at a temperature $T=0.05$.  The spectrum used to compute $G(i\omega_n)$ from \eqref{eq:G_omega_n_K} is the sum of two Gaussians. It is shown in green in Fig. \ref{fig:example_1}(a). In Fig. \ref{fig:example_1}(b), we observe clearly the three regimes discussed in section \ref{sec:alpha} for $\chi^2$ as a function of $\alpha$ on a log-log plot. If we take the optimal $\alpha$ at the position of the maximum in the curvature of $\log\chi^2$ as a function of $\gamma\log\alpha$, where $\gamma=0.2$, shown in Fig. \ref{fig:example_1}(c), we obtain the spectrum in red triangles shown in (a). 



Figure \ref{fig:example_1}(d) shows, as a function of $\alpha$, the spectral function at the frequencies corresponding to the two maxima and to the minimum between them in Fig.\ref{fig:example_1}(a). As discussed in section \ref{sec:diagn_tools}, the plateaus in the curves indicate convergence of the spectrum at those frequencies, which is expected for sufficiently precise data.  As seen from the optimal spectrum shown in Fig.\ref{fig:example_1}(a), those plateaus occur at values that coincide with the exact spectrum with very good accuracy. Other useful information is obtained by comparing the curves to each other. A good overlap of the plateaus over some range of $\alpha$ indicates that the different parts of the spectrum reach their optimal value around the same $\alpha$. When this is not the case, then the ratios of the standard deviations $\sigma(i\omega_n)$ between different frequency regions are probably incorrect,\footnote{An overall factor in the definition of $ \sigma$ just scales the value of $\alpha$ and does not cause any problem.} causing the regions with an underestimated $\sigma$ to converge faster. Note however that, for a given precision, some parts of the spectrum may not converge, while others do. Sharp features, or parts of the spectrum that are very different from the default model, typically reach stability at lower values of $\alpha$ and thus require higher precision to converge. For instance, we note in Fig. \ref{fig:example_1}(d) that the peak at $\omega=1$ converges at $\alpha\approx 10^5$, while the spectrum at the other peak converges at $\alpha\approx 10^7$. Something similar happens in the next example discussed below. Finally, below some value of $\alpha$ the spectrum eventually behaves in an unpredictable manner as a function of $\alpha$ since the conditioning of the linear system becomes bad at very low $\alpha$. Indeed, in the brackets of expression \eqref{eq:delta_A_j_main} the term proportional to $\alpha$ that regularizes the linear system gradually vanishes as $\alpha$ decreases.

\begin{figure}
\includegraphics[width=0.49\columnwidth]{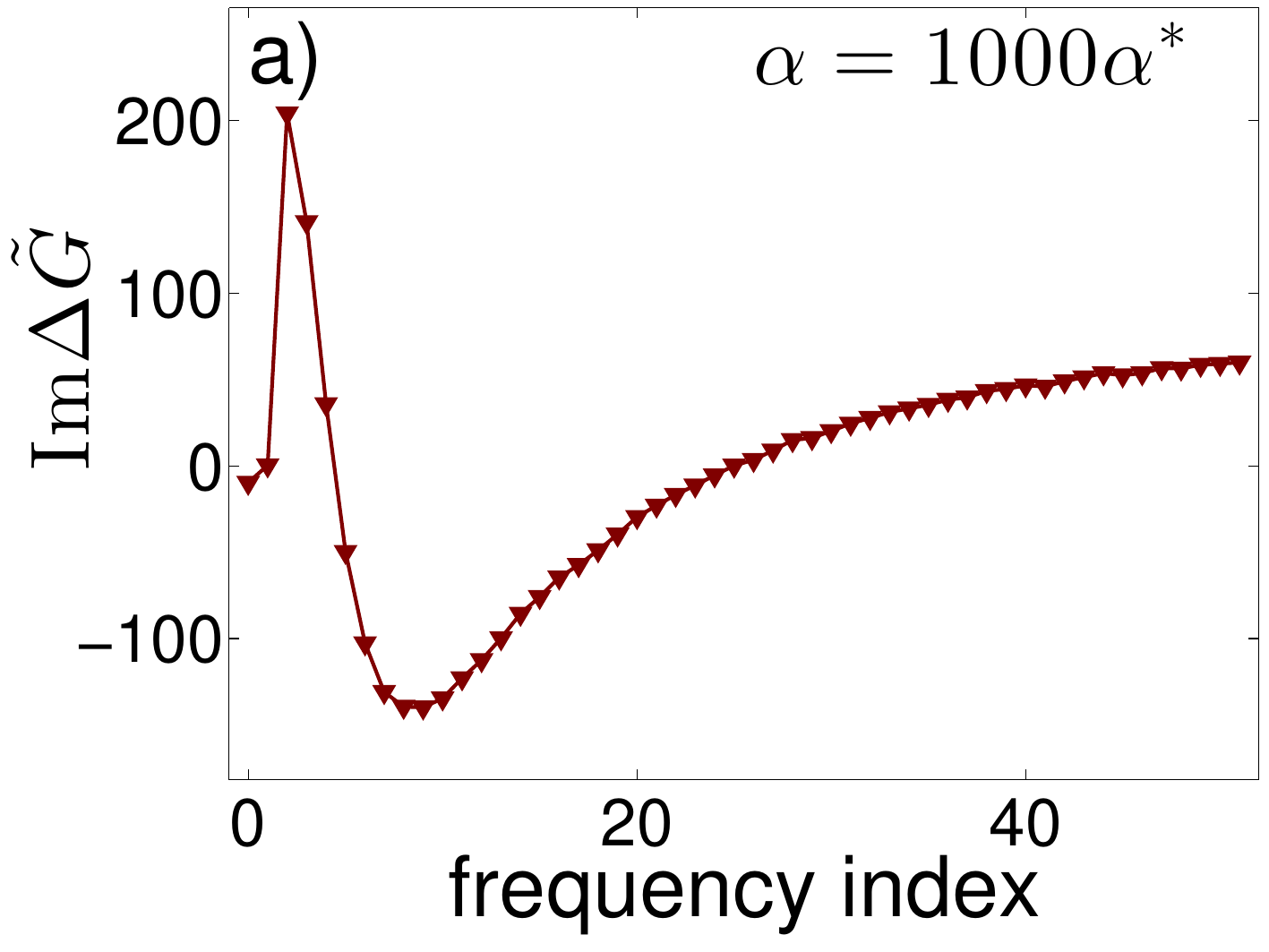}
\includegraphics[width=0.49\columnwidth]{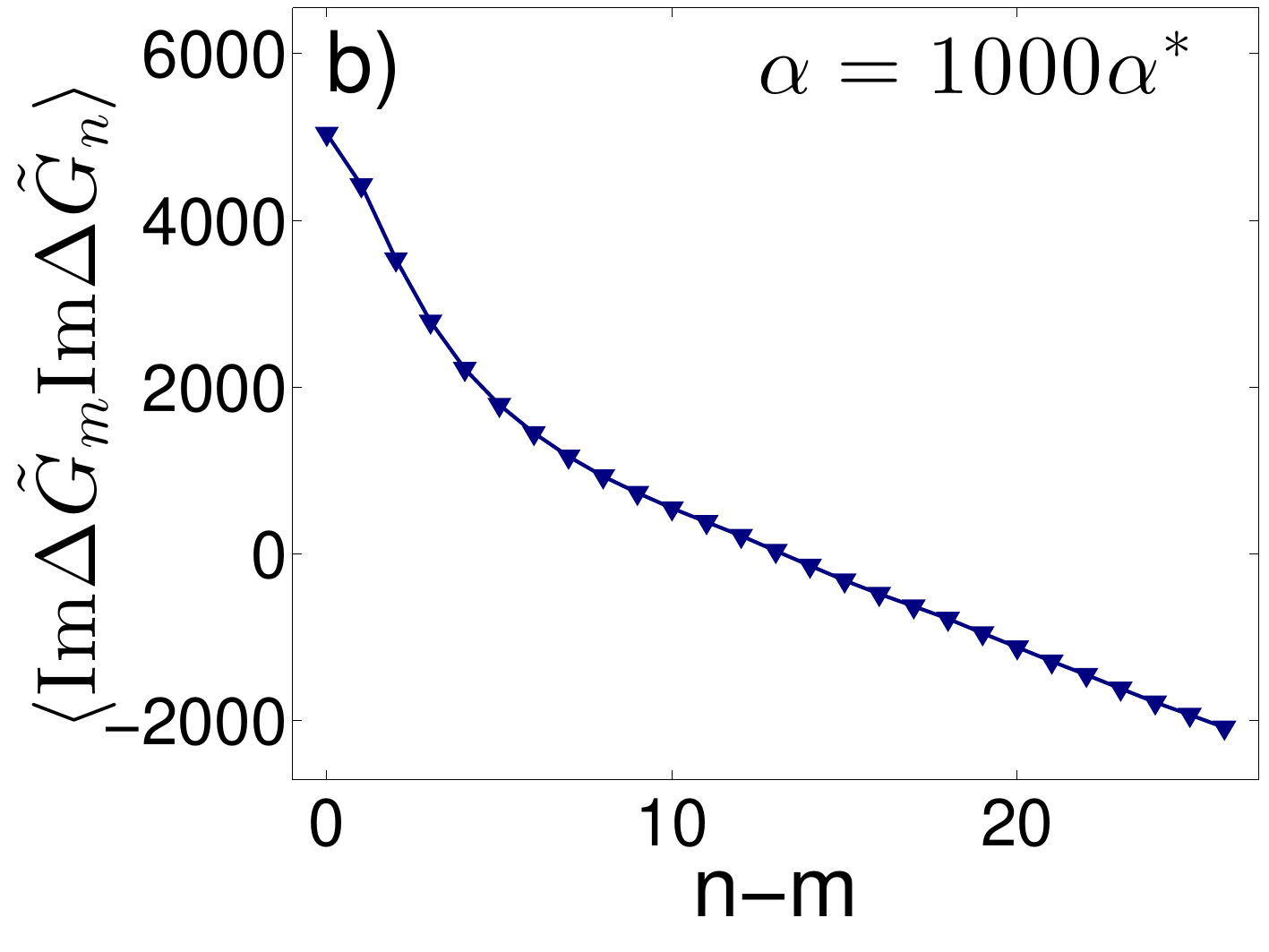}\\
\includegraphics[width=0.49\columnwidth]{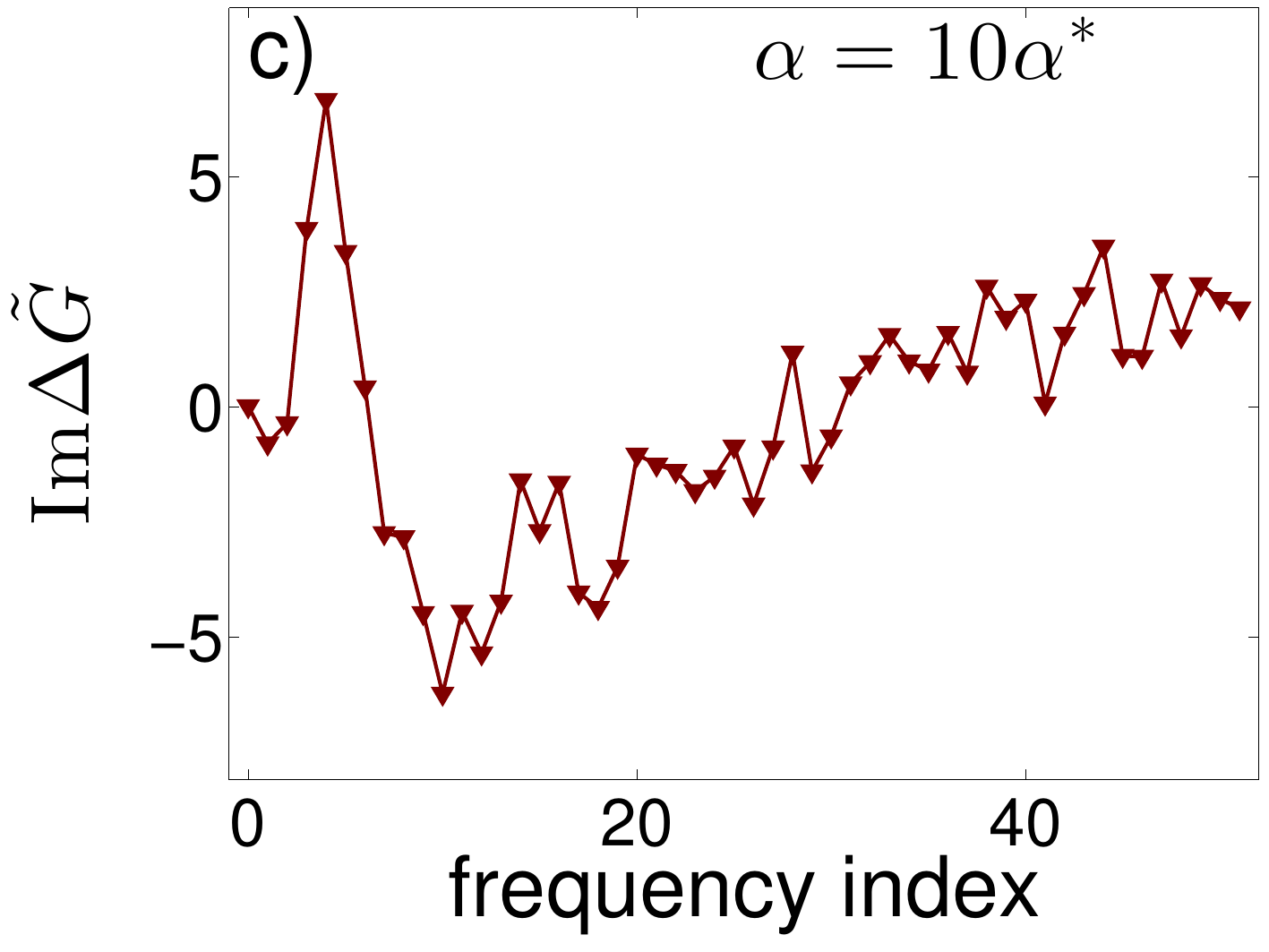}
\includegraphics[width=0.49\columnwidth]{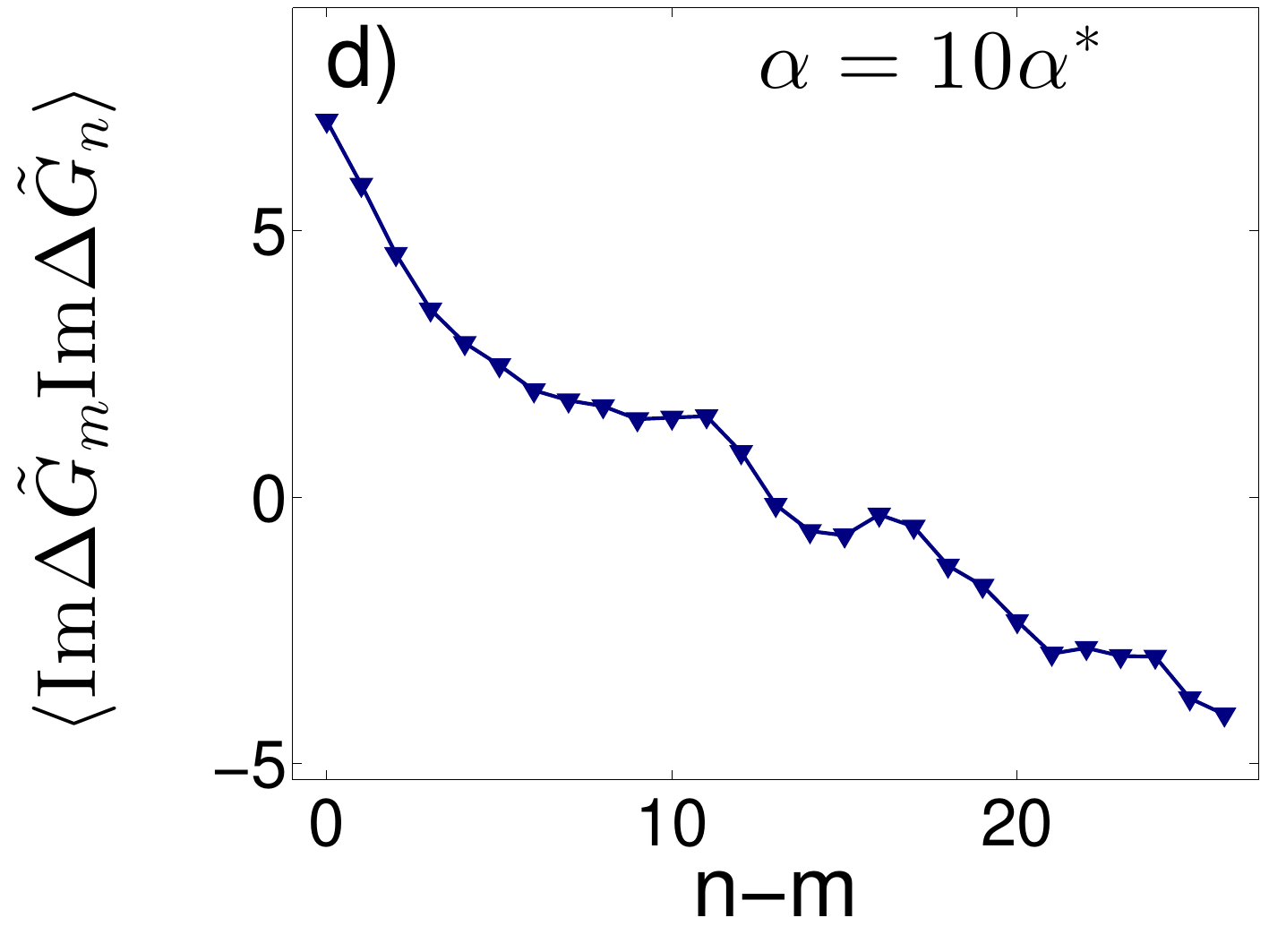}\\
\includegraphics[width=0.49\columnwidth]{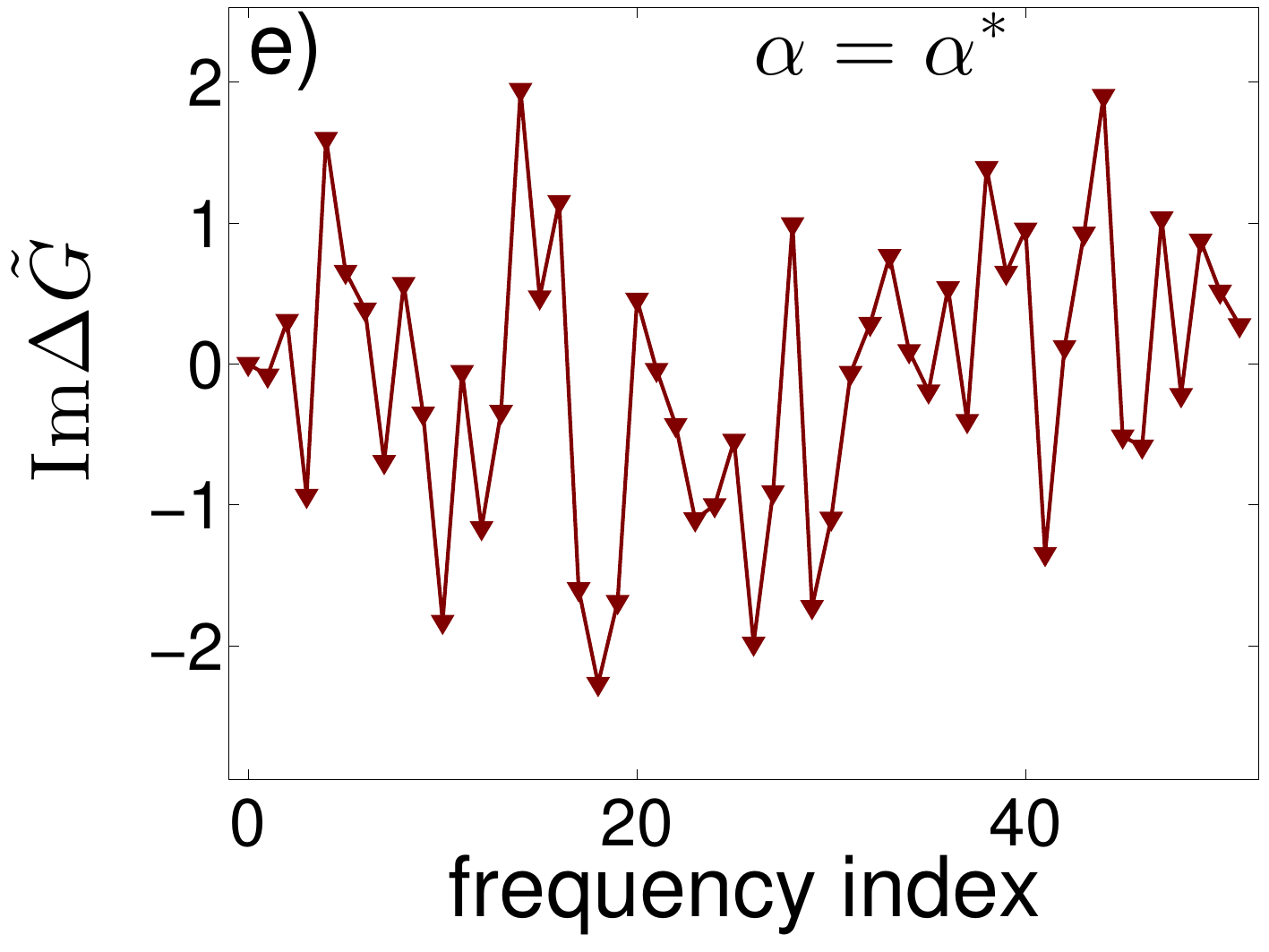}
\includegraphics[width=0.49\columnwidth]{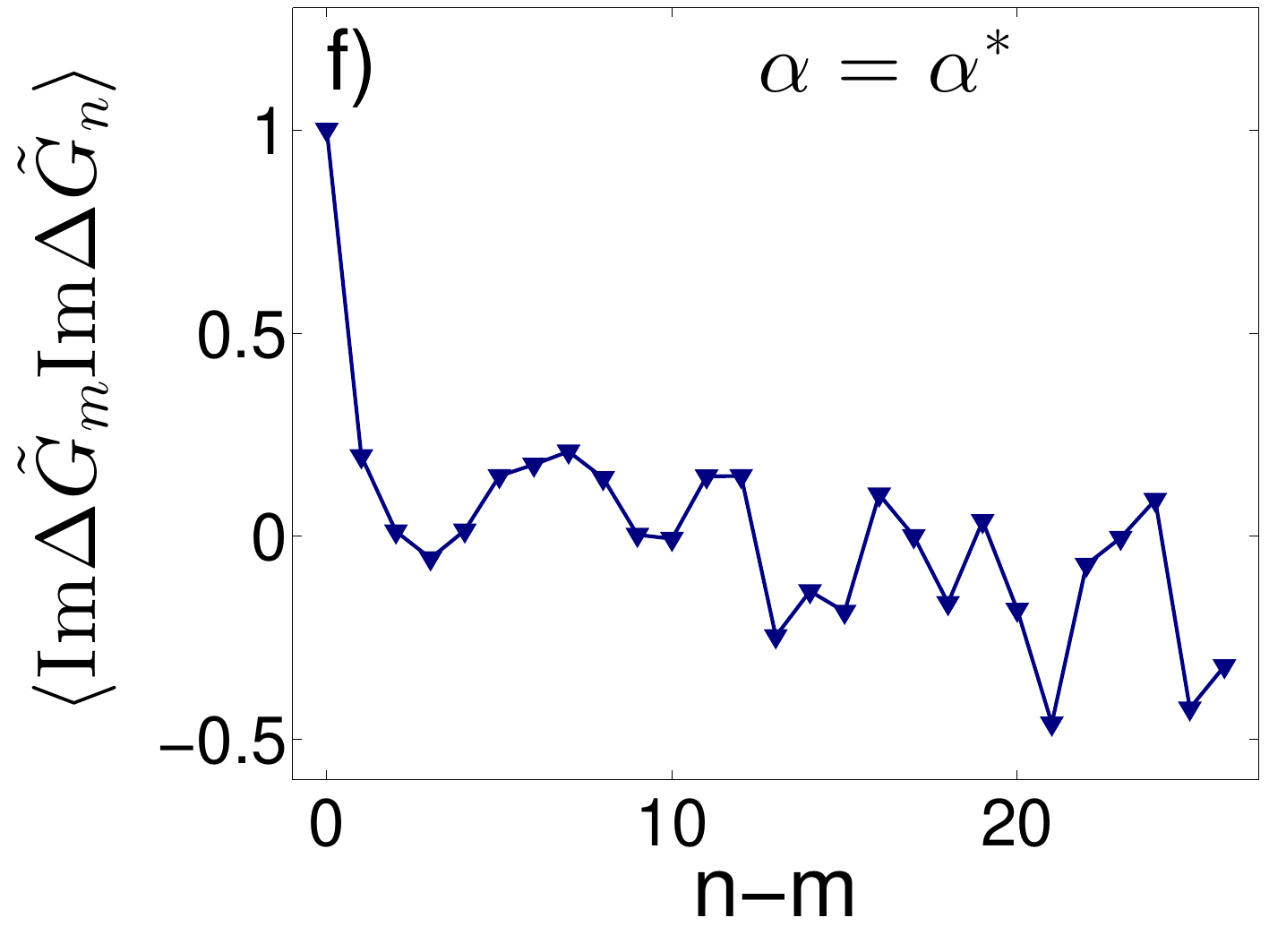}
\caption{Normalized errors and auto-correlations as a function of frequency index at three different values of $\alpha$ for the example given in Fig. \ref{fig:example_1}. a) error and b) autocorrelation at $\alpha=1000\alpha^*$, c) error and d) autocorrelation at $\alpha=10\alpha^*$, e) error and f) autocorrelation at $\alpha^*$.}
\label{fig:errors_alpha_r_alpha_opt}
\end{figure}
Continuing the analysis of the example shown in Fig. \ref{fig:example_1}, consider in Fig. \ref{fig:errors_alpha_r_alpha_opt} the normalized distance $\Delta \tilde{G}(\omega_n)$ between the input Green function $G_{in}$ and output $G_{out}$, as well as the auto-correlation $\Delta \tilde{G}^2(\Delta n)$. Those functions are plotted for three different values of $\alpha$, namely two values above $\alpha^*$ as well as the optimal value $\alpha^*$. In (a), far above $\alpha^*$, $\Delta \tilde{G}(\omega_n)$ is a smooth function since the difference between $G_{out}$ and $G_{in}$ is much larger than the noise amplitude in $G_{in}$. The corresponding auto-correlation in (b) is also smooth and broad since the correlations between different frequencies of $\Delta \tilde{G}$ are large. Slightly above $\alpha^*$, in (c), $\Delta \tilde{G}$ becomes noisy as $G_{in}-G_{out}$ becomes comparable to the noise in $G_{in}$. The auto-correlation is still broad, since  $\Delta \tilde{G}$ still has well defined structures, but it is becoming slightly noisy. At $\alpha^*$, $\Delta \tilde{G}$, in (e), is essentially noise, as expected from the discussion in section \ref{sec:alpha}. This is confirmed by the noisy Kronecker $\delta$ shape of $\Delta \tilde{G}^2$, in (f). In addition, $\Delta \tilde{G}^2(0)$ is close to 1, as expected since it is equal to $\chi^2/N$, where $N$ is the number of terms in $\chi^2$, and the standard deviation for the data is known exactly in that example (see the discussion on the \textit{historic} approach in section \ref{sec:Bayes_ME}).

\begin{figure}
\includegraphics[width=0.8\columnwidth]{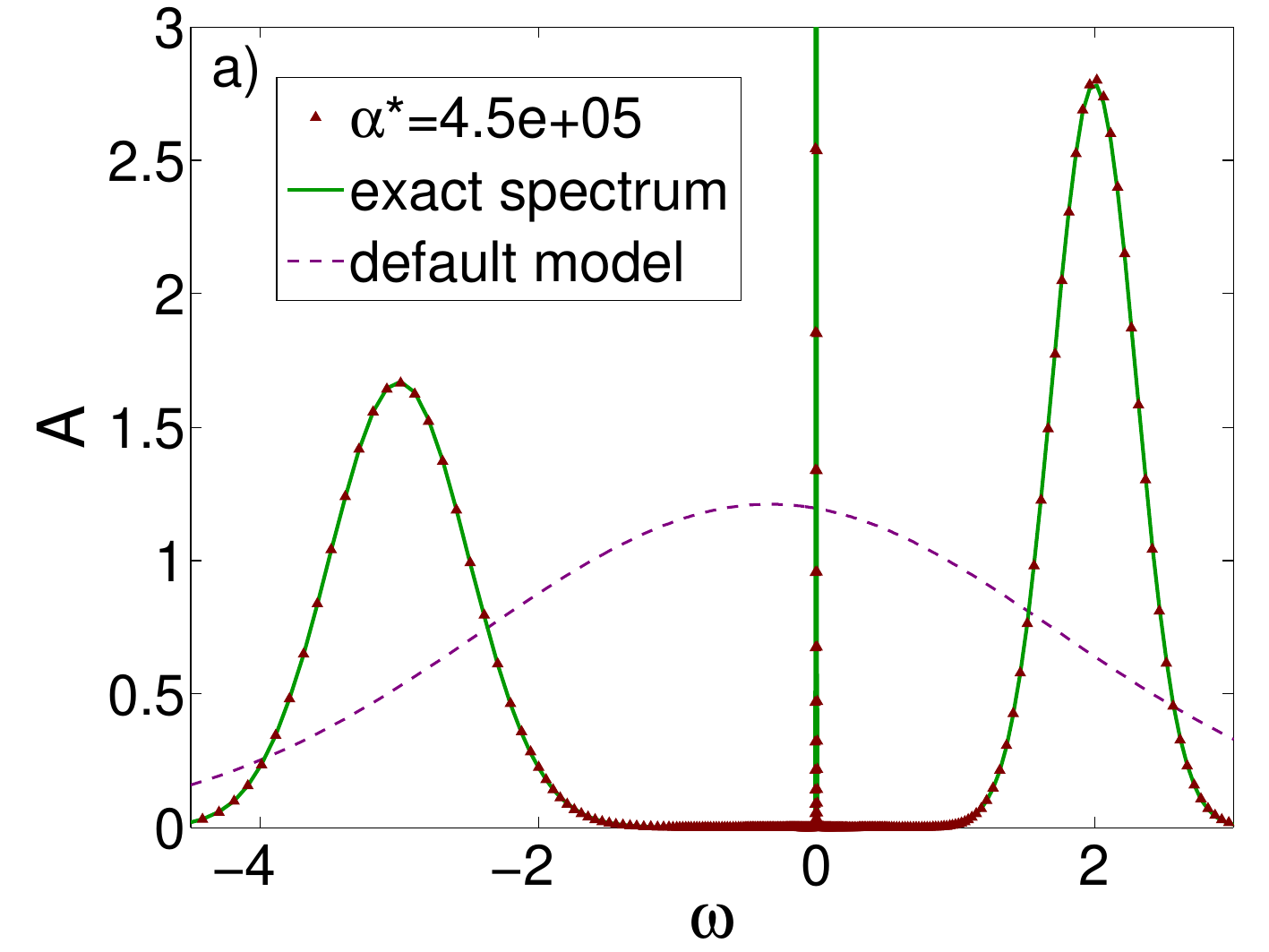}\\
\includegraphics[width=0.8\columnwidth]{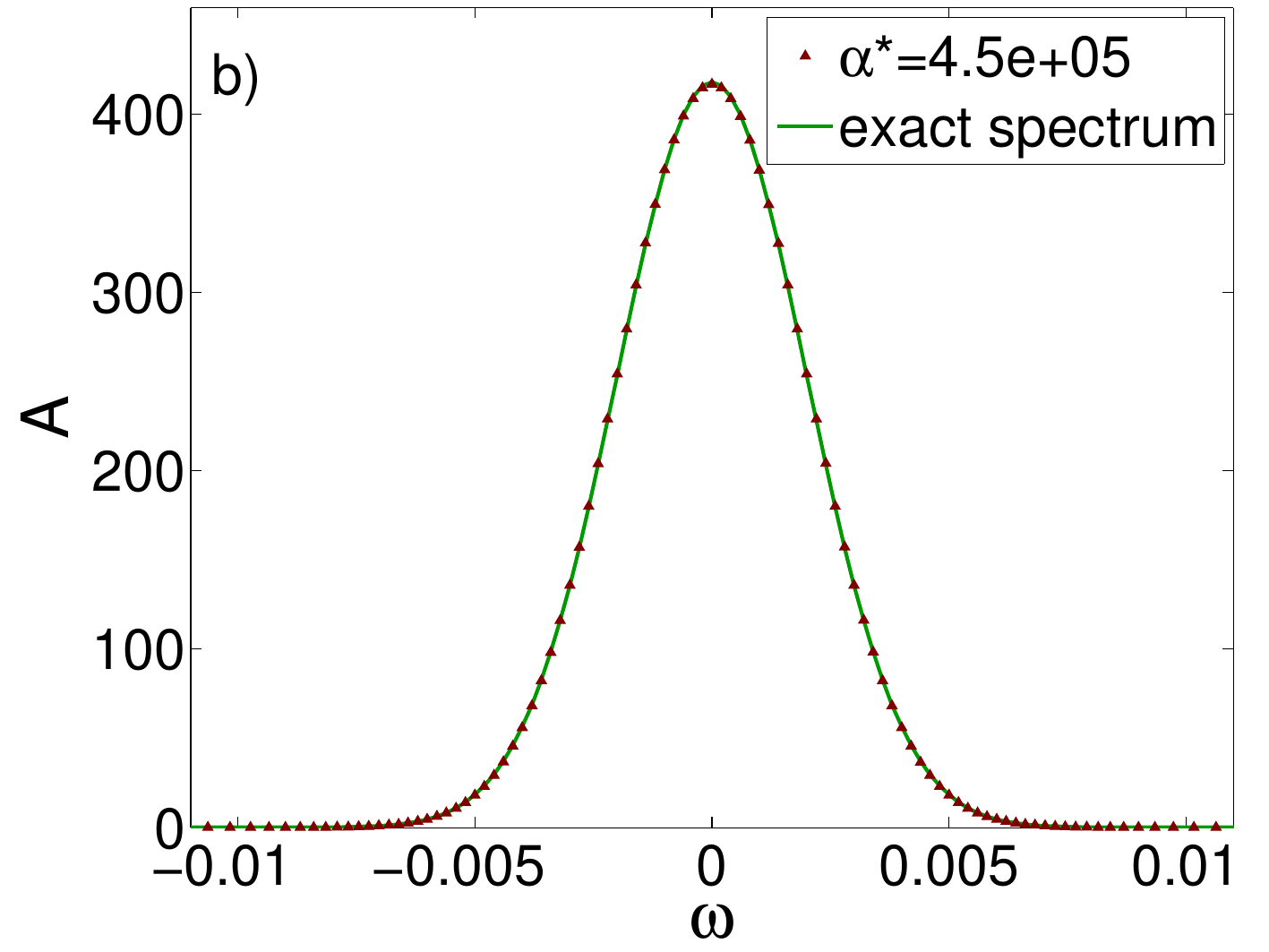}\\
\includegraphics[width=0.8\columnwidth]{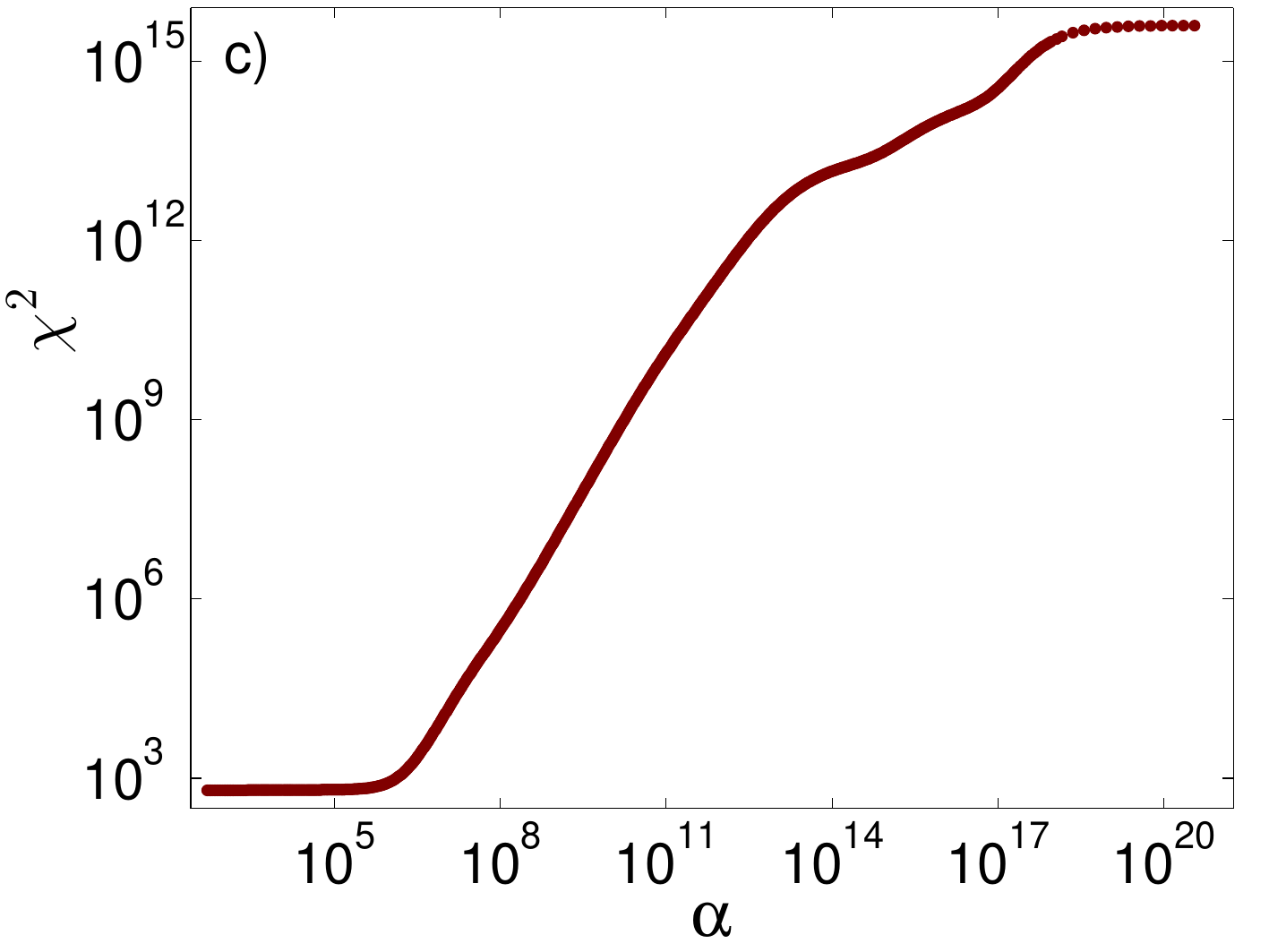}\\
\hspace{0.03\columnwidth}\includegraphics[width=0.84\columnwidth]{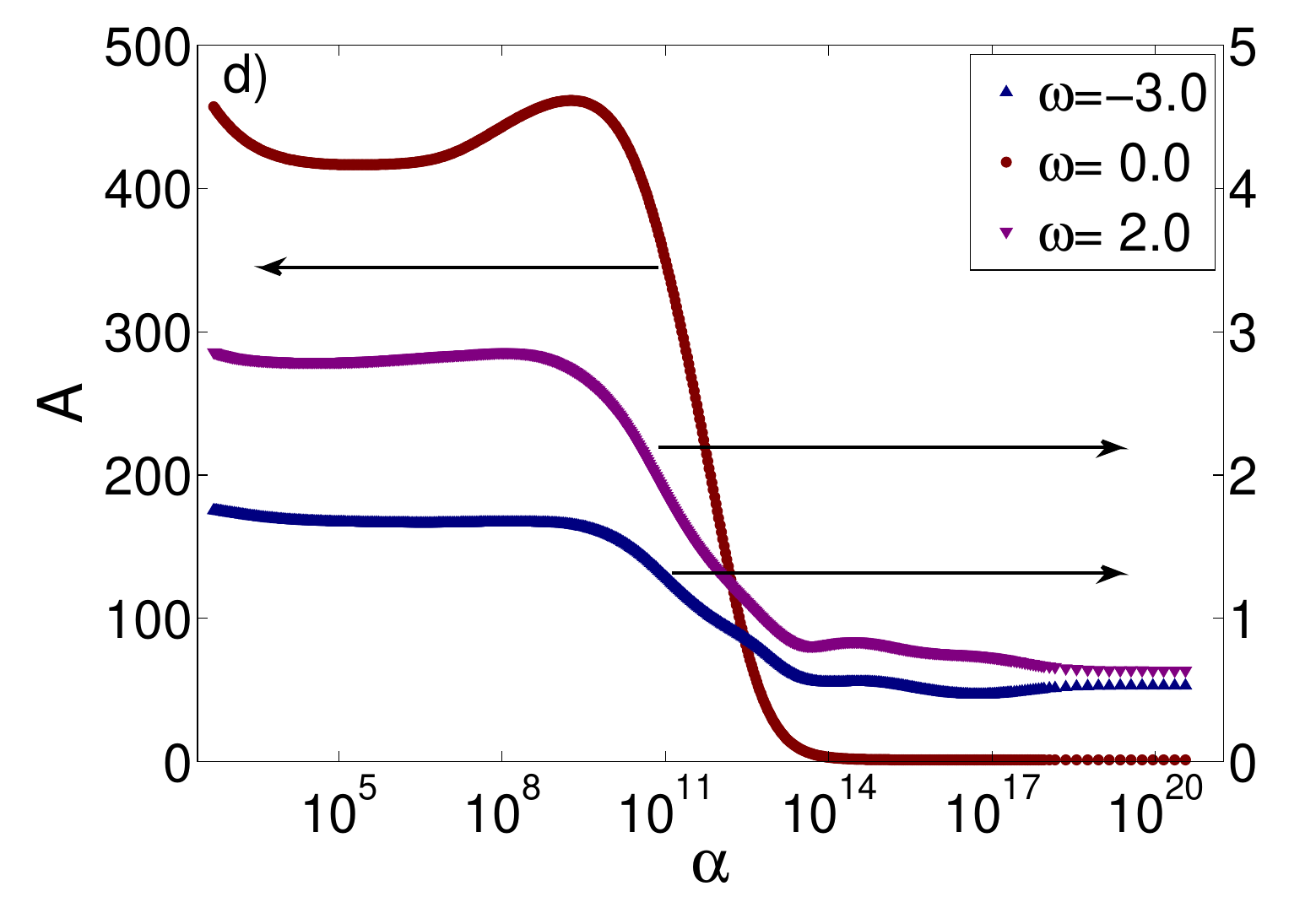}
\caption{a) Spectrum obtained with the program at a temperature $T=0.002$, with a constant relative diagonal error of $10^{-6}$, b) magnification of the low frequency peak, c) $\chi^2$ as a function of $\alpha$ and d) spectrum as a function of $\alpha$ at sample frequencies corresponding to the maxima. The value for the $\omega=0$ peak is given on the left axis while the values for the two other peaks are given on the right axis.}
\label{fig:example_2}
\end{figure}

For the second example, we show in Fig. \ref{fig:example_2} the analytic continuation of another fermionic Green function computed with a spectrum that is the sum of three Gaussians, on which a constant relative diagonal error of $10^{-6}$ has been added. The scale in Fig. \ref{fig:example_2}(a) is adapted to the two broad peaks while the scale in Fig. \ref{fig:example_2}(b) is adapted to the  very sharp central peak around $\omega=0$. Note in Fig. \ref{fig:example_2}(c) the presence of the three regimes in $\chi^2$ as a function of $\alpha$, and in Fig. \ref{fig:example_2}(d) the plateaus indicating good quantitative precision in the result. The much smaller width and much larger amplitude of the central peak compared to the two others makes this case difficult from a computational point of view. Indeed, if the grid in the main spectral region was uniform, while fine enough to resolve the central peak, it would contain thousands of points, increasing significantly the computation time. Therefore, in addition to the non-uniform grid used automatically by the program outside the main spectral region, and described in Appendix \ref{sec:non_uniform_omega_grid}, a non-uniform grid within the main spectral region has also been used. For that purpose, as explained in section \ref{eq:grids}, a vector of the form \eqref{Eq:Grid} was given to the program, which has generated a grid with a step varying smoothly between the provided intervals. For the example in Fig. \ref{fig:example_2}, the step around $\omega=0$ is $\Delta\omega=2\times 10^{-4}$ while $\Delta\omega=0.1$ for the peak centered on $\omega=-3$ and $\Delta\omega=0.05$ for the one at $\omega=2$. 

The standard deviation of the central peak is equal to the temperature $T=0.002$, putting the magnitude of the first Matsubara frequency $\pi T$ outside its frequency range. This might seem to make the analytic continuation difficult for that part of the spectrum since essentially only information on the moments of that peak is contained in the Matsubara Green function. The central peak is nevertheless very well reproduced, as shown in Fig. \ref{fig:example_2}(b). It converges for a value of $\alpha$ only slightly lower than the two broad peaks, as shown in Fig. \ref{fig:example_2}(d). This is analog to what is observed in the example of Fig. \ref{fig:example_1}.


The temperature in the example of Fig. \ref{fig:example_2} is very small compared to the total width of the spectrum, therefore the number of Matsubara frequencies that are within the spectral range, and which have to be taken into account, is large. If all the frequencies below the onset frequency of asymptotic behavior $\omega_{as}$ were used, the computation time would be quite high since the minimization of $Q$ scales as $O(N_{\omega_n}^2N_\omega)$ if $N_{\omega_n}<N_{\omega}$, or $O(N_{\omega_n}N_\omega^2)$ if $N_{\omega_n}>N_{\omega}$ (see Appendix \ref{sec:minimize_Q}). However, although the whole frequency range below $\omega_{as}$ contains relevant information on the spectrum, using all the Matsubara frequencies in that range is not necessary (see Sec. \ref{eq:grids}). In that example, the spectrum was obtained with high accuracy using a non-uniform Matsubara grid, of the type described in Appendix \ref{sec:non_uniform_Matsubara_grid_def}, which contains 321 frequencies instead of about 1000 below $\omega_{as}$.

Finally, we notice in figures \ref{fig:example_1}(d) and \ref{fig:example_2}(d) that there is a range of values of $\alpha$ where the values of the spectrum at the maxima are higher than both the exact values and the optimal ones (both essentially identical). The structures in the spectrum in that range of $\alpha$ are therefore sharper than those of the exact spectrum. This means that, if the data's standard deviations were comparable to the values of $|\Delta G|$ in that range of $\alpha$, the final MaxEnt result would have sharper structures than the exact spectrum. Therefore, contrary to a common belief about maximum entropy, even for a very smooth and featureless default model, the maximum entropy results are not systematically smoother than the exact result. Note that this ``overshoot'' of the values at the maxima as a function of $\alpha$ is not a numerical artefact. The spectra in that range are the actual solutions of the equation $\nabla Q=0$ at each $\alpha$. This can also be verified by starting the calculation at the lowest $\alpha$, using the spectrum obtained at that value as the initial spectrum, and recomputing the spectrum for increasing values of $\alpha$ instead of the opposite. If there is no hysteresis in the sample-frequency curves, as we did observe during the tests, then we can have very high confidence in the results.


\subsection{Analytic continuation for single-site dynamical mean-field theory}\label{sec:DMFT}

\begin{figure}
\includegraphics[width=0.8\columnwidth]{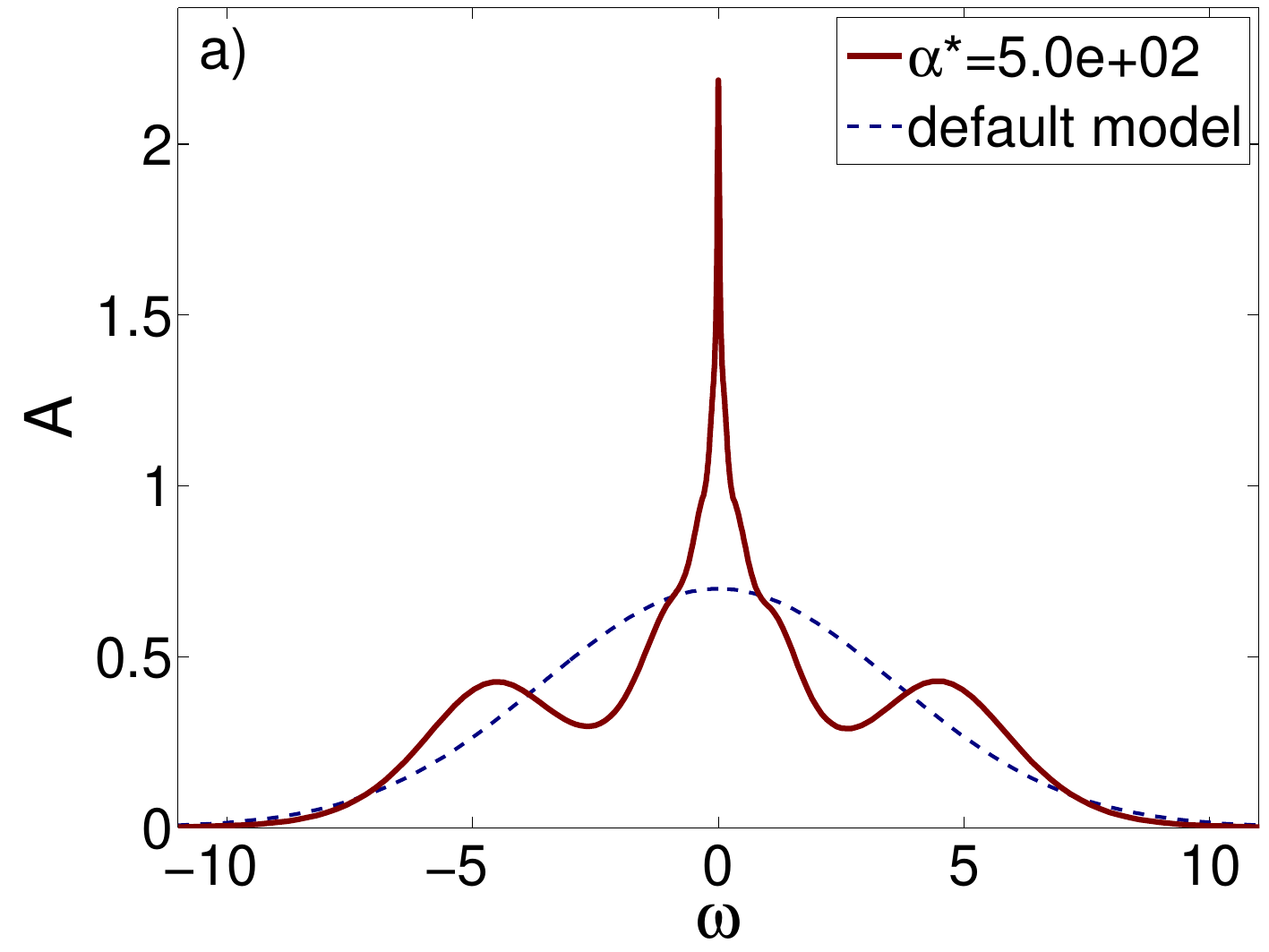}\\
\includegraphics[width=0.8\columnwidth]{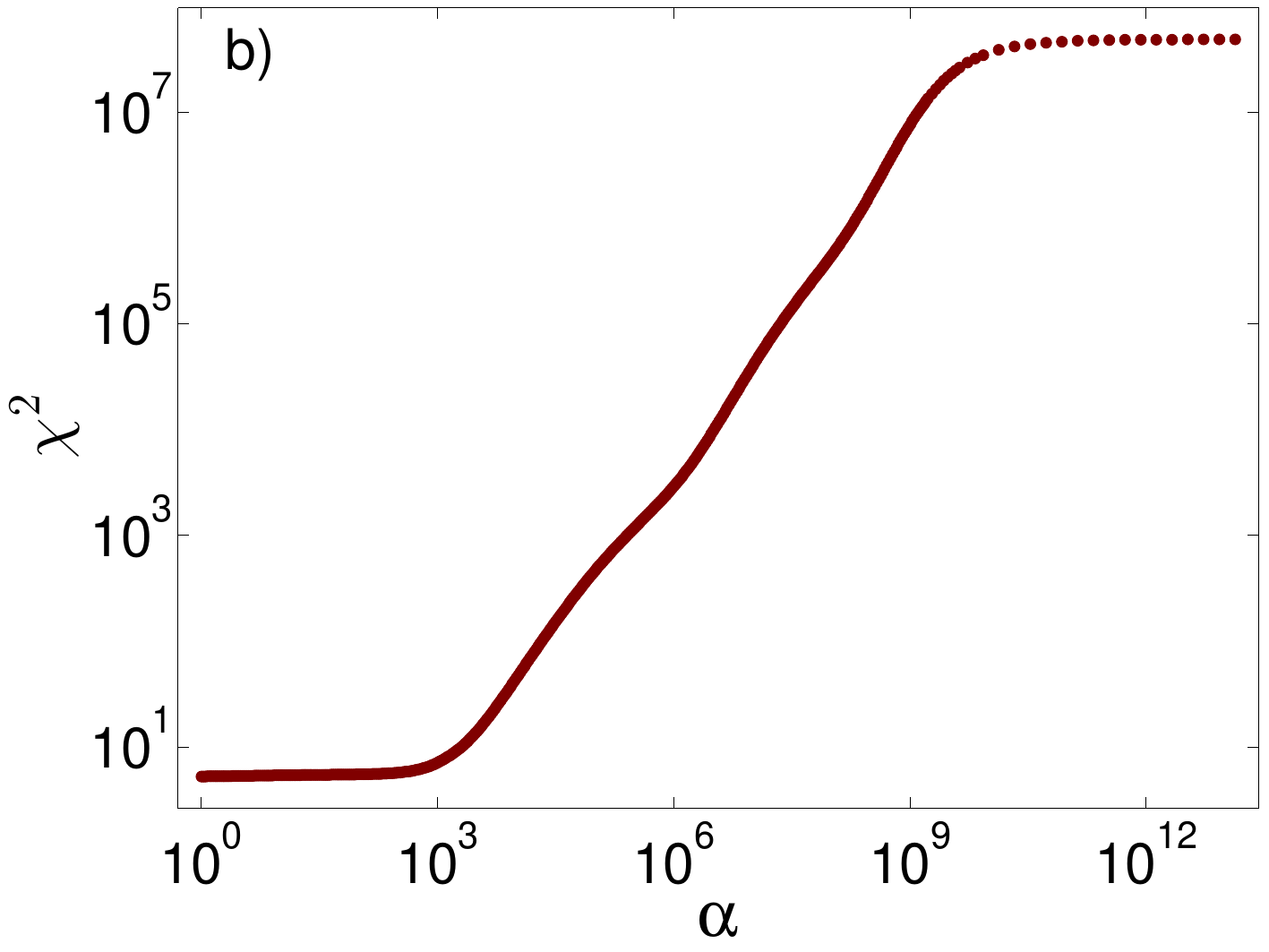}\\
\includegraphics[width=0.8\columnwidth]{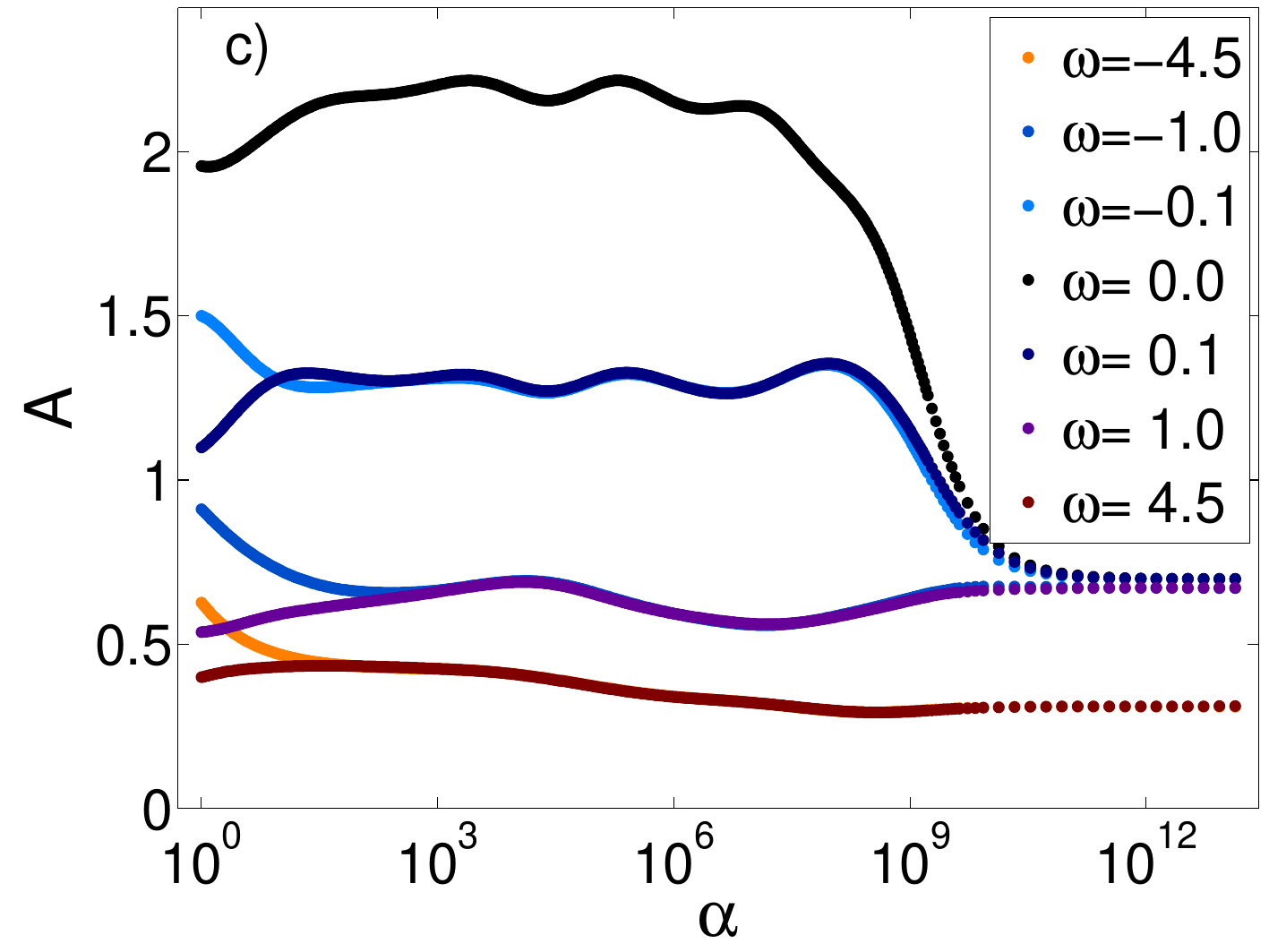}
\caption{a) Spectrum for single-site DMFT on a square lattice at $U=6$, filling $n=1$ and temperature $T=0.01$, b) $\chi^2$ as a function of $\alpha$ for the above example, c) spectrum as a function of $\alpha$ at sample frequencies. The relative error in the data is of order $5\times 10^{-4}$ at most.}
\label{fig:example_appl}
\end{figure}

Finally, we present in Fig. \ref{fig:example_appl} the result of analytic continuation of single-site dynamical mean field theory (DMFT) data obtained with a continuous-time quantum Monte Carlo solver \cite{WernerMillis:2006} for the one-band Hubbard model on the half-filled bipartite square lattice at $U=6t$ and temperature $T=0.01t$, where $t$ is the nearest neighbor hopping\footnote{We are grateful to Giovanni Sordi and Patrick S\'emon for providing the quantum Monte Carlo data used in this section.}. The covariance matrix was computed using the Green functions of the last 24 DMFT converged iterations. The relative error is of order $5\times 10^{-4}$ at most. This type of problem was studied in detail at the beginning of DMFT when quantum Monte Carlo algorithms did not allow low temperature and high precision.\cite{Georges:1996} However, the qualitative form of the spectrum is well known. Here we find that, as expected, the peak in the density of states at the Fermi level becomes very sharp at low temperature. This can be obtained even in the presence of broad incoherent features at high frequency.

As shown in Fig. \ref{fig:example_appl}(b), $\chi^2$ versus $\alpha$ has the same general shape as in the previous examples. The spectrum shown in \ref{fig:example_appl}(a) is the result at the maximum of the curvature in $\log\chi^2$ versus $0.2\log\alpha$, as in the previous examples. That example is particularly interesting for benchmarks because particle-hole symmetry makes it clear when only information is fitted and when noise is also fitted. Indeed, since the noise does not respect particle-hole symmetry (the real part of the data does not vanish completely), noise fitting starts when the spectrum stops being symmetric as $\alpha$ decreases. This value is easy to find in Fig. \ref{fig:example_appl}(c), where the spectrum is plotted at symmetric frequencies on each side of $\omega=0$, as a function of $\alpha$. We note, by comparing figures \ref{fig:example_appl}(b) and \ref{fig:example_appl}(c) that the highest $\alpha$ where the values of the spectrum at two symmetric frequencies depart from each other is effectively around the maximum curvature in (b). Figure \ref{fig:example_appl}(c) also suggests a relatively good quantitative accuracy of the result in \ref{fig:example_appl}(a), considering that the input data here are the result of an actual physical simulation, and not an ideal case like in the previous examples. Indeed the curves at sample frequencies have a relatively stable region as a function of $\alpha$. However, the absence of flat plateaus suggests a precision of the order of $10\%$ at most around $\omega=0$.

We should mention that when the covariance is not diagonal, $\Delta \tilde{G}$ is not a smooth function anymore at large $\alpha$ because the diagonalization routine sorts eigenvalues in decreasing order. This makes the difference between the information and noise fitting regimes less clear to estimate from that quantity, which therefore becomes slightly more difficult to analyze. On the other hand, the difference between the auto-correlations $\Delta \tilde{G}^2(\Delta n)$  at $\alpha>\alpha^*$ and $\alpha\leq \alpha^*$ is still clear, with one exception, namely for a particle-hole symmetric case like the example we just considered in Fig. \ref{fig:example_appl}. Indeed, in that case the real part of the Green function should vanish, but it is mostly noise because of intrinsic computational errors. In addition, because there are correlations in general between the errors on the real and on the imaginary parts, then the transformation \eqref{eq:G_K_diag} mixes the real and imaginary parts of the Green function. This can cause $\Delta \tilde{G}$ and $\Delta \tilde{G}^2(\Delta n)$ to look noisy for all values of $\alpha$. Therefore, those quantities are much less useful for the analysis in the particle-hole symmetric case, including the example of Fig. \ref{fig:example_appl}. This does not affect however the behavior of $\chi^2$ and $A(\omega^{sample}_i)$ as a function of $\alpha$, and therefore it does not influence how the optimal $\alpha$ is determined. In the general case where both the real and imaginary parts have structure, $\Delta \tilde{G}$ and $\Delta \tilde{G}^2(\Delta n)$ are useful, although $\Delta \tilde{G}$ can be more difficult to read for non-diagonal covariance.

\section{Discussion}\label{sec:discussion}

The approach presented here relies critically on the algorithms discussed in section \ref{sec:algos}. Indeed, because of the ill-conditioned nature of the problem and the bad scaling of the minimization problem with grid size, all aspects of the calculation must be optimized to compute the spectrum in a reasonable time, and without sacrificing accuracy, for the hundreds of values of $\alpha$ required for the analysis.

First, the real frequency grid size can be minimized because we combine non-uniform real frequency grids with hybrid cubic splines in $\omega$ and $u=1/(\omega-\omega_\mu)$ that are well adapted to each other. Cubic splines are the best interpolation approximation for smooth functions on arbitrary grids. They allow minimization of the grid density for a given precision of the integrals. In addition, the hybrid splines are preferable because they give much better accuracy than ordinary splines on the type of highly non-uniform grids we need. Second, using the Matsubara frequency spectral representation of the Green function also helps minimize both the real and imaginary frequency grid sizes. Indeed, because the kernel times the spectral weight can be integrated analytically when $A(\omega)$ is modeled as a piecewise polynomial function, the real frequency grid does not need to be adapted to the kernel, but only to the spectrum. In addition, working with Matsubara frequency functions instead of imaginary-time ones allows a straightforward optimization of the grid on the imaginary time axis, first by replacing the frequencies in the asymptotic region, $i\omega_n > i\omega_{as}$, by constraints on moments and, second, by using an adapted non-uniform Matsubara frequency grid below the cutoff $ i\omega_{as}$ determined during the moment-fitting procedure. Those carefully chosen approximations and optimizations lead to a computational complexity that is only weakly dependent on temperature and spectrum complexity.

The minimization algorithm is another key aspect of the implementation. Since the spectrum changes only slightly between consecutive values of $\alpha$, the minimization algorithm based on Newton's approach is very well adapted. In addition, the singular value decomposition takes advantage of the specific structure of the linear system to solve it more efficiently. The resulting routine is very fast. For example, on a 2GHz laptop, computing the spectrum for about 600 values of $\alpha$ with a grid of 110 real and 95 Matsubara frequencies takes 15 seconds, and the same number of values of $\alpha$ for a grid of 550 real and 510 Matsubara frequencies takes about 8 minutes to compute. As a comparison, the computation time with a generic routine based on the \textit{trust-region-reflective} algorithm that we used in a previous work was a few hours for a few hundred values of $\alpha$ and frequencies.\cite{Bergeron:2011} Speed can be very important for example in the context of modern \textit{ab initio} methods that combine density functional methods with dynamical mean-field theory.\cite{KotliarRMP:2006} The stochastic approach would not be a good choice in that context since the time required for analytical continuation can become longer than the computation time of the Matsubara data.\footnote{Markus Aichhorn, private communication.}

As discussed in section \ref{sec:alpha}, it is a consistency requirement of the maximum entropy approach that $\alpha$ be chosen in the region where $d\log\chi^2/d\log\alpha$ drops as $\alpha$ decreases. This is where the likelihood \eqref{eq:P_G_si_A} is actually valid, where most of the information has been fitted and where noise starts to be fitted as well. However, we still need a criterion to choose precisely $\alpha$ within that crossover region. Since both information and noise are being fitted in that region, that criterion is not obvious in general. For data accurate enough however, the actual value of $\alpha$ within the crossover region is irrelevant since the spectrum does not change within that region. This might apply quite routinely to deterministic numerical calculation, which typically has only small round-off errors, but it will apply more rarely to stochastic results. For this type of data, when the covariance is reasonably accurate, the crossover region is narrow, and the choice of $\alpha$ is still relatively easy. Indeed, as discussed in section \ref{sec:diagn_tools}, local minima in $|dA(\omega)/d\log\alpha|$ are usually present in the crossover region and thus the spectrum will only have small variations over a narrow range of $\log\alpha$. Choosing the maximum of curvature in $\log\chi^2$ as a function of $\gamma\log\alpha$ will thus give very similar results for different values of $\gamma$. Given that the spectrum cannot have large variations when the crossover region is narrow and that there is still some information to fit within that region, it might be preferable to choose $\alpha^*$, the optimal $\alpha$, closer to the noise-fitting region to ensure fitting as much information as possible. The choice $\gamma=0.2$ serves that purpose.

It is always preferable to look at the diagnostic functions to verify that they have the correct properties. At the optimal $\alpha$, the function $\Delta\tilde{G}$ should look like noise and its autocorrelation should have a Kronecker $\delta$ shape, while the curves $A(\omega_i^{samp})$ should have either an extremum or an inflexion point at $\alpha^*$. It may happen that $\Delta\tilde{G}$ seems to contain only noise, but its autocorrelation has a finite width. This indicates that the covariance matrix is not well estimated, in particular if it has been assumed to be diagonal, then a non-diagonal covariance should be used. Another possibility is that $\Delta\tilde{G}$ never becomes completely noisy. This is generally because the grid is not dense enough to produce a good fit, and it must be refined until $\Delta\tilde{G}$ has the correct properties at $\alpha^*$.

The diagnostic tools become even more useful as the error estimates are less accurate and the region where both information and noise are fitted simultaneously becomes wider. They can then be used to choose $\alpha$ such that the most relevant part of the spectrum has converged. For example, if we are more interested in the spectrum around $\omega=0$, one can choose a value of $\alpha$ for which $\Delta\tilde{G}$ is noisy at low Matsubara frequency and $A(\omega)$ is the most stable around $\omega=0$ as a function of $\log\alpha$. To allow the user to choose a different value of $\alpha$ than the one selected automatically, the program saves the results for a certain range of $\alpha$ around that value (by default, one decade above and below). For very bad estimates of the errors however, some compromise must be made because some parts of the spectrum might become quite distorted because too much noise has been fitted, while other parts have not yet converged. The error estimate is therefore a very important part of the data. As for the input data itself, ensuring accuracy of the error estimate is the responsibility of the user. As discussed in section \ref{sec:Bayes_ME} and emphasized in the classic review,~\cite{Jarrell:1996} the maximum entropy formalism assumes that the data obeys Gaussian statistics with accurate estimates of the covariance. This can be checked by histogram methods and by binning procedures such as those available in the ALPS software.~\cite{ALPS:2007} In practice, the data are not perfectly Gaussian and the error estimate is often not very accurate. The diagnostic tools provide at the same time a means to evaluate if the error estimate is good and, to some extent, to compensate if it is not.

The question of whether the spectrum should be taken at a single value of $\alpha$ or averaged over a certain range is certainly relevant for poor estimates of the covariances. Since the software saves the spectrum for a range of $\alpha$ around the selected optimal value, averaging is a possible option. This can however produce unpredictable results since overfitted spectra are mixed with underfitted ones. On the other hand, when choosing the best $\alpha$ for a particular spectral frequency range, one neglects the rest of the spectrum, but ensures that the part of the spectrum that seems the most interesting is as accurate as possible.

The program can be used both in an interactive way, where the user can obtain results in real time and modify parameters to see the effect on the results, or in batch calculations. The level of automatization is sufficient to generate series of calculations without the need to interact with the program, assuming that realistic standard deviations or covariance matrices are provided.

\section{Summary and conclusion}\label{sec:conclusion}

With the entropic prior, only the correlations present in the data are significant,~\cite{Jarrell:1996} while positivity and smoothness of the spectrum are enforced. Given that the maximum entropy approach is very tractable computationally, it is therefore the best practical choice for analytic continuation of noisy numerical data, whether the noise comes from a Monte Carlo procedure or from numerical round-off. The main challenge in that approach is to determine $\alpha$, or a range of values of $\alpha$, for which the result is the most accurate possible. As we have illustrated with examples, consistency with the assumptions of the approach is often sufficient to determine the optimal value of $\alpha$. This is possible because, for realistic estimates of the standard deviation (or covariance), there exist well separated regions where the behavior of $\chi^2$  as a function of $\alpha$ differs, depending on whether only information is being fitted or only noise is added to the fit. Those regions can be clearly identified when $\chi^2(\alpha)$ is plotted on a log-log scale. The optimal $\alpha$ is in the transition region between the information- and the noise-fitting regions.


The information- and noise-fitting regimes also have distinctive signatures (a) in the normalized distance between the data and the fit expressed in the eigenbasis of the covariance matrix, (b) in the autocorrelation of that function, and (c) in sample frequencies of the spectrum plotted with respect to $\log\alpha$. Those functions can therefore be used as diagnostic tools to assess the quality of the fit and also to fine tune the value of the optimal $\alpha$. The spectrum as a function of $\log(\alpha)$ at sample frequencies also informs us on the accuracy of the spectrum. Indeed, high accuracy is indicated by high stability of the spectrum as a function of $\log\alpha$ around the optimal $\alpha$.

The accuracy and efficiency of our implementation is optimized using approximations adapted to the specific numerical challenges of the problem. This includes the use of non-uniform frequency grids, moments constraints, hybrid spline interpolation and piecewise analytical integration of the spectral representation. Those optimizations, combined with a good minimization routine, make it possible to obtains high quality results very quickly, and therefore to check the sensitivity of the results to the grid, the default model, or other parameters, in real time. The level of automatization of the program is also sufficient for batch calculations. Unlike the classic and Bryan approaches, our method to choose the optimal $\alpha$ does not require the default model to be close to the final answer. An annealing procedure is therefore not necessary to obtain reliable results.


To illustrate our approach, we have shown three examples with fermionic Matsubara frequency Green functions: Two fictitious examples with diagonal covariance and one example with real data from a physical simulation, with general covariance. The examples demonstrate that our approach can produce very accurate spectra when the Matsubara data are sufficiently, but realistically, precise. They also show that our implementation can resolve very sharp and very incoherent features at the same time without difficulty, including a spectrum possessing a low-energy peak of width smaller than $\pi T$.


Our maximum entropy implementation is freely available under the GNU \textit{general public license} as a software with a complete user interface. It is written in C++, but uses Python to automatically plot the results and diagnostic functions at the end of the calculation. It comes with a detailed user guide and provides default options that make it easy to use for beginners. The numerous options make the software flexible for the experienced user. The software is fast and can handle fermionic and bosonic input Green functions, self-energies, and correlation functions, with arbitrary covariance, and whether the data are provided in Matsubara frequency or in imaginary time.\cite{OmegaMaxEnt_url}

\acknowledgments
We thank G. Sordi and P. S\'emon for providing the DMFT data used for the example of Fig. \ref{fig:example_appl}. We are also indebted to M. Aichhorn, Erik Koch, and Wei Wu for useful discussions, to S. Allen for expert help with some aspects of numerical analysis and to S. Fuchs and M. Ferrero for sharing some of their data. This work has been supported, by the Natural Sciences and Engineering Research Council of Canada (NSERC), and by the Tier I Canada Research Chair Program (A.-M.S.T.). Some of the simulations were performed on computers provided by Canadian Foundation for Innovation, Minist\`ere de l'Enseignement du Loisir et du Sport (Qu\'ebec), Calcul Qu\'ebec and Compute Canada.

\appendix

\section{Heuristic derivation of the Maximum entropy method}\label{sec:Heuristics}

In this appendix we present a heuristic derivation of the maximum entropy
method that suggests a connection with the stochastic analytical
continuation approach.\cite{Beach:2004} It follows a derivation of statistical ensembles
often found in textbooks and based on ideas from information theory.\cite%
{Jaynes:1957}

Since the spectral weight $A^{\prime }\left( \omega _{i}\right) $ on a
frequency grid $\omega _{i}$ of $M$ points is normalized, it can be thought
of as a probability Here we take%
\begin{equation}
\sum_{i-1}^{M}A^{\prime }\left( \omega _{i}\right) \Delta \omega _{i}=1.
\end{equation}%
In the main text, $A\left( \omega _{i}\right) $ is normalized to $2\pi $
instead$.$ 

Define a stochastic process with statistically independent trials where $%
q_{i}$ is the \textit{a priori} probability to fall on the interval $i$. By repeating
the trials $N$ times with $N$ large, we can discover empirically the
probabilities $q_{i}.$ Let the empirical probability $P_{i}$ for interval $i$
be
\begin{equation}
P_{i}\equiv \frac{n_{i}\left( N\right) }{N}=A^{\prime }\left( \omega
_{i}\right) \Delta \omega _{i}
\end{equation}%
where $n_{i}$ is the number of times that the result $i$ is obtained in the
trials ($i$ is between $1$ and $M$)$.$ The central limit theorem suggests
that in the limit $N\rightarrow \infty $ $\left( N\gg M\right) ,$ the most
probable value of  $n_{i}$ and its average value will be identical and the
corresponding $P_{i}$ will converge to $q_{i},$ in other words we expect 
\begin{equation}
\lim_{N\rightarrow \infty }P_{i}=\lim_{N\rightarrow \infty }\frac{%
	n_{i}\left( N\right) }{N}=q_{i}.  \label{ValueOfP}
\end{equation}%
This can be checked as follows.

The probability $\Gamma \left( n_{1},n_{2},\ldots n_{M}\right) $ to obtain $%
n_{1}$ times the result $1,$ $n_{2}$ times the result $2,$ and so on, is
given by
\begin{equation}
\Gamma \left( n_{1},n_{2},\ldots n_{M}\right) =\frac{N!}{n_{1}!n_{2}!\ldots
	n_{M}!}\left( q_{1}\right) ^{n_{1}}\left( q_{2}\right) ^{n_{1}}\ldots \left(
q_{M}\right) ^{n_{M}}
\end{equation}%
subject to the constraint 
\begin{equation}
\sum_{i=1}^{M}n_{i}=N.  \label{Somme_N}
\end{equation}%
The most probable values can be obtained by maximizing the logarithm.
Stirling's formula is then extremely accurate since $N$ tends to infinity.
Rewritten, it is%
\begin{eqnarray}
\nonumber \ln \Gamma \left( n_{1},n_{2},\ldots \right)  &=&\ln N!-\sum_{i}\ln
n_{i}!+\sum_{i}n_{i}\ln q_{i} \\ 
&\rightarrow &-\sum_{i}n_{i}\ln \frac{n_{i}}{Nq_{i}} \\
&=&-N\sum_{i}P_{i}\ln \left( \frac{P_{i}}{q_{i}}\right) \,,
\end{eqnarray}%
which is called the relative entropy. We just need to maximize
this quantity while satisfying the constraint Eq.(\ref{Somme_N}) which we
rewrite in the form 
\begin{equation}
N\sum_{r}P_{i}=N.
\end{equation}%
Using a Lagrange multiplier, it is easy to find the expected solution $%
P_{i}=q_{i}$ Eq.\ (\ref{ValueOfP}). 

In statistical mechanics, the canonical ensemble is obtained by assuming a)
that $i$ labels microstates that are equiprobable (all $q_{i}$ are
identical) and b) that, in addition to the normalization constraint, there
is another constraint, namely the average energy is fixed, 
\begin{equation}
\sum_{i}E_{i}n_{i}=N\sum_{i}E_{i}P_{i}=N\overline{E}.  \label{ContrainteE}
\end{equation}%
The inverse temperature $\beta $ is the Lagrange multiplier and the final
result for $P_{i}$ is given by the canonical distribution. Note that $N$
drops out of the final answers. It is just a conceptual device.

Coming back to our problem, we must identify the constraints and choose the
\textit{a priori} probabilities $q_{i}.$ There are two constraints, normalisation and
the value of $\chi ^{2}$ which is quadratic in $P_{i}$. The parameter $N/\alpha $ plays the role of the Lagrange
multiplier for the constraint on $\chi^2$. In this case however, we do not know the value of $\chi ^{2}$,
hence we must find a way to determine $\alpha$. In the \textit{historic} approach,
one assumes that $\chi ^{2}$ equals the number of Matsubara frequencies (not
to be confused with $N$), which fixes $\alpha$, but there are several alternatives and our paper
proposes a new one. Concerning the \textit{a priori} probabilities, if we assume that
all real-frequency grid points are equally probable, the $q_{i}$ are all
equal to $1/M$, with $M$ the number of grid points, and the relative
entropy becomes%
\begin{equation}
\ln \Gamma \left( n_{1},n_{2},\ldots \right) =-N\sum_{i}\Delta \omega
_{i}A^{\prime }\left( \omega _{i}\right) \ln \left( \frac{\Delta \omega
	_{i}A^{\prime }\left( \omega _{i}\right) }{1/M}\right) .
\end{equation}%
In the language of maximum entropy methods, our default model is then $%
D\left( \omega _{i}\right) =1/(M\Delta \omega _{i})$ and the choice of grid
determines the default model.\cite{Koch:2014} It is clearly preferable to
choose $q_{i}=D\left( \omega _{i}\right) \Delta \omega _{i}$ which decouples
the choice of grid and of default model. In this case we have,   
\begin{equation}
\ln \Gamma \left( n_{1},n_{2},\ldots \right) =-N\sum_{i}\Delta \omega
_{i}A^{\prime }\left( \omega _{i}\right) \ln \left( \frac{A^{\prime }\left(
	\omega _{i}\right) }{D\left( \omega _{i}\right) }\right)\,,
\end{equation}
which is the form used in the maximum entropy method before imposing the normalization constraint with a Lagrange multiplier. As mentioned in section \ref{sec:kernel}, instead of using a Lagrange multiplier, a constraint on normalization in $\chi^2$ can also be used.

\section{Extracting moments from $G$}\label{eq:extract_moments_app}
This appendix explains the procedure to extract the moments from either $G(i\omega_n)$ or $G(\tau)$.
\subsection{From $G(i\omega_n)$}\label{eq:fit_M_G_omega_n}

For frequencies $\omega_n>W$, where $A(|\omega|>W)\approx 0$, the spectral form of the Green function,
\begin{equation}
G(i\omega_n)=\int_{-\infty}^{\infty}\frac{d\omega}{2\pi}\,\frac{A(\omega)}{i\omega_n-\omega}
\end{equation}
can be written as
\begin{equation}\label{eq:forme_spectrale_G_HF}
\begin{split}
G(i\omega_n)&=\int_{-\infty}^{\infty}\frac{d\omega}{2\pi}\,A(\omega)\left(\frac{1}{i\omega_n}+\frac{\omega}{(i\omega_n)^2}+\frac{\omega^2}{(i\omega_n)^3}+\ldots\right)\\
&=\frac{1}{i\omega_n}\int_{-\infty}^{\infty}\frac{d\omega}{2\pi}\,A(\omega)+\frac{1}{(i\omega_n)^2}\int_{-\infty}^{\infty}\frac{d\omega}{2\pi}\,\omega A(\omega)\\
&\qquad+\frac{1}{(i\omega_n)^3}\int_{-\infty}^{\infty}\frac{d\omega}{2\pi}\,\omega^2 A(\omega)+\ldots\,,
\end{split}
\end{equation}
or
\begin{equation}\label{eq:as_exp_G_app}
G(i\omega_n)=\frac{M_0}{i\omega_n}+\frac{M_1}{(i\omega_n)^2}+\frac{M_2}{(i\omega_n)^3}+\ldots\,,\\
\end{equation}
where $M_j$ is the $j^{th}$ moment of $A(\omega)$. If the Green function is known for a certain frequency range in that asymptotic region, the first $N$ moments can be deduced by fitting the function
\begin{equation}\label{eq:G_asympt}
G_{as}(i\omega_n)=\frac{M_0}{i\omega_n} + \frac{M_1}{(i\omega_n)^2} + \ldots + \frac{M_{N-1}}{(i\omega_n)^N}
\end{equation}
to that part of $G_{in}$, assuming that the terms $M_{N}/(i\omega_n)^{N+1}$ and $M_{N+1}/(i\omega_n)^{N+2}$ can be neglected for frequencies larger than a certain $\omega_L$.

If we write $G_{as}(i\omega_n)$ as
\begin{small}
\begin{multline}
\begin{bmatrix}
\text{Re} [G_{as}(i\omega_L)]\\
\text{Im} [G_{as}(i\omega_L)]\\
\text{Re} [G_{as}(i\omega_{L+1})]\\
\text{Im} [G_{as}(i\omega_{L+1})]\\
\vdots
\end{bmatrix}=\\
\begin{bmatrix}
0 & -\frac{1}{\omega_L^2}  & 0 & \frac{1}{\omega_L^4} & \ldots & 0 & (-1)^{N}\frac{1}{\omega_L^{2N}}\\
-\frac{1}{\omega_L} & 0 & \frac{1}{\omega_L^3} & 0 & \ldots & (-1)^{N}\frac{1}{\omega_L^{2N-1}} & 0\\
0 & -\frac{1}{\omega_{L+1}^2}  & 0 & \frac{1}{\omega_{L+1}^4} & \ldots & 0 & (-1)^{N}\frac{1}{\omega_{L+1}^{2N}}\\
-\frac{1}{\omega_{L+1}} & 0 & \frac{1}{\omega_{L+1}^3} & 0 & \ldots & (-1)^{N}\frac{1}{\omega_{L+1}^{2N-1}} & 0\\
\vdots & \vdots & \vdots & \vdots & \vdots & \vdots & \vdots
\end{bmatrix}\\\times
\begin{bmatrix}
M_0\\
M_1\\
\vdots\\
M_{2N-1}\\
M_{2N}
\end{bmatrix}
\end{multline}
\end{small}
or
\begin{equation}
\mathbf{G}_{as}=\mathbf{X}\mathbf{M}\,,
\end{equation}
the moments are obtained by minimizing
\begin{equation}
\chi_M^2=(\mathbf{G}_{in}^{HF}-\mathbf{X}\mathbf{M})^T \mathbf{C}_{HF}^{-1} (\mathbf{G}_{in}^{HF}-\mathbf{X}\mathbf{M})
\end{equation}
where the superscript or subscript $HF$, for high-frequency, indicates that $\omega_n\geq \omega_L$. To minimize $\chi_M^2$, we have to solve the system $\partial\chi_M^2/\partial M_j=0$, which, using the fact that $\mathbf{C}$ is symmetric, is written as
\begin{equation}\label{eq:sol_M}
\mathbf{X}^T \mathbf{C}_{HF}^{-1} \mathbf{X} \mathbf{M}=\mathbf{X}^T\mathbf{C}_{HF}^{-1}\mathbf{G}_{in}^{HF}\,.
\end{equation}
This linear system can be solved if only moments of low order are taken into account. Otherwise the matrix $\mathbf{X}^T \mathbf{C}_{HF}^{-1} \mathbf{X}$ becomes too ill-conditioned. This is of no consequence in practice however since only the information on a few moments is present in noisy data. To find the onset $\omega_L$, the fit is done repeatedly, while sweeping the starting frequency from small to high until the result of the fit is stable.

We also need the covariance for the moments. It can be found from the covariance of the Green functions
\begin{equation}
\mathbf{C}_{HF}=\mn{(\mathbf{G}_{in}^{HF}-\mn{\mathbf{G}_{in}^{HF}})(\mathbf{G}_{in}^{HF}-\mn{\mathbf{G}_{in}^{HF}})^T}\,.
\end{equation}
By multiplying the above by $\mathbf{X}^T\mathbf{C}_{HF}^{-1}$ from the left and by the transpose of that same matrix from the right and then using \eqref{eq:sol_M} to relate $M$ to $\mathbf{G}_{in}^{HF}$ we obtain
\begin{equation}
\mathbf{C}_{M}=\mn{(\mathbf{M}-\mn{\mathbf{M}})(\mathbf{M}-\mn{\mathbf{M}})^T}=\left(\mathbf{X}^T\mathbf{C}_{HF}^{-1}\mathbf{X}\right)^{-1}\,.
\end{equation}

\subsection{From $G(\tau)$}\label{eq:fit_M_G_tau}

Using the spectral form
\begin{equation}\label{eq:spectr_G_bosons}
G(\tau)=-\int \frac{d\omega}{2\pi} \frac{e^{-\omega\tau} A(\omega)}{1\pm e^{-\beta\omega}}\,,
\end{equation}
where $+$ is for fermions and $-$ is for bosons, we easily obtain the moments
\begin{equation}\label{eq:moments_diffGtau}
\begin{split}
M_0&=-\left[G(0)\pm G(\beta)\right],\\
M_1&=\left[G'(0)\pm G'(\beta)\right],\\
&\vdots\\
M_j&=(-1)^{j+1}\left[G^{(j)}(0)\pm G^{(j)}(\beta)\right],\\
&\vdots
\end{split}
\end{equation}
where $\pm$ has the same correspondence with fermions and bosons as in \eqref{eq:spectr_G_bosons}, and where we defined
\begin{equation}
G^{(j)}(\tau')=\frac{d^j G(\tau)}{d \tau^j}\Big|_{\tau'}\,.
\end{equation}

Now, when $\tau<1/W$, the exponential in the integrand of \eqref{eq:spectr_G_bosons} can be expanded around $\tau=0$, and similarly around $\tau=\beta$ when $\beta-\tau<1/W$ since the Taylor series of $G(\tau)$ converges in those regions of $\tau$. For $\tau<1/W$, we thus have
\begin{equation}\label{eq:Gtau_0}
G(\tau)=G(0)+G'(0)\tau+\frac{G^{(2)}(0)}{2}\tau^2 + \frac{G^{(3)}(0)}{6}\tau^3 + \ldots \,,
\end{equation}
and for $\beta-\tau<1/W$,
\begin{multline}
G(\tau)=G(\beta)-G'(\beta)(\beta-\tau)+\frac{G^{(2)}(\beta)}{2}(\beta-\tau)^2\\
 - \frac{G^{(3)}(\beta)}{6}(\beta-\tau)^3 + \ldots
\end{multline}
or
\begin{equation}\label{eq:Gtau_beta}
G(\beta-\tau)=G(\beta)-G'(\beta)\tau+\frac{G^{(2)}(\beta)}{2}\tau^2- \frac{G^{(3)}(\beta)}{6}\tau^3 + \ldots
\end{equation}
Now, adding and subtracting \eqref{eq:Gtau_0} and \eqref{eq:Gtau_beta} gives
\begin{multline}
G(\tau)+G(\beta-\tau)=\left[G(0)+G(\beta)\right]+\left[G'(0)-G'(\beta)\right]\tau\\
+\frac{1}{2}\left[G^{(2)}(0)+G^{(2)}(\beta)\right]\tau^2 + \frac{1}{6}\left[G^{(3)}(0)-G^{(3)}(\beta)\right]\tau^3 + \ldots \,,
\end{multline}
and
\begin{multline}
G(\tau)-G(\beta-\tau)=\left[G(0)-G(\beta)\right]+\left[G'(0)+G'(\beta)\right]\tau\\
+\frac{1}{2}\left[G^{(2)}(0)-G^{(2)}(\beta)\right]\tau^2 + \frac{1}{6}\left[G^{(3)}(0)+G^{(3)}(\beta)\right]\tau^3 + \ldots \,.
\end{multline}
From \eqref{eq:moments_diffGtau} we see that the moments are obtained from $(n!)$, where $n$ is a moment order, times the coefficients of the polynomial fits to $G(\tau)+G(\beta-\tau)$ and $G(\tau)-G(\beta-\tau)$ in the range $\tau<1/W$. For fermions, we use the even powers of $\tau$ in the fit to $G(\tau)+G(\beta-\tau)$, and the odd powers in the fit to $G(\tau)-G(\beta-\tau)$. For bosons, it is simply the opposite.

As in the fit of the asymptotic form \eqref{eq:G_asympt} to $G(i\omega_n)$, a weighted least-squares fit can also be used here. However, the procedure is slightly heavier because both the order of the polynomial and the number of points must be varied to ensure a stable result is found. If there are enough $G(\tau_i)$ with $\tau_i<1/W$, accurate stationary values of the moments are obtained.

For the least square fit we need the covariances of $G(\tau)+G(\beta-\tau)$ and $G(\tau)-G(\beta-\tau)$. For $G(\tau)+G(\beta-\tau)$,
\begin{equation}
\begin{split}
C_{ij}^+&=\mn{\left[\Delta G(\tau_i)+\Delta G(\beta-\tau_i)\right]\left[\Delta G(\tau_j)+\Delta G(\beta-\tau_j)\right]}\\
&=\mn{\Delta G(\tau_i)\Delta G(\tau_j)}+\mn{\Delta G(\tau_i)\Delta G(\beta-\tau_j)}\\
&\qquad+\mn{\Delta G(\beta-\tau_i)\Delta G(\tau_j)}+\mn{\Delta G(\beta-\tau_i)\Delta G(\beta-\tau_j)}\\
\end{split}
\end{equation}
and for $G(\tau)-G(\beta-\tau)$,
\begin{equation}
\begin{split}
C_{ij}^-&=\mn{\left[\Delta G(\tau_i)-\Delta G(\beta-\tau_i)\right]\left[\Delta G(\tau_j)-\Delta G(\beta-\tau_j)\right]}\\
&=\mn{\Delta G(\tau_i)\Delta G(\tau_j)}-\mn{\Delta G(\tau_i)\Delta G(\beta-\tau_j)}\\
&\qquad-\mn{\Delta G(\beta-\tau_i)\Delta G(\tau_j)}+\mn{\Delta G(\beta-\tau_i)\Delta G(\beta-\tau_j)}\\
\end{split}
\end{equation}
where $\Delta G(\tau_i)=G(\tau_i)-\mn{G(\tau_i)}$. Now $G(\beta)$ is related to $G(0)$, so that elements containing $\Delta G(\beta)$ can be expressed using $\Delta G(0)$.
  
For fermions, we have
\begin{equation}
G_{AB}(\beta)=-\langle \{A,B\}\rangle -G_{AB}(0^+)\,,
\end{equation}
and for bosons,
\begin{equation}
G_{AB}(\beta)=\langle [A,B]\rangle +G_{AB}(0^+)\,.
\end{equation}
Since the (anti-)commutators are canceled when $\mn{G(\beta)}$ is subtracted from $G(\beta)$, we obtain, for $\tau_j\neq \beta$, 
\begin{equation}
\begin{split}
\mn{\Delta G(\beta)\Delta G(\tau_j)}=-\mn{\Delta G(0)\Delta G(\tau_j)}
\end{split}
\end{equation}
for fermions and
\begin{equation}
\begin{split}
\mn{\Delta G(\beta)\Delta G(\tau_j)}=\mn{\Delta G(0)\Delta G(\tau_j)}
\end{split}
\end{equation}
for bosons.

\section{Computing $G(i\omega_n)$ from $G(\tau)$}\label{sec:TF_Gtau}

To obtain $G(i\omega_n)$ from a $G(\tau)$ known only on a discrete set of points, we need an interpolating method. Simply using a discrete Fourier transform would produce a periodic function in $\omega_n$, while Im$[G(i\omega_n)]$ must decrease like $1/\omega_n$ and  Re$[G(i\omega_n)]$ like $1/\omega_n^2$ at high $\omega_n$. We can use a cubic spline as the interpolating method. This has the advantages of recovering automatically the asymptotic behavior for the Fourier transform and of giving good accuracy in the whole frequency range.~\cite{Georges:1996,Bergeron:2011} 

Suppose we have $G(\tau_i)$, for $i=0\ldots N$, we need $4N$ equations to define the spline. If $S_i(\tau)$ is the cubic polynomial in the $i^{th}$ interval, the equations
\begin{equation}
\begin{split}
S_i(\tau_{i-1})&=G(\tau_{i-1})\,,\\
S_i(\tau_{i})&=G(\tau_{i})\,,\\
S_i'(\tau_{i-1})&=S_{i-1}'(\tau_{i-1})\,,\\
S_i''(\tau_{i-1})&=S_{i-1}''(\tau_{i-1})\,,
\end{split}
\end{equation}
for $i=1\ldots N$, give $4N-2$ equations. To obtain the last two equations, we can use the first moment $M_1$ and the second $M_2$ of the spectral function. Using the definitions \eqref{eq:moments_diffGtau}, we have
\begin{equation}\label{eq:spline_moments_eqs}
\begin{split}
S_1'(0)\pm S_N'(\beta)&=M_1\\
S_1''(0)\pm S_N''(\beta)&=-M_2\,.
\end{split}
\end{equation}
where the $+$ sign is for fermions and the $-$ for bosons.

Knowing the spline, the Fourier transform, 
\begin{equation}\label{eq:TF_Gtau}
\begin{split}
G(i\omega_n)&=\int_0^\beta d\tau\, e^{i\omega_n\tau} G(\tau)\\
&\approx \sum_{j=1}^N \int_{\tau_{j-1}}^{\tau_j} d\tau\, e^{i\omega_n\tau} S_j(\tau)
\end{split}
\end{equation}
becomes, after integration by parts three times,
\begin{multline}
G(i\omega_n)=-\frac{G(0)\pm G(\beta)}{i\omega_n}+\frac{S_1'(0)\pm S_N'(\beta)}{(i\omega_n)^2}\\-\frac{S_1''(0)\pm S_N''(\beta)}{(i\omega_n)^3}+\frac{1-e^{i\omega_n\beta/N}}{(i\omega_n)^4}\sum_{j=0}^{N-1}e^{i\omega_n\tau_j} S_{j+1}^{(3)}\,,
\end{multline}
where $S_{j}^{(3)}$ is the third derivative of $S_{j}(\tau)$ and we assume that the imaginary time step is constant. Using \eqref{eq:moments_diffGtau} and \eqref{eq:spline_moments_eqs}, this expression can also be written as 
\begin{multline}\label{eq:Gwn_spline}
G(i\omega_n)=\frac{M_0}{i\omega_n}+\frac{M_1}{(i\omega_n)^2}+\frac{M_2}{(i\omega_n)^3}\\+\frac{1-e^{i\omega_n\beta/N}}{(i\omega_n)^4}\sum_{j=0}^{N-1}e^{i\omega_n\tau_j} S_{j+1}^{(3)}\,,
\end{multline}
The sum in the last term is a discrete Fourier transform that can be computed with a fast Fourier transform (FFT) routine. Using a cubic spline as the interpolation function therefore allows to set the norm and the first two moments of the Fourier transform to the correct values. It is also very efficient because we end up with a FFT.

We also want to obtain the covariance matrix of $G(i\omega_n)$ from the one for $G(\tau)$, which is assumed to be provided as input data. We start with
\begin{equation}
C(\tau_i,\tau_j)=\mn{\left[G(\tau_i)-\bar{G}(\tau_i)\right]\left[G(\tau_j)-\bar{G}(\tau_j)\right]}\,,
\end{equation}
where $\bar{G}(\tau_i)$ is the expected value of $G(\tau_i)$. In this case we replace the continuous Fourier transform by a discrete one. We then have
\begin{equation}
\begin{split}
C_{RR}(\omega_l,&\omega_m)=\\
&\mn{\text{Re}\left[G(i\omega_l)-\bar{G}(i\omega_l)\right]\text{Re}\left[G(i\omega_m)-\bar{G}(i\omega_m)\right]}\\
&\quad\;\;\,=\left(\frac{\beta}{N}\right)^2\sum_{ij} \cos(\omega_l\tau_i)\cos(\omega_m\tau_j) C(\tau_i,\tau_j)\,,
\end{split}
\end{equation}
\begin{equation}
\begin{split}
C_{RI}(\omega_l,&\omega_m)=\\
&\mn{\text{Re}\left[G(i\omega_l)-\bar{G}(i\omega_l)\right]\text{Im}\left[G(i\omega_m)-\bar{G}(i\omega_m)\right]}\\
&\quad\;\;\,=\left(\frac{\beta}{N}\right)^2\sum_{ij} \cos(\omega_l\tau_i)\sin(\omega_m\tau_j) C(\tau_i,\tau_j)\,,
\end{split}
\end{equation}
and
\begin{equation}
\begin{split}
C_{II}(\omega_l,&\omega_m)=\\
&\mn{\text{Im}\left[G(i\omega_l)-\bar{G}(i\omega_l)\right]\text{Im}\left[G(i\omega_m)-\bar{G}(i\omega_m)\right]}\\
&\quad\;\;\,=\left(\frac{\beta}{N}\right)^2\sum_{ij} \sin(\omega_l\tau_i)\sin(\omega_m\tau_j) C(\tau_i,\tau_j)\,.
\end{split}
\end{equation}
Those matrices can also be obtained more efficiently by first calculating 
\begin{equation}
C(\tau_i,\omega_m)=\left(\frac{\beta}{N}\right)\sum_j e^{i\omega_m\tau_j} C(\tau_i,\tau_j)\,,
\end{equation}
then
\begin{equation}
C_{R}(\omega_l,\omega_m)=\left(\frac{\beta}{N}\right)\sum_i e^{i\omega_l\tau_i} \text{Re} \left[C(\tau_i,\omega_m)\right]
\end{equation}
and
\begin{equation}
C_{I}(\omega_l,\omega_m)=\left(\frac{\beta}{N}\right)\sum_i e^{i\omega_l\tau_i} \text{Im} \left[C(\tau_i,\omega_m)\right]\,,
\end{equation}
so that
\begin{equation}
\begin{split}
C_{RR}(\omega_l,\omega_m)&=\text{Re}\left[C_{R}(\omega_l,\omega_m)\right]\,,\\
C_{RI}(\omega_l,\omega_m)&=\text{Re}\left[C_{I}(\omega_l,\omega_m)\right]\,,\\
C_{II}(\omega_l,\omega_m)&=\text{Im}\left[C_{I}(\omega_l,\omega_m)\right]\,.\\
\end{split}
\end{equation}

\section{Estimating the width and weight of a quasiparticle peak}\label{sec:qp_peak_width}

If the spectrum has a well defined peak around $\omega=0$, it is possible, under two conditions, to estimate its width and weight. Those conditions are
\begin{enumerate}
\item There is a frequency interval $W_{qp}<|\omega |<W_{inc}$ where $A(\omega)\approx 0$, such that the contribution to $G(i\omega_n)$ from this part of the spectrum is negligible.
\item A sufficient number of Matsubara frequencies satisfying the condition $W_{qp}<\omega_n<W_{inc}$ exists, such that $G(i\omega_n)$ is represented by a unique Laurent series in that interval.
\end{enumerate}

The treatment is however slightly different for fermions and bosons. In the latter case, what we call the spectrum is in fact $A(\omega)/\omega$. Let us start with the fermionic case.

If the spectrum is made of a well defined peak around zero frequency and a finite frequency part that is usually incoherent, the Green function can be written as
\begin{equation}
\begin{split}
G(i\omega_n)&=\int_{-\infty}^{\infty}\frac{d\omega}{2\pi}\,\frac{A_{qp}(\omega)}{i\omega_n-\omega}+\int_{-\infty}^{\infty}\frac{d\omega}{2\pi}\,\frac{A_{inc}(\omega)}{i\omega_n-\omega}\\
&=G_{qp}(i\omega_n)+G_{inc}(i\omega_n)\,.
\end{split}
\end{equation}
Then, since $A_{qp}(|\omega |>W_{qp})\approx 0$, for frequencies $\omega_n>W_{qp}$, a large $\omega_n$ expansion can be used for the kernel in $G_{qp}(i\omega_n)$, 
\begin{equation}\label{eq:G_qp}
\begin{split}
G_{qp}(i\omega_n)&=\int_{-\infty}^{\infty}\frac{d\omega}{2\pi}\,\frac{A_{qp}(\omega)}{i\omega_n-\omega}\\
&=\int_{-\infty}^{\infty}\frac{d\omega}{2\pi}\,A_{qp}(\omega)\left(\frac{1}{i\omega_n}+\frac{\omega}{(i\omega_n)^2}+\frac{\omega^2}{(i\omega_n)^3}+\ldots\right)\\
&=\frac{M_0^{qp}}{i\omega_n}+\frac{M_1^{qp}}{(i\omega_n)^2}+\frac{M_2^{qp}}{(i\omega_n)^3}+\ldots\,,\\
\end{split}
\end{equation}
where $M_j^{qp}$ is the $j^{th}$ quasiparticle peak moment.

On the other hand, the contribution to $G$ from the incoherent part of the spectrum is
\begin{equation}\label{eq:forme_spectrale_G}
\begin{split}
G_{inc}(i\omega_n)&=\int_{-\infty}^{\infty}\frac{d\omega}{2\pi}\,\frac{A_{inc}(\omega)}{i\omega_n-\omega}\\
&=\int_{-\infty}^{-W_{inc}}\frac{d\omega}{2\pi}\,\frac{A_{inc}(\omega)}{i\omega_n-\omega}+\int_{W_{inc}}^{\infty}\frac{d\omega}{2\pi}\,\frac{A_{inc}(\omega)}{i\omega_n-\omega}\,.
\end{split}
\end{equation}
Now, for $\omega_n<W_{inc}$ the kernel can be expanded in powers of $\omega_n/\omega$,
\begin{multline}
G_{inc}(i\omega_n)=\\
-\int_{-\infty}^{-W_{inc}}\frac{d\omega}{2\pi}\,\frac{A_{inc}(\omega)}{\omega}\left[1+\frac{i\omega_n}{\omega}+\left(\frac{i\omega_n}{\omega}\right)^2+\ldots\right]\\
-\int_{W_{inc}}^{\infty}\frac{d\omega}{2\pi}\,\frac{A_{inc}(\omega)}{\omega}\left[1+\frac{i\omega_n}{\omega}+\left(\frac{i\omega_n}{\omega}\right)^2+\ldots\right],
\end{multline}
which gives
\begin{equation}
G_{inc}(i\omega_n)=-M_{-1}^{inc}-M_{-2}^{inc} (i\omega_n)-M_{-3}^{inc} (i\omega_n)^2+\ldots\,,
\end{equation}
where the $M_{j}^{inc}$s are inverse moments of $A_{inc}(\omega)$.

We finally obtain the following Laurent series form for $W_{qp}<\omega_n<W_{inc}$,
\begin{multline}\label{eq:G_laurent}
G_{LS}(i\omega_n)\approx\ldots-M_{-3}^{inc} (i\omega_n)^2-M_{-2}^{inc} (i\omega_n)-M_{-1}^{inc}\\
+\frac{M_0^{qp}}{i\omega_n}+\frac{M_1^{qp}}{(i\omega_n)^2}+\ldots
\end{multline}

By fitting this form to $G(i\omega_n)$ in the frequency range $W_{qp}<\omega_n<W_{inc}$ using the same method as the one used to fit the moments in Appendix \ref{eq:fit_M_G_tau}, we can extract the moments of the quasiparticle peak. The weight of the peak will then be given by $M_0^{qp}$, its position, by $M_1^{qp}/M_0^{qp}$, and its width by $\sqrt{M_2^{qp}/M_0^{qp}-(M_1^{qp}/M_0^{qp})^2}$.

There is no way however to be sure that the conditions 1 and 2 are fulfilled by looking at $G(i\omega_n)$ only, without trying to fit the form \eqref{eq:G_laurent}. When the weight $M_0^{qp}$ of the peak is large enough however, the imaginary part will seem to diverge at small $\omega_n$. Therefore we know if there is a peak, but we cannot know from the shape of $G(i\omega_n)$ if it is isolated enough from the rest of the spectrum so that the above procedure will work.

In addition, if conditions 1 and 2 are satisfied, the optimal numbers $N_{qp}$ and $N_{inc}$ of moments $M_j^{qp}$ and $M_j^{inc}$, respectively, and the frequency range $[W_{qp},W_{inc}]$ satisfying conditions 1 and 2 are not known in advance. Those parameters are to be determined during the fitting procedure. In our code, the fit is done repeatedly by varying those parameters, and the optimal values are determined by a stationary point. If the fit is attempted and no stationary point is found, then conditions 1 and 2 are not satisfied. In particular, if there are not enough Matsubara frequencies satisfying $W_{qp}<\omega_n<W_{inc}$, the stationary character becomes ill-defined, and no unique result can be found, hence the importance of condition 2.

For bosons, we consider the positive function $\Lambda (\omega)=A(\omega)/\omega$ as the spectrum. If conditions 1 and 2 are satisfied, the spectral form of $G$ thus becomes
\begin{equation}
\begin{split}
G(i\omega_n)&=\int_{-\infty}^{\infty}\frac{d\omega}{2\pi}\,\frac{\omega\Lambda_{qp}(\omega)}{i\omega_n-\omega}+\int_{-\infty}^{\infty}\frac{d\omega}{2\pi}\,\frac{\omega\Lambda_{inc}(\omega)}{i\omega_n-\omega}\\
&=G_{qp}(i\omega_n)+G_{inc}(i\omega_n)\,.
\end{split}
\end{equation}

From that expression, we deduce that the order of the moment associated with a given power of $i\omega_n$ in the Laurent series is increased by one with respect to the fermionic case. We thus obtain directly
\begin{multline}\label{eq:G_laurent_bosons}
G_{LS}^B(i\omega_n)\approx\ldots-M_{-2}^{inc} (i\omega_n)^2-M_{-1}^{inc} (i\omega_n)-M_{0}^{inc}\\
+\frac{M_1^{qp}}{i\omega_n}+\frac{M_2^{qp}}{(i\omega_n)^2}+\ldots
\end{multline}
Note that we still use the $qp$ notation but the low frequency-peak in the case of the conductivity, for example, would be a Drude peak. 

In that case, we do not obtain $M_0^{qp}$ from the fit since for the $qp$ part we are in the large $\omega_n$ limit. However, since
\begin{equation}
G(i\omega_n=0)=-\int_{-\infty}^{\infty}\frac{d\omega}{2\pi}\,\Lambda_{qp}(\omega)\,,
\end{equation}
we have
\begin{equation}
M_0^{qp}=-G(i\omega_n=0)-M_{0}^{inc}\,,
\end{equation}
and we have all the parameters necessary to compute the peak position, width and weight.

\section{Defining the kernel matrix}\label{sec:Kernel_matrix_app}

We want to approximate the form
\begin{equation}\label{eq:forme_spectrale_G_kernel}
G(i\omega_n)=\int_{-\infty}^{\infty}\frac{d\omega}{2\pi}\,\frac{A(\omega)}{i\omega_n-\omega}
\end{equation}
when $A(\omega)$ is known only on a discrete set of frequencies $\omega_j$. To do that we use a cubic spline model for $A(\omega)$ and integrate analytically \eqref{eq:forme_spectrale_G_kernel}. However, instead of using the same form for the polynomials $S_j(\omega)$ in the whole frequency range, we take
\begin{equation}\label{eq:spline_center}
S_j(\omega)=
a_j(\omega-\omega_j)^3+b_j(\omega-\omega_j)^2+c_j(\omega-\omega_j)+d_j
\end{equation}
for $\omega_l<\omega<\omega_r$ and
\begin{equation}\label{eq:spline_left_right}
S_j(u)=
a_j'(u-u_j)^3+b_j'(u-u_j)^2+c_j'(u-u_j)+d_j'
\end{equation}
for $\omega<\omega_l$ and $\omega>\omega_r$, 
\begin{equation}\label{eq:def_u_app}
u=
\begin{cases}
\frac{1}{\omega-\omega_{0l}}\,,\quad &\omega<\omega_l\\
\frac{1}{\omega-\omega_{0r}}\,,\quad & \omega>\omega_r\,.
\end{cases}
\end{equation}
where $\omega_l$ and $\omega_r$ are defined in Appendix~\ref{sec:non_uniform_omega_grid}. We use $\omega$ as the integration variable in the central region and $u$ on the left- and right-hand side regions. Thus, \eqref{eq:forme_spectrale_G_kernel} becomes 
\begin{multline}
G(i\omega_n)=-\frac{1}{2\pi}\int_{0}^{u_l}\frac{du}{u^2}\,\frac{A(u)}{i\omega_n-1/u-\omega_{0l}}\\
+\frac{1}{2\pi}\int_{\omega_l}^{\omega_r}\frac{d\omega}{2\pi}\,\frac{A(\omega)}{i\omega_n-\omega}-\frac{1}{2\pi}\int_{u_r}^{0}\frac{du}{u^2}\,\frac{A(u)}{i\omega_n-1/u-\omega_{0r}}\,,
\end{multline}
where $u_l=1/(\omega_l-\omega_{0l})$ and $u_r=1/(\omega_r-\omega_{0r})$. Let us assume for now that the coefficients in \eqref{eq:spline_center} and \eqref{eq:spline_left_right} are known. On the left-hand side we have 
\begin{widetext}
\begin{equation}\label{eq:integ_G_spectr_left}
\begin{split}
G_{l}(i\omega_n)=-\frac{1}{2\pi}\sum_{j=1}^{L}&\int_{u_j}^{u_{j+1}} du\,\frac{a_j u^3+b_j u^2+c_j u +d_j}{(i\omega_n-\omega_{0l})u^2-u}\\
= -\frac{1}{2\pi}\sum_{j=1}^{L}&\left[\frac{u}{(i\omega_n-\omega_{0l})^2} +\frac{ u^2}{2 (i\omega_n-\omega_{0l})} + \frac{\ln\left[1 - u (i\omega_n-\omega_{0l})\right]}{(i\omega_n-\omega_{0l})^3}\right]_{u_j}^{u_{j+1}}a_j\\
+&\left[\frac{u}{i\omega_n-\omega_{0l}} +\frac{\ln\left[1 - u (i\omega_n-\omega_{0l})\right]}{(i\omega_n-\omega_{0l})^2}\right]_{u_j}^{u_{j+1}}b_j\\
+& \frac{\ln\left[1 - u (i\omega_n-\omega_{0l})\right]}{i\omega_n-\omega_{0l}}\Bigg|_{u_j}^{u_{j+1}}c_j\\
+&\Big(- \ln(u) + \ln\left[1 - u (i\omega_n-\omega_{0l})\right]\Big)\Bigg|_{u_j}^{u_{j+1}}d_j\,,
\end{split}
\end{equation}
where $L$ is the number of intervals on the left and $a_j$, $b_j$, $c_j$ and $d_j$ are the coefficients obtained after expanding \eqref{eq:spline_left_right}. In the center, we obtain
\begin{equation}\label{eq:integ_G_spectr_center}
\begin{split}
G_{c}(i\omega_n)&=\frac{1}{2\pi}\sum_{j=L+1}^{L+M} \int_{\omega_j}^{\omega_{j+1}} \frac{d\omega}{2\pi} \frac{a_j(\omega-\omega_j)^3+b_j(\omega-\omega_j)^2+c_j(\omega-\omega_j)+d_j}{i\omega_n-\omega}\\
&=\frac{1}{2\pi}\sum_{j=L+1}^{L+M} \int_{0}^{\omega_{j+1}-\omega_j} \frac{d\omega}{2\pi} \frac{a_j\omega^3+b_j\omega^2+c_j\omega+d_j}{i\omega_n-\omega_j-\omega}\\
&=\frac{1}{2\pi}\sum_{j=1}^{M}  \left[-(i\omega_n-\omega_j)^2 \omega - (i\omega_n-\omega_j) \frac{\omega^2}{2} - \frac{\omega^3}{3} - (i\omega_n-\omega_j)^3 \ln[-i\omega_n+\omega_j + \omega]\right]_{0}^{\omega_{j+1}-\omega_j}a_j\\
&\quad+\left[-(i\omega_n-\omega_j) \omega - \frac{\omega^2}{2} - (i\omega_n-\omega_j)^2 \ln[-i\omega_n+\omega_j + \omega]\right]_{0}^{\omega_{j+1}-\omega_j}b_j\\
&\quad+\Bigg[-\omega - (i\omega_n-\omega_j) \ln[-i\omega_n+\omega_j + \omega]\Bigg]_{0}^{\omega_{j+1}-\omega_j}c_j\\
&\quad -\ln[-i\omega_n+\omega_j + \omega]\Bigg|_{0}^{\omega_{j+1}-\omega_j}d_j\,,
\end{split}
\end{equation}
\end{widetext}
where $M$ is the number of intervals in the central region. Between the first and the second line, we have made the change of variable $(\omega-\omega_j)\rightarrow \omega$ in each interval. For the right-hand side, the expression for $G_r$ is the same as \eqref{eq:integ_G_spectr_left}, except that $\omega_{0l}$ is replaced with $\omega_{0r}$ and the sum goes from $L+M+1$ to $L+M+N$, where $N$ is the number of intervals in that region. Then, we have $G(i\omega_n)=G_l(i\omega_n)+G_c(i\omega_n)+G_r(i\omega_n)$ and, forming a vector $\Gamma$ with the coefficients $a_j$, $b_j$, $c_j$ and $d_j$ of all the intervals, we obtain, in matrix form,
\begin{equation}
G=\mathcal{K}\Gamma
\end{equation}
where $\mathcal{K}$ is the matrix obtained from expressions \eqref{eq:integ_G_spectr_left}, \eqref{eq:integ_G_spectr_center} and the expression similar to \eqref{eq:integ_G_spectr_left} for the right-hand side of the grid.

The coefficients are the solution to the equations
\begin{equation}\label{eq:eqs_spline}
\begin{split}
S_j(\omega_j)&=A(\omega_j)\\
S_j(\omega_{j+1})&=A(\omega_{j+1})\\
S_j'(\omega_{j+1})&=S_{j+1}'(\omega_{j+1})\\
S_j''(\omega_{j+1})&=S_{j+1}''(\omega_{j+1})\\
j&=1,\ldots, \mathcal{N}\,,
\end{split}
\end{equation}
which provides $4\mathcal{N}-2$ equations, where $\mathcal{N}=L+M+N$ is the total number of intervals. The last two equations can be taken as
\begin{equation}\label{eq:spline_frontiere_u}
\begin{split}
\frac{\partial A(u)}{\partial u}\Big|_{u=0^+}=0\,,\\
\frac{\partial A(u)}{\partial u}\Big|_{u=0^-}=0\,,
\end{split}
\end{equation}
which correspond to
\begin{equation}\label{eq:spline_frontiere}
\begin{split}
\lim_{\omega\rightarrow \infty}\omega^2\partial A(\omega)/\partial\omega&=0\,,\\
\lim_{\omega\rightarrow -\infty}\omega^2\partial A(\omega)/\partial\omega&=0\,.
\end{split}
\end{equation}
The linear system that gives the spline coefficients in terms of the spectral weight has the form $B\Gamma=TA$, where $TA$ is a vector with elements that are equal either to values of $A$ or to zero. Finally,
\begin{equation}
G=\mathbf{K}A\,,
\end{equation}
where
\begin{equation}\label{eq:G_KA}
\mathbf{K}=\mathcal{K}B^{-1}T\,.
\end{equation}
is the kernel matrix.

\section{Minimizing $\frac{1}{2}\chi^2-\alpha S$}\label{sec:minimize_Q}

We want to minimize
\begin{equation}\label{eq:def_Q_app}
\begin{split}
Q&=\frac{\chi^2}{2}-\alpha S\\
&=\frac{1}{2}\left(G-\mathbf{K}A\right)^T \mathbf{C}^{-1}\left(G-\mathbf{K}A\right)\\
&\qquad\qquad+\alpha \sum_i \Delta\omega_i A(\omega_i) \ln \frac{A(\omega_i)}{D(\omega_i)} \,.
\end{split}
\end{equation}
This form of $\chi^2$ requires $O(N_{\omega_n}^2)$ operations to compute, which is not optimal for numerical calculation. Instead, if we diagonalize the covariance $\mathbf{C}$ to obtain
\begin{equation}
\tilde{\mathbf{C}}=\mathbf{U}^\dagger\mathbf{C} \mathbf{U}\,,
\end{equation}
we can rewrite $\chi^2$ as
\begin{equation}\label{eq:chi2_short}
\chi^2=(\tilde{G}-\tilde{\mathbf{K}}A)^T(\tilde{G}-\tilde{\mathbf{K}}A)\,,
\end{equation}
where
\begin{equation}\label{eq:G_K_diag}
\begin{split}
\tilde{G}&=\sqrt{\tilde{\mathbf{C}}^{-1}}\mathbf{U}^\dagger G\,,\\
\tilde{\mathbf{K}}&=\sqrt{\tilde{\mathbf{C}}^{-1}}\mathbf{U}^\dagger \mathbf{K}\,,
\end{split}
\end{equation}
and $\chi^2$ is computed in $O(N_{\omega_n})$ operations instead. Thus,
\begin{equation}\label{eq:Q_start_minim}
\begin{split}
Q&=\frac{1}{2}(\tilde{G}-\tilde{\mathbf{K}}A)^T(\tilde{G}-\tilde{\mathbf{K}}A)\\
&\qquad +\alpha \sum_i \Delta\omega_i A(\omega_i) \ln \frac{A(\omega_i)}{D(\omega_i)}\,,
\end{split}
\end{equation}
is the function to minimize with respect to elements of $A$. This is done by solving $\nabla Q=0$, namely
\begin{equation}\label{eq:sol_A_non-lin}
-\tilde{\mathbf{K}}^T (\tilde{G}-\tilde{\mathbf{K}}A)+\alpha\left(\mathbf{\Delta\omega} \ln \left(\mathbf{D}^{-1}A\right)+\Delta\omega\right)=0\,,
\end{equation}
where $\mathbf{\Delta\omega}$ and $\mathbf{D}$ are the diagonal matrices with $\Delta\omega$ and $D$ as their diagonal, respectively. 

Now, suppose we know the spectrum $A_{j-1}$ at $\alpha_{j-1}$ and want to obtain the spectrum $A_j$ at $\alpha_j$ such that $A_j$ differs only slightly from $A_{j-1}$. We can write $A_j=A_{j-1}+\delta A_j$ and we have
\begin{equation}\label{eq:lin_approx_logA}
\ln \left(\frac{A_j(\omega_i)}{D(\omega_i)}\right)\approx \ln \left(\frac{A_{j-1}(\omega_i)}{D(\omega_i)}\right)+\frac{\delta A_j(\omega_i)}{A_{j-1}(\omega_i)}\,.
\end{equation}
Then, \eqref{eq:sol_A_non-lin} can be written as
\begin{multline}\label{eq:lin_sys_dA_0}
\left[\tilde{\mathbf{K}}^T \tilde{\mathbf{K}}+\alpha_j\mathbf{\Delta\omega}\mathbf{A}_{j-1}^{-1}\right]\delta A_{j}=\tilde{\mathbf{K}}^T (\tilde{G}-\tilde{\mathbf{K}}A_{j-1})\\
-\alpha_j\left(\mathbf{\Delta\omega} \ln \left(\mathbf{D}^{-1}A_{j-1}\right)+\Delta\omega\right)
\end{multline}
where $\mathbf{A}_{j-1}$ is the diagonal matrix made from $A_{j-1}$. This equation can be solved iteratively to obtain $A_j$, corresponding to $\alpha_j$, starting from $A_{j-1}$, corresponding to $\alpha_{j-1}$. This means that, after the first iteration, $A_{j-1}$ in Eq.~\eqref{eq:lin_sys_dA_0} is replaced with $A_j^n+\delta A_j^n$ obtained at the $n^{th}$ iteration. The iterations stop when $\delta A_j$ is vanishingly small.

However, even though the original expression for the entropy part prevents $A$ from being negative, the solution to \eqref{eq:lin_sys_dA_0} can produce negative values because of the approximation \eqref{eq:lin_approx_logA}. Since, in any case, the argument of the $\log$ on the right-hand side of \eqref{eq:lin_sys_dA_0} cannot be smaller than the smallest representable floating point number $f_{min}$ on the computer, one solution to that problem is to smoothly continue the $\log$ by a quadratic function whenever $A_j(\omega_i)/D(\omega_i)<f_{min}$. Although it will not completely prevent negative values of $A(\omega_i)$ from appearing, the parameters of the quadratic function can be chosen to strongly penalize those values, so that only rare negative values of very small magnitude will appear in regions where $A(\omega_i)$ should vanish, values that can thus be considered indeed as vanishing. Therefore, we replace the entropy part in \eqref{eq:Q_start_minim} with
\begin{equation}\label{eq:Sbar}
\bar{S}=\sum_i s_i
\end{equation}
where
\begin{widetext}
\begin{equation}
s_i=
\begin{cases}
- \Delta\omega_iA(\omega_i)\ln \left(\frac{A(\omega_i)}{D(\omega_i)}\right) & A(\omega_i)>A_{min}(\omega_i)\,,\\
-\Big(\frac{c_2(\omega_i)}{2}[A(\omega_i)-A_{min}(\omega_i)]^2+c_1(\omega_i)[A(\omega_i)-A_{min}(\omega_i)]+c_0(\omega_i)\Big) & A(\omega_i)<A_{min}(\omega_i)\,,
\end{cases}
\end{equation}
with $A_{min}(\omega_i)=f_{min}D(\omega_i)$, and thus
\begin{equation}\label{eq:grad_Si_Aj}
\frac{\partial s_i}{\partial A(\omega_j)}=
\delta_{ij}
\begin{cases}
- \left(\Delta\omega_i\ln \left(\frac{A(\omega_i)}{D(\omega_i)}\right)+\Delta\omega_i\right) & A(\omega_i)>A_{min}(\omega_i)\,,\\
- \Big(c_2(\omega_i)[A(\omega_i)-A_{min}(\omega_i)]+c_1(\omega_i)\Big) & A(\omega_i)<A_{min}(\omega_i)\,.
\end{cases}
\end{equation}
\end{widetext}
The matching of the derivatives at $A(\omega_i)=A_{min}(\omega_i)$ gives $c_1(\omega_i)=\Delta\omega_i (\ln(f_{min})+1)$. As for $c_2(\omega_i)$, it must be large enough to avoid negative values of $A$, but not too large, otherwise it would degrade the system's conditioning.

Using \eqref{eq:grad_Si_Aj}, the system \eqref{eq:lin_sys_dA_0} becomes
\begin{equation}\label{eq:lin_sys_dA_1}
\left[\tilde{\mathbf{K}}^T \tilde{\mathbf{K}}+M_j^S\right]\delta A_{j}=\tilde{\mathbf{K}}^T (\tilde{G}-\tilde{\mathbf{K}}A_{j-1})+\alpha_jS_{j-1}'\,,
\end{equation}
where
\begin{equation}
\left(M_j^{S}\right)_{lm}=\alpha_j\delta_{lm}
\begin{cases}
\frac{\Delta\omega_l}{A_{j-1}(\omega_l)} & A(\omega_l)>A_{min}(\omega_l)\,,\\
c_2 &  A(\omega_l)<A_{min}(\omega_l)\,,
\end{cases}
\end{equation}
and $S_{j-1}'=\nabla \bar{S}_{j-1}$ is the gradient of the entropy \eqref{eq:Sbar} evaluated at $A_{j-1}$. 

Because the elements in $M_j^S$ can differ by several orders of magnitude, the system \eqref{eq:lin_sys_dA_1} seems ill-conditioned. This problem can however be overcome by a preconditioning step. If we write the right-hand side as $B_j$, \eqref{eq:lin_sys_dA_1} can be written as
\begin{multline}\label{eq:lin_sys_dA_2}
\left[\sqrt{\left(M_j^S\right)^{-1}}\tilde{\mathbf{K}}^T \tilde{\mathbf{K}}\sqrt{\left(M_j^S\right)^{-1}}+I\right]\sqrt{M_j^S}\delta A_{j}\\
=\sqrt{\left(M_j^S\right)^{-1}}B_j\,,
\end{multline}
where $I$ is the identity matrix. Then, we use a singular value decomposition (SVD) to write
\begin{equation}\label{eq:SVD_Kj}
\tilde{\mathbf{K}}\sqrt{\left(M_j^S\right)^{-1}}=\mathbf{U}_j\kappa_j\mathbf{V}_j^T\,,
\end{equation}
where $\mathbf{U}_j$ and $\mathbf{V}_j$ are orthogonal matrices and $\kappa_j$ is diagonal, so that \eqref{eq:lin_sys_dA_2} can be written
\begin{equation}\label{eq:diag_sys_delta_A_j}
\left[\kappa_j^T \kappa_j+I\right]\mathbf{V}_j^T\sqrt{M_j^S}\delta A_{j}=\mathbf{V}_j^T\sqrt{\left(M_j^S\right)^{-1}}B_j\,.
\end{equation}
Finally, the solution
\begin{equation}\label{eq:delta_A_j_ini}
\delta A_{j}=\sqrt{\left(M_j^S\right)^{-1}}\mathbf{V}_j\left[\kappa_j^T \kappa_j+I\right]^{-1}\mathbf{V}_j^T\sqrt{\left(M_j^S\right)^{-1}}B_j
\end{equation}
is easy to obtain since $\left[\kappa_j^T \kappa_j+I\right]$ is diagonal.

The value of $A_j$ is obtained by iterating \eqref{eq:delta_A_j_ini} until $\delta A_{j}\approx0$. The convergence criteria we use is $\sum_i \Delta\omega_i  |\delta A_{j}|<\text{tol}_{\delta A}$, where $\text{tol}_{\delta A}=10^{-12}$ by default.

Assuming that $N_{\omega_n}<N_\omega$, where $N_{\omega_n}$ is the number of Matsubara frequencies and $N_\omega$, the number real frequencies, the SVD solves the system \eqref{eq:lin_sys_dA_2} in $O(N_{\omega_n}^2N_\omega)$ operations. This is more efficient than a more direct method, like Gauss elimination, that would take $O(N_\omega^3)$ operations. Having the condition $N_{\omega_n}<N_\omega$ is preferable in general, to ensure that there are enough degrees of freedom in the spectrum $A$ to capture all the structures contained in the Green function $G$. If $N_\omega<N_{\omega_n}$, then the SVD computes the solution in $O(N_{\omega_n}N_\omega^2)$.

For the above method to work, the initial value of $\alpha$ must be chosen such that the entropy term dominates. Using the SVD for the default-model region
\begin{equation}
\tilde{\mathbf{K}}\sqrt{e^{-1}\mathbf{\Delta\omega}^{-1}\mathbf{D}}=\mathbf{U}_D\kappa_D\mathbf{V}_D^T\,,
\end{equation}
we use $\alpha_{init}=100\, max(\kappa_D^T\kappa_D)$. This ensures that the solution is very close to $e^{-1}D$ when the computation starts, and that $\alpha_{init}$ is not too high, which would uselessly make the computation longer.

We know that the computation is over when $d\log\chi^2/d\log\alpha$ is very small compared to its typical value in the information-fitting region. The condition we use to stop the computation is when $d\log\chi^2/d\log\alpha$ becomes smaller than a certain fraction, say 0.01, of its maximum value.

\section{Recursive solution for the spectrum $A$ in the limit of small changes in $\alpha$}\label{sec:alpha_opt}

In this appendix we derive Eq.~\eqref{eq:delta_A_j_main} which is useful to understand the changes of $\chi^2$ in the noise-fitting region. This expression is not used in the code itself. Defining
\begin{equation}
\alpha_j=\alpha_{j-1}-\Delta\alpha_j\,,
\end{equation}
expression \eqref{eq:lin_sys_dA_0} can be rewritten as
\begin{multline}
\left[\tilde{\mathbf{K}}^T \tilde{\mathbf{K}}+\alpha_j\mathbf{\Delta\omega}\mathbf{A}_{j-1}^{-1}\right]\delta A_{j}=\tilde{\mathbf{K}}^T (\tilde{G}-\tilde{\mathbf{K}}A_{j-1})\\
-\alpha_{j-1}\left(\mathbf{\Delta\omega} \ln \left(\mathbf{D}^{-1}A_{j-1}\right)+\Delta\omega\right)\\
+\Delta\alpha_{j}\left(\mathbf{\Delta\omega} \ln \left(\mathbf{D}^{-1}A_{j-1}\right)+\Delta\omega\right)\,.
\end{multline}
Now, if $A_{j-1}$ is the spectrum that minimizes \eqref{eq:def_Q_app} when $\alpha=\alpha_{j-1}$, from \eqref{eq:sol_A_non-lin}, the first two terms on the right-hand side cancel each other and we obtain
\begin{widetext}
\begin{equation}\label{eq:delta_A_j}
\delta A_{j}=\Delta\alpha_j\left[\tilde{\mathbf{K}}^T \tilde{\mathbf{K}}+\alpha_j\mathbf{\Delta\omega}\mathbf{A}_{j-1}^{-1}\right]^{-1}\left(\mathbf{\Delta\omega} \ln \left(\mathbf{D}^{-1}A_{j-1}\right)+\Delta\omega\right)\,.
\end{equation}
Hence, in the limit of small $\Delta\alpha$, the variation in $A$ does not depend explicitly on the input Green function, and the spectrum at $\alpha_j$ is given by
\begin{equation}\label{eq:A_j_recursive}
A_j=A_{j-1}+\Delta\alpha_j\left[\tilde{\mathbf{K}}^T \tilde{\mathbf{K}}+\alpha_j\mathbf{\Delta\omega}\mathbf{A}_{j-1}^{-1}\right]^{-1}\left(\mathbf{\Delta\omega} \ln \left(\mathbf{D}^{-1}A_{j-1}\right)+\Delta\omega\right)\,.
\end{equation}
\end{widetext}

The usefulness of this result is that it shows that the variation $\delta A$ is a smooth function of $\omega$ if $\Delta\omega$, $D$ and the current spectrum $A$ are themselves smooth. Therefore $A$ will have the tendency to remain a smooth function when $\alpha$ decreases, even in the overfitting regime. This behavior in the evolution of $A$ as a function of $\alpha$ is very useful in practice. Indeed, on one hand, a smooth spectrum is generally desirable and, on the other hand, because it prevents fitting the noise easily, it provides an easy and efficient method for choosing the optimal value of $\alpha$, as discussed in Sec. \ref{sec:alpha}.

\section{Non-uniform real frequency grid}\label{sec:non_uniform_omega_grid}

Let us define three regions for the grid: from $-\infty$ to $\omega_l$, from $\omega_l$ to $\omega_r$ and from $\omega_r$ to $\infty$. Then, let us assume that the step in the central region varies smoothly between $\Delta\omega_l$ and $\Delta\omega_r$. Now for $\omega\leq\omega_l$ we define
\begin{equation}\label{eq:def_u_l}
u=\frac{1}{\omega-\omega_{0l}}
\end{equation}
and, for $\omega\geq\omega_r$,
\begin{equation}
u=\frac{1}{\omega-\omega_{0r}}.
\end{equation}
Let us consider the left region. Assuming a constant step $\Delta u$, and that $\omega_{min}$ is the first finite frequency of the grid on the left, we want to determine $\omega_{0l}$, and the number of values of $u$, $N_{u}$, such that $\Delta\omega$ is equal on the left and on the right side of $\omega_l$. Those conditions give the step
\begin{equation}\label{eq:Delta_u_l}
\begin{split}
\Delta u&=\frac{1}{\omega_{l}-\omega_{0l}}-\frac{1}{\omega_l-\Delta\omega_l-\omega_{0l}}\\
&=-\frac{\Delta\omega_l}{\left(\omega_l-\Delta\omega_l-\omega_{0l}\right)\left(\omega_{l}-\omega_{0l}\right)}\,,
\end{split}
\end{equation}
and the grid
\begin{equation}\label{eq:def_u_l_Delta_u}
u=\Delta u,2\Delta u,\ldots, N_{u}\Delta u\,,
\end{equation}
so that, given that the first finite value of $u$ is $\Delta u$,
\begin{equation}\label{eq:def_omega_min}
\begin{split}
\omega_{min}&=\frac{1}{\Delta u}+\omega_{0l}\\
&=-\frac{\left(\omega_l-\Delta \omega_l-\omega_{0l}\right)\left(\omega_l-\omega_{0l}\right)}{\Delta\omega_l}+\omega_{0l}\,,
\end{split}
\end{equation}
and thus,
\begin{equation}\label{eq:def_omega_0_l_p}
\omega_{0l}=\omega_l+\sqrt{\Delta\omega_l(\omega_l-\omega_{min})}\,,
\end{equation}
which is the solution such that $u$ stays finite for all $\omega<\omega_l$. For now, $\omega_{0l}$ is a temporary value. For $N_{u}$, from \eqref{eq:def_u_l_Delta_u}, the definition of $u$ \eqref{eq:def_u_l}, and the fact that $\Delta\omega$ is equal on each side of $\omega_l$, we have
\begin{equation}
N_{u}\Delta u=\frac{1}{\omega_l-\Delta\omega_l-\omega_{0l}}\,,
\end{equation}
which gives, using \eqref{eq:Delta_u_l} for $\Delta u$,
\begin{equation}
N_{u}=\frac{\omega_{0l}-\omega_l}{\Delta\omega_l}\,.
\end{equation}
Now, if we substitute \eqref{eq:def_omega_0_l_p} in that expression, $N_{u}$ in general is not an integer. Therefore, we define $N_{u}$ as
\begin{equation}
N_{u}=\text{ceil}\left(\frac{\omega_{0l}-\omega_l}{\Delta\omega_l}\right)\,.
\end{equation}
We then redefine $\omega_{0l}$ as
\begin{equation}
\omega_{0l}=\omega_l+N_{u}\Delta\omega_l\,,
\end{equation}
and $\omega_{min}$ as well using \eqref{eq:def_omega_min}. The procedure is the same for the right region of the grid.

\begin{figure}
\includegraphics[width=0.9\columnwidth]{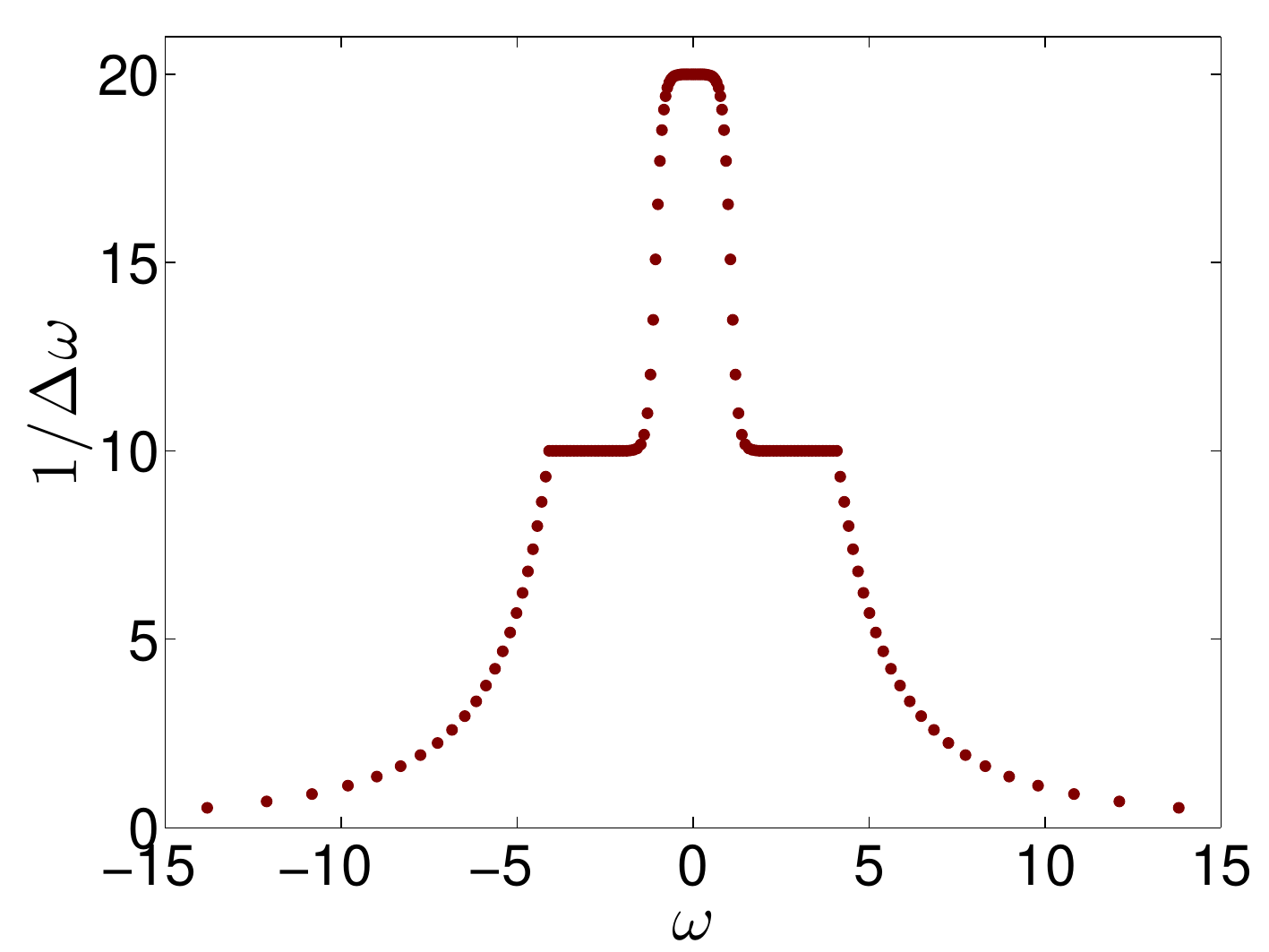}\\
\caption{Example of frequency grid density $1/\Delta \omega$ as a function of $\omega$ around the central part of the grid. In this example, the user has provided parameters defining a grid with different constant frequency steps in different intervals in the form of Eq.~\eqref{Eq:Grid}. The program matches smoothly the steps in the different regions using a hyperbolic tangent shape. Then the distance $\Delta u =1/(\omega_{i+1}-\omega_{\mu})-1/(\omega_{i}-\omega_{\mu})$, (with $\omega_{\mu}$ fixed by the cutoff), is constant outside the central interval $[-5,5]$. }
\label{fig:Delta_omega}
\end{figure}

Figure \ref{fig:Delta_omega} shows $1/\Delta\omega$ as a function of $\omega$ around the central region for such a grid. The central region is also non-uniform, with a step that varies smoothly between the subregions of constant step defined by the user in the form given by Eq.~\eqref{Eq:Grid}. In practice it is recommended that the ratio between step sizes of consecutive frequency ranges does not exceed a factor of four. Large ratios can lead to spurious oscillations in the spectral weight.

This is the natural grid corresponding to the spline used to model $A(\omega)$, and described in Appendix \ref{sec:Kernel_matrix_app}. It keeps the number of points very low while keeping the integrals of $A(\omega)$ very accurate for smooth spectra. Using a continuous step is also necessary to avoid spurious oscillations in the results.


\section{Non-uniform Matsubara frequency grid}\label{sec:non_uniform_Matsubara_grid_def}

Let us assume an initial number of Matsubara frequencies $N_0+1$, where $N_0=2^r$ with $r$ an integer. A non-uniform Matsubara frequency grid can be generated as follows. Define

$N_1=\frac{N_0}{2^m}$ : $N_1+1$ is the number of adjacent frequencies close to the first frequency $i\omega_{0}$, where $m$ is an integer that satisfies $0 \leq m < r$,

$N_2=\frac{N_1}{2}$ : number of frequencies in each subinterval with a fixed spacing between Matsubara frequencies, and

$N=N_1+1+mN_2$ :  total number of frequencies. 

\begin{figure}
\includegraphics[width=0.92\columnwidth]{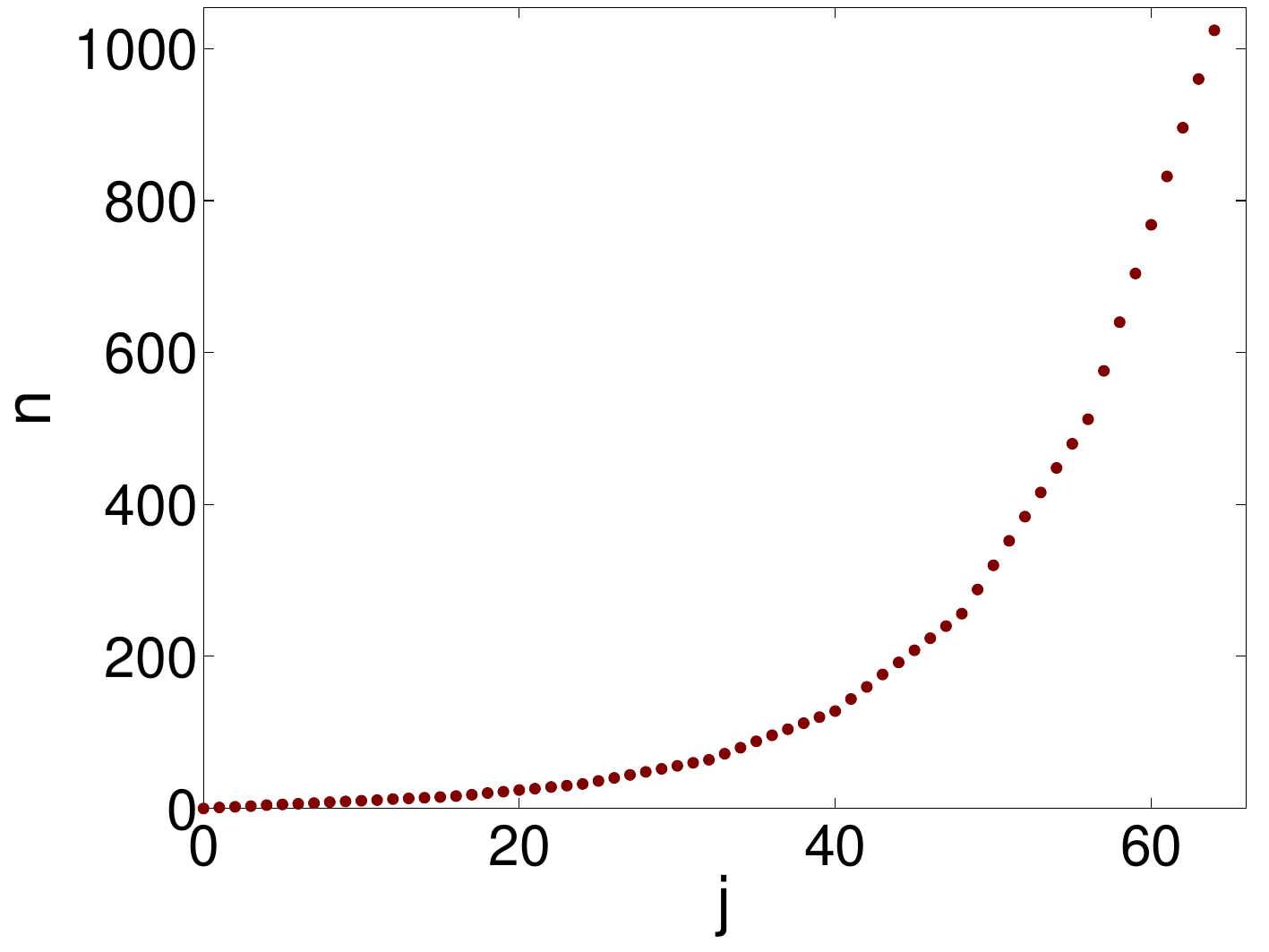}\\
\caption{Example of non-uniform Matsubara frequency index grid. The vertical axis is the index of the Matsubara frequency and the horizontal axis numbers the frequencies that are kept.}
\label{fig:n_omega}
\end{figure}

Then, the Matsubara frequencies $i\omega_{n(j)}$ are defined using
\begin{equation}
n(j)=
\begin{cases}
j\,,\qquad j=0,\ldots, N_1-1,\\
\,\\
2^{l_j+1}mod(j-N_1,N_2) + N_12^{l_j}\,,\\
\text{where } l_j=floor\left(\frac{j-N_1}{N_2}\right)\,,\quad j=N_1,\ldots,  N-1\,,
\end{cases}
\end{equation}
Figure \ref{fig:n_omega} shows $n(j)$ for $N_0=1024$ and $N_1=16$, which gives a total of 65 frequencies. 

The above non-uniform grid is used if $N_{\omega_n}$, the number of Matsubara frequencies in the data, exceeds a maximum default value $N_d$ chosen typically between a few hundreds and a thousand. This default value can be modified by the user. To generate the non-uniform grid, we first choose $r$ such that $(N_0=2^r) \geq N_{\omega_n}-1$ is satisfied. For a given $m$, if $N_0+1$ is strictly larger than $ N_{\omega_n}$ then the grid must be truncated to obtain a maximum frequency $i\omega_{max}$ smaller or equal to the maximum frequency available in the data $i\omega_{N_{\omega_n}-1}$. Given that procedure, $m$ is chosen such that the number of Matsubara frequencies does not exceed $N_d$.

%

\end{document}